\documentclass[manuscript,screen,nonacm]{acmart}
\usepackage{url,textcomp,placeins}
\usepackage{tabularx,multirow,threeparttable,makecell}
\usepackage{mathtools,amsmath,amsfonts,array,interval}
\usepackage{adjustbox}
\usepackage[inline]{enumitem}
\usepackage[ruled]{algorithm2e}
\usepackage[italic]{mathastext}
\usepackage[normalem]{ulem}
\usepackage[most]{tcolorbox}
\usepackage{xcolor}
\usepackage[capitalise]{cleveref} % noabbrev
\usepackage{marginnote}

\Crefname{figure}{Fig.}{Figs.}
\Crefname{equation}{Eq.}{Eq.}

\usepackage[acronym,nogroupskip]{glossaries}

\renewcommand{\glossarysection}[2][]{}
\setacronymstyle{long-short}

\newacronym{ours}{CoCoMagic}{\underline{Co}operative \underline{Co}-evolutionary differential and \underline{M}etamorphic \underline{A}utomated \underline{G}enerator for \underline{I}nterpretable \underline{C}ases in autonomous systems}
\newacronym{ads}{ADS}{Autonomous Driving System}
\newacronym{aed}{AED}{Average Execution Distance}
\newacronym{ai}{AI}{Artificial Intelligence}
\newacronym{as}{AS}{Autonomous System}
\newacronym{auc}{AUC}{Area Under the Curve}
\newacronym{ccea}{CCEA}{Cooperative Co-Evolutionary Algorithm}
\newacronym{dl}{DL}{Deep Learning}
\newacronym{ds}{DS}{Distinct Solutions}
\newacronym{dt}{DT}{Differential Testing}
\newacronym{ea}{EA}{Evolutionary Algorithm}
\newacronym{gbdt}{GBDT}{Gradient Boosted Decision Tree}
\newacronym{mae}{MAE}{Mean Absolute Error}
\newacronym{ml}{ML}{Machine Learning}
\newacronym{mr}{MR}{Metamorphic Relation}
\newacronym{mt}{MT}{Metamorphic Testing}
\newacronym{rf}{RF}{Random Forest}
\newacronym{ri}{RI}{Runtime Initialization}
\newacronym{rq}{RQ}{Research Question}
\newacronym{rs}{RS}{Random Search}
\newacronym{sga}{SGA}{Standard Genetic Algorithm}

\theoremstyle{definition}
\newtheorem{definition}{Definition}

\newtcolorbox[auto counter]{finding}{enhanced,
  attach boxed title to top text left={yshift=-2mm},
  fonttitle=\bfseries, title=Answer to RQ\thetcbcounter}
\newcommand{\Finding}[1]{\begin{finding}#1\end{finding}}

\newcolumntype{Y}{>{\centering\arraybackslash}X}

\newif\ifannotated%
\annotatedfalse%  % Set this to \annotatedfalse for a clean version
\newcommand{\annotate}[2][0cm]{%
  \ifannotated%
    \checkoddpage%
    \ifoddpage%
      \marginnote{\scriptsize\bfseries\color{blue}#2}[#1]%
    \else
      \reversemarginpar%
      \marginnote{\scriptsize\bfseries\color{blue}#2}[#1]%
      \normalmarginpar%
    \fi
  \fi
}
\newcommand{\addedtext}[1]{%
  \ifannotated{%
    \begingroup%
    \renewcommand*{\glstextformat}[1]{\textcolor{blue}{##1}}%
    \hypersetup{citecolor=blue,linkcolor=blue}%
    \color{blue}#1%
    \hypersetup{citecolor=black,linkcolor=black}%
    \endgroup%
  }%
  \else#1\fi%
}
\newcommand{\deletedtext}[1]{%
  \ifannotated%
    \begingroup%
    \renewcommand*{\glstextformat}[1]{\textcolor{red}{##1}}%
    \hypersetup{citecolor=red,linkcolor=red}%
    \color{red}\sout{#1}%
    \hypersetup{citecolor=black,linkcolor=black}%
    \endgroup%
  \fi%
}

\newif\ifannotatedsec%
\annotatedsecfalse%  % Set this to \annotatedsecfalse for a clean version

\newcommand{\addedtextsec}[1]{%
  \ifannotatedsec{%
    \begingroup%
    \renewcommand*{\glstextformat}[1]{\textcolor{blue}{##1}}%
    \hypersetup{citecolor=blue,linkcolor=blue}%
    \color{blue}#1%
    \hypersetup{citecolor=black,linkcolor=black}%
    \endgroup%
  }%
  \else#1\fi%
}
\newcommand{\deletedtextsec}[1]{%
  \ifannotatedsec%
    \begingroup%
    \renewcommand*{\glstextformat}[1]{\textcolor{red}{##1}}%
    \hypersetup{citecolor=red,linkcolor=red}%
    \color{red}\sout{#1}%
    \hypersetup{citecolor=black,linkcolor=black}%
    \endgroup%
  \fi%
}

\begin{document}

\title{Constrained Co-evolutionary Differential Metamorphic Testing for Autonomous Systems with an Interpretability Approach}

\author{Hossein Yousefizadeh}
\email{hyous028@uottawa.ca}
\affiliation{
  \institution{School of EECS, University of Ottawa}
  \city{Ottawa}
  \state{Ontario}
  \country{Canada}
}
\authornote{Hossein Yousefizadeh and Shenghui Gu contributed equally to this work.}

\author{Shenghui Gu}
\email{sgu2@uottawa.ca}
\affiliation{
  \institution{School of EECS, University of Ottawa}
  \city{Ottawa}
  \state{Ontario}
  \country{Canada}
}
\authornotemark[1]

\author{Lionel C. Briand}
\email{lbriand@uottawa.ca}
\affiliation{
  \institution{School of EECS, University of Ottawa}
  \city{Ottawa}
  \state{Ontario}
  \country{Canada}
}
\affiliation{
  \institution{Lero SFI centre for Software Research, University of Limerick}
  \city{Limerick}
  \country{Ireland}
}

\author{Ali Nasr} %Correct
\email{ali.nasr2@huawei.com} %Correct
\affiliation{
  \institution{Waterloo Research Center of Huawei Technologies Canada Co., Ltd.} %Correct
  \city{Waterloo} %Correct
  \state{Ontario} %Correct
  \country{Canada} %Correct
}

\renewcommand{\shortauthors}{Yousefizadeh et al.}

\begin{abstract}
    Autonomous systems, such as autonomous driving systems, evolve rapidly through frequent updates, risking unintended behavioral degradations. Effective system-level testing is challenging due to the vast scenario space, the absence of reliable test oracles, and the need for practically applicable and interpretable test cases. We present \emph{CoCoMagic}, a novel automated test case generation method that combines metamorphic testing, differential testing, and advanced search-based techniques to identify behavioral divergences between versions of autonomous systems.
    \emph{CoCoMagic} formulates test generation as a constrained cooperative co-evolutionary search, evolving both source scenarios and metamorphic perturbations to maximize differences in violations of predefined metamorphic relations across versions. Constraints and population initialization strategies guide the search toward realistic, relevant scenarios.
    An integrated interpretability approach aids in diagnosing the root causes of divergences.
    We evaluate \emph{CoCoMagic} on an end-to-end autonomous driving system, \textsc{InterFuser}, within the \textsc{Carla} virtual simulator.
    Results show significant improvements over baseline search methods, identifying more distinct high-severity behavioral differences while maintaining scenario realism.
    The interpretability approach provides actionable insights for developers, supporting targeted debugging and safety assessment.
    \emph{CoCoMagic} offers an efficient, effective, and interpretable way for the differential testing of evolving autonomous systems across versions.
\end{abstract}




\maketitle

\section{Introduction}

\Glspl{as} are complex software-hardware systems designed to autonomously make real-time operating decisions in open contexts by processing data from a multitude of inputs. These systems are mostly built with \gls{ai} units that learn from vast datasets and may, in the future, continuously evolve through reinforcement learning.
As \glspl{as} continue to advance, the need for systematic testing becomes increasingly important, to maintain public trust and to satisfy regulatory requirements~\cite{tang2023survey,khan2023safety,lou2022testing}.
Concurrently, \gls{as} development relies on iterative refinement, incorporating new real-world data, optimized network architectures, hyperparameter tuning, and algorithmic tweaks to boost \gls{as} performance across diverse operating conditions~\cite{yang2020iterative,li2022robust,yao2024enhancing}.
However, each update can inadvertently introduce changes that undermine previously validated behaviors or functionalities.
Assumptions about system behavior or responses that once appeared benign may no longer hold under new or even familiar conditions, leading to unsafe outcomes.
Without systematic testing, such behavioral degradations can remain hidden, compromising system reliability and safety.
Consequently, there is an urgent need for effective testing methods that systematically identify critical violations triggered by \gls{as} updates, particularly as modern \glspl{as} are increasingly built on complex \gls{ai} models whose behavior can be difficult to predict and verify.

Meeting this need is complicated by the fact that the behavior of \glspl{as} across diverse scenarios is both complex and unpredictable, making it challenging to detect degradations and trace their root causes.
% The complexity and unpredictability of \gls{as} behavior across diverse scenarios make it challenging to detect behavioral degradations and trace their root causes.
Variations in environmental conditions, task parameters, and system inputs interact with the uncertainty inherent in \gls{ml} or \gls{dl} models to produce a vast array of possible \gls{as} responses.
In practice, a single update, whether to the model, the control policy, or the training data, can affect hundreds of interconnected components\annotate{Comment 3.1}\deletedtext{ so that}\addedtext{. Consequently,} a failure observed in one scenario\addedtext{, whether a safety violation or a subtle or indirect unsafe behavior,} may \deletedtext{originate}\addedtext{arise} from a subtle modification elsewhere.
Isolating which change triggered the degradation, therefore, requires re-creating and re-evaluating numerous complex test cases, a process that demands extensive computational resources and careful methodological design.

In addition, the main obstacle in testing \glspl{as} is the lack of dependable test oracles.
This challenge arises because, in many cases, especially in complex or open-ended scenarios, formal specifications or expected outcomes are either absent or difficult to establish~\cite{fraser2013whole,riccio2020testing}.
Unlike traditional software, where a function's input and output can be precisely defined, \glspl{as} often operate in environments where a spectrum of behaviors may be acceptable for a given situation.
For instance, a robot navigating a crowded space may choose different but equally valid paths, or an automated control system may apply varying control signals within a safe operating range.
Human experts sometimes disagree on the correct maneuver in novel situations, and encoding every nuance of safe behavior into a fixed rule set is generally infeasible.
Without clear pass/fail criteria, automated testing cannot definitively flag deviations as degradations, making it nearly impossible to scale testing across the full range of real-world operating conditions.

Furthermore, even when degradations can be detected without a formal oracle, the practical relevance of such detections depends heavily on the realism of the scenarios in which they occur.
While artificially constructed edge cases can expose system weaknesses, failures observed in \annotate{Comment 3.3}\deletedtext{realistic }scenarios \addedtext{that closely resemble conditions the system has recently or commonly encountered during real-world operation} are often of greater practical importance\deletedtext{, especially scenarios resembling situations commonly encountered during system execution}.
Such scenarios are inherently more likely to increase the probability that any detected failure will manifest after deployment.
\addedtext{We refer to such scenarios as ``realistic'' throughout this paper.}
Therefore, prioritizing failure discovery in realistic settings enhances the \addedtext{practical} relevance of the testing process and helps focus engineering efforts on risks more likely to impact users.
% \addedtext{It is worth noting that the degree of realism prioritization is configurable, allowing practitioners to adjust this trade-off based on their testing objectives.}

To address these challenges, we propose \emph{\gls{ours}}, an effective and efficient automated test case generation method that integrates \gls{dt} with \gls{mt}~\cite{chen2018metamorphic} to expose behavioral divergences between \gls{as} versions.
\annotate{Comments 1.1 and 2.1}\addedtext{In \emph{\gls{ours}}, \gls{dt} serves as the primary mechanism for efficiently detecting diverse instances of behavioral divergence by comparing the outputs of two \glspl{as} under identical test inputs. However, \gls{dt} alone cannot address the oracle problem, as it does not assess the impact of detected behavioral differences on system safety. \Gls{mt} is therefore integrated into \emph{\gls{ours}} to provide a mechanism for assessing the safety implications of such behavioral changes.}
\deletedtext{
\Gls{mt} offers a practical approach for evaluating system behavior through MRs, which define expected relationships between inputs and their corresponding outputs.
Instead of verifying absolute correctness, \gls{mt} checks for consistency across related test cases, making it particularly well-suited for domains such as autonomous driving, where correct behavior is often context-dependent and ambiguous~\cite{tang2023survey, zhou2019metamorphic}, and where labeled data may be insufficient, since it does not require ground-truth outputs for every test case.
}
\addedtext{
\Gls{mt} provides a practical approach to evaluating system behavior using \glspl{mr}, which define expected relationships between related inputs and outputs rather than verifying the absolute correctness of individual test outputs.
As a result, it is particularly well-suited for domains such as autonomous driving, where correct behavior is often context-dependent and ambiguous~\cite{tang2023survey, zhou2019metamorphic}, and where labeled test data may be impractical or too expensive.
}
\annotate{Comments 1.1 and 3.2}\addedtext{Furthermore, unlike threshold-based safety metrics, which can only detect specific types of safety violations, \glspl{mr} can capture a broader range of unsafe behaviors, including subtle and indirect ones that do not necessarily lead to immediate safety violations but indicate underlying instability or degradation in system performance that could eventually compromise safety.}
% \addedtext{In summary, \gls{dt} assesses the conditions where the two versions behave differently, while
% \gls{mt} ensures these differences are safety-relevant under expected behavioral standards.}
\deletedtext{\Gls{dt} exposes discrepancies by executing multiple versions of a system on the same input and comparing their outputs.}

In the context of evolving \glspl{as}, where system updates may subtly alter behavior, \gls{dt} coupled with \gls{mt} can help reveal deviations that violate the expected output relations defined by \glspl{mr}.
Such violations may indicate degradations or unexpected side effects introduced during model fine-tuning or other updates.
We cast test case generation as a search problem and seek inputs in which the extent of violation of the expected output relation differs markedly across versions.
\addedtextsec{
This paired formulation is important because the goal is not merely to find severe violations in either version, but to identify test conditions under which the update changes the system's safety-relevant behavior.
}
By systematically identifying such inputs, our method yields effective yet diverse test cases that highlight safety risks and unintended behaviors introduced by the continuous evolution of \glspl{as}.
We also incorporate constraints and a population initialization strategy to guide the search toward realistic test cases that reflect real-world conditions.
\annotate{Comment 3.5}\addedtext{Specifically, the population is initialized with scenarios derived from real-world execution data, providing a realistic starting point, while a penalty term in the fitness function discourages the generation of test cases that deviate significantly from observed operational conditions.}
This design ensures that the generated cases are relevant and applicable to practical scenarios, particularly in contexts where the updated system is expected to operate.
\addedtext{The extent to which realism is prioritized can also be adjusted, allowing practitioners to balance realism against other testing objectives based on their specific needs.
}

To further enhance the utility of \emph{\gls{ours}}, we integrate an interpretability approach to support the analysis and understanding of the root causes of the identified behavioral divergences.
This approach allows us to analyze the generated test cases and gain insights into the underlying factors contributing to the observed discrepancies, thereby helping developers address potential safety risks more effectively,
% and assess the overall impact of updates on system behavior,
producing cases that are not only effective at exposing divergences but also understandable and interpretable.

% They are representative, at the very end of the complexity spectrum, of the more general category of autonomous systems operating in open contexts.
% Though we focus on \glspl{ads} in this paper, no element of \emph{\gls{ours}} is specific to that domain and we expect our approach to apply to other autonomous systems.
While our method is designed for general applicability to \glspl{as} operating in open and complex environments, this work focuses on \glspl{ads} as a representative and challenging case study.
End-to-end \glspl{ads} exemplify one of the most complex forms of \glspl{as}, positioned at the higher end of the complexity spectrum~\cite{burton2020mind}.
To evaluate the effectiveness and efficiency of \emph{\gls{ours}}, we conducted experiments using the \textsc{Carla} simulator~\cite{dosovitskiy17carla} with the \textsc{InterFuser} \gls{ads}~\cite{shao2022safety}.
The results show that \emph{\gls{ours}} significantly outperforms baseline methods in identifying severe behavioral discrepancies, with consistent advantages across a wide range of search budgets and configurations.
This superior performance highlights \emph{\gls{ours}} as a promising and scalable solution for testing \glspl{ads}.
The use of constraints and population initialization effectively guides the search toward realistic and practically relevant test cases, thereby enhancing the applicability of the generated test cases.
Moreover, the interpretability approach integrated into \emph{\gls{ours}} enables a detailed analysis of behavioral divergences, offering valuable insights into the underlying factors contributing to the observed discrepancies.

The major contributions of this paper are summarized as follows:

\begin{itemize}
    \item We propose \emph{\gls{ours}}, a novel method for automated test case generation that combines \gls{mt} and \gls{dt} to expose behavioral divergences between different versions of \glspl{as}.
          We also introduce constraints and a population initialization strategy to guide the search toward realistic test cases that reflect real-world conditions.
    \item We incorporate an interpretability approach into \emph{\gls{ours}} to facilitate the analysis and understanding of the root causes behind the identified behavioral divergences.
    \item We apply \emph{\gls{ours}} to a complex case study utilizing an industry-grade simulator and a high-performing \gls{ads}.
    \item We provide a comprehensive evaluation of \emph{\gls{ours}} through large-scale experiments and comparisons against baseline methods, showing its ability to identify behavioral discrepancies and highlight potential safety risks introduced by updates.
\end{itemize}

\emph{\Gls{ours}} builds on our previous method, \emph{CoCoMEGA}~\cite{yousefizadeh2025using}, and mainly differs from it by focusing on efficiently and effectively testing and analyzing updates across system versions.
The key differences are as follows:

\begin{itemize}
    \item \emph{\Gls{ours}} is specifically designed to systematically detect behavioral divergences between different versions of \glspl{as}, rather than focusing solely on testing a single version.
    \item It incorporates constraints and a population initialization strategy to guide the search toward realistic test cases that reflect real-world conditions.
    \item It integrates an interpretability approach that facilitates understanding of the root causes of observed behavioral divergences by analyzing the generated test cases.
\end{itemize}

The remainder of this paper is organized as follows.
\Cref{sec:background} introduces the necessary background concepts.
\Cref{sec:problem} formally defines the problem and outlines the key challenges.
\Cref{sec:related_work} reviews related research and highlights existing gaps.
\Cref{sec:approach} presents a detailed description of \emph{\gls{ours}}.
\Cref{sec:evaluation} details the design of our empirical evaluation.
\Cref{sec:results} presents and interprets the experimental results.
\Cref{sec:discussion} analyzes the findings and discusses potential threats to validity.
Finally, \cref{sec:conclusion} summarizes the paper and outlines future work directions.

\section{Background}\label{sec:background}

In this section, we provide background information on the key concepts and our previous work that underpin our proposed framework for testing evolving \glspl{as}.

\subsection{Metamorphic Testing}

\Acrfull{mt}, first introduced by Chen et al.~\cite{chen2020metamorphic}, addresses the oracle problem in software testing by leveraging \glspl{mr}.
Instead of requiring a ground-truth output for every test case, \gls{mt} begins with a set of source test inputs and their observed outputs, then applies transformations to generate new follow-up inputs.
The expected relationship between the outputs of the original and transformed inputs is defined by an \gls{mr}.
If the program's outputs violate this relationship, a defect is detected.
For example, consider testing a program \(f\) intended to compute \(\sin(x)\).
Normally, one can check the correctness directly by verifying if \(f(x) = \sin(x)\) for any input \(x\).
However, assume that the ground-truth value \(\sin(x)\) is unknown, then such direct checking is not possible.
\Gls{mt} provides an alternative approach by testing the program using an \gls{mr}.
For instance, using the mathematical property \(\sin(x) = \sin(x + 2\pi)\), we can define an \gls{mr} stating that \(f(x)\) should equal \(f(x + 2\pi)\).
Testing then reduces to checking whether this relation holds for any input \(x\).
To ground these ideas more formally, we now introduce the precise definition of an \gls{mr} and explain how it is represented in our framework.

An \gls{mr} generally specifies a relation \(\mathcal{R}\) that defines a relationship between a sequence of inputs and their respective outputs~\cite{chen2018metamorphic}.
\annotate{Comment 1.3}\deletedtext{In the most common form, an \gls{mr} consists of two components: an input relation \(ir\) and an output relation \(or\).}
\addedtext{A common category of \glspl{mr} consists of two components: an input relation \(ir\) and an output relation \(or\).}
The input relation \(ir\) specifies how two inputs \(x_1\) and \(x_2\) are related, while the output relation \(or\) describes the expected relationship between their outputs \(f(x_1)\) and \(f(x_2)\)~\cite{deng2023declarative}, where \(f\) represents the function or system under test.
The relation \(\mathcal{R}\) holds when both \(ir\) and \(or\) are satisfied, and can be formally expressed as:
\[
    \mathcal{R}\left(x_1, x_2, f(x_1), f(x_2)\right) \coloneq ir\left(x_1, x_2\right) \annotate{Comment 1.3}\deletedtext{\land}\addedtext{\to} or\left(f(x_1),f(x_2)\right)
\]
This formulation allows an \gls{mr} to be interpreted as a consistency rule that governs how the system's output should change in response to a defined transformation of its input.
In our work, we formally define an \gls{mr} as a tuple \(mr \coloneq (ir, or)\).
Given the input relation \(ir\) providing an abstract specification of how a source test input can be transformed, we can derive concrete metamorphic transformations that modify the attributes of a source input to produce a follow-up input.
A violation occurs when the follow-up input satisfies \(ir\), but the corresponding outputs fail to satisfy \(or\), indicating that the \gls{mr} has been violated.
In our search framework, we employ a set of \glspl{mr} that share a common output relation \(or\) to guide the search, which can be defined as \(MR \coloneq \{mr_i = (ir_i, or) \mid i = 1, \ldots, k \} \), \addedtext{where \(k\) denotes the number of \glspl{mr} in the set.}

\subsection{Differential Testing}

\Acrfull{dt} is a testing technique in which multiple comparable implementations of a system are exercised on the same automatically generated inputs, and any divergence in their outputs is treated as evidence of a potential bug~\cite{mckeeman1998differential}.
In practice, \gls{dt} works by executing large numbers of inputs across multiple implementations and automatically flagging discrepant executions for analysis.
\Gls{dt} has been successfully applied to a variety of software systems~\cite{yang2011finding,lidbury2015manycore,petsios2017nezha}, whenever multiple independent implementations or versions are available for comparison on shared inputs.

\Gls{dt} is well suited to iterative development workflows in which models, training data, and control policies are continually revised. By running the same test inputs across successive versions of an \gls{as} and comparing their behavior, it directly targets update-induced changes, surfacing cases where system behavior improves or degrades. This comparison enables efficient identification of \deletedtext{regressions}\addedtext{degradations} and other unintended behavioral changes across system updates.

\subsection{Cooperative Co-evolutionary Algorithms}

\Glspl{ccea} form a particular type of \glspl{ea}, a family of algorithms inspired by biological evolution and widely used to solve optimization problems that are difficult to handle with standard mathematical optimization techniques~\cite{luke2013metaheuristics}.
In \glspl{ea}, candidate solutions are represented as individuals within a population, and their quality is evaluated by a fitness function.
Individuals with higher fitness are more likely to be selected as parents, and new individuals are generated through crossover (combining parts of parent solutions) and mutation (introducing random variation in an individual).
Through iterative cycles of selection, crossover, and mutation, the population evolves toward increasingly effective solutions.

However, many real-world problems involve extremely high-dimensional search spaces, making standard \glspl{ea} inefficient~\cite{ma2019survey,luke2013metaheuristics}.
To address this challenge, \glspl{ccea}, originally proposed by Potter and De Jong~\cite{potter1994cooperative}, decomposes the original problem into a set of lower-dimensional and tractable subproblems, each of which can be solved in a separately evolving subpopulation as in conventional \glspl{ea}.
As individuals from each subpopulation must work together to form a complete solution, an individual's fitness is evaluated based on the joint fitness of that complete solution, which is created by pairing the individual with representative collaborators from other populations or from population archives such as those used in \emph{iCCEA}~\cite{panait2006archive}.
By leveraging decomposition and collaborative evaluation, \glspl{ccea} can effectively search high-dimensional spaces, making them well-suited to our task of identifying scenarios in a complex, open-ended environment.

\subsection{RuleFit}

The \emph{RuleFit} algorithm, introduced by Friedman and Popescu~\cite{friedman2008predictive}, combines decision trees with linear models to construct predictive models that are both expressive and interpretable.
Linear regression models are simple and interpretable, but cannot naturally capture feature interactions or nonlinear relationships.
In contrast, decision trees can model complex interactions, yet they become increasingly difficult to interpret as their depth and number of leaf nodes grow.
\emph{RuleFit} bridges this gap by learning a sparse linear model that incorporates both the original features and a set of automatically generated decision rules.
These rules are derived from decision trees and explicitly encode feature interactions.
Each path from the root to a leaf in a tree is transformed into a binary rule by conjoining the split conditions along that path.
The trees themselves are trained to predict the target outcome, ensuring that the generated rules are relevant to the learning task.
\emph{RuleFit} is flexible with respect to the tree-generation process, and any method that produces an ensemble of trees, such as \glspl{rf} or \glspl{gbdt}, can be used.
The resulting rules are then combined with the original features in a sparse linear model, typically using Lasso~\cite{tibshirani1996regression}, which promotes sparsity and selects a compact set of informative rules.

To illustrate, consider the example from the original \emph{RuleFit} paper~\cite{friedman2008predictive}, which predicts the median house value of a Boston neighborhood.
The input features include the average number of rooms per dwelling, the proportion of owner-occupied units built before 1940, and weighted distances to five Boston employment centers, while the output is the median house value in thousands of dollars.
One rule generated by \emph{RuleFit} is: ``IF the number of rooms \(> 6.64\) AND the concentration of nitric oxides \(< 0.67\), THEN increase the predicted median house value by 1.30''.
This rule captures an interaction between two features that a standard linear model would not represent without manual specification.
By automatically introducing interaction terms, \emph{RuleFit} alleviates the need to manually engineer feature interactions and improves linear models' ability to capture nonlinear effects.
At the same time, the resulting model remains interpretable, as each rule is a simple binary condition indicating whether it applies to a given instance.
\emph{RuleFit} supports both regression and classification tasks, making it a versatile tool for predictive modeling where interpretability and expressiveness are equally important.
In our framework, we leverage \emph{RuleFit} to derive such interpretable rules from the identified test cases that expose the behavioral differences between \glspl{as}.
These rules help engineers understand and communicate the specific scenario conditions under which an updated \gls{as} exhibits different behavior from a previous version, thereby facilitating debugging and improvement efforts.

\subsection{CoCoMEGA}\label{sec:cocomega}

\emph{CoCoMEGA}~\cite{yousefizadeh2025using} is a search-based testing framework that combines \gls{mt} with an archive-based \gls{ccea} to generate system-level test cases for \glspl{as}. The key idea is to decompose the high-dimensional test generation problem into two coevolving populations: one representing source scenarios and the other representing perturbations derived from a predefined set of \glspl{mr}. Each perturbation encodes a concrete metamorphic transformation (e.g., adding a pedestrian, changing weather, or adjusting vehicle speed) that can be applied to a source scenario to produce a follow-up scenario. By collaborating individuals from the two populations, \emph{CoCoMEGA} forms complete solutions (i.e., scenario-perturbation pairs) that are executed in simulation to assess the extent to which the corresponding \gls{mr} is violated.
In the following, we summarize the key concepts and components of \emph{CoCoMEGA}.

\subsubsection{Representations}

There are two co-evolving populations involved in \emph{CoCoMEGA}, i.e., one for source scenarios and one for perturbations.

A scenario \(s\) can be formally represented as a vector composed of real and integer values.
In practice, the specific vectorization scheme depends (in part) on the observable and controllable elements of the underlying system or simulator.
For example, in the context of \glspl{ads}, a scenario typically includes the ego vehicle, its trajectory, environmental factors (e.g., weather), and both dynamic (e.g., other vehicles and pedestrians) and static (e.g., obstacles) objects~\cite{haq2021offline,sharifi2023identifying}.
Each of these elements is characterized by multiple attributes of different types, including continuous values and categorical variables encoded as integers.
In this context, let \(ego\) be the ego vehicle and its attributes, \(WP\) be a set of waypoints that characterizes the trajectory, \(ATTR\) be a set of global attributes, e.g., weather and brightness, \(STAT\) be a set of static objects, e.g., traffic signs, and \(DYNA\) be a set of dynamic objects, e.g., pedestrians and vehicles.
The representation of a scenario can be defined as the tuple \(s \coloneq (ego, WP, ATTR, STAT, DYNA)\), where:

\begin{align*}
    ego  & \coloneq (mod_{ego}, pos_{ego}, rot_{ego}, v_{ego})                                  \\ % chktex 35
    WP   & \coloneq \{wp_i \mid i \in \mathbb{N}^\ast \}                                        \\
    ATTR & \coloneq \{attr_i \mid i \in \mathbb{N}^\ast \}                                      \\
    STAT & \coloneq \{stat_i \coloneq (mod_i, pos_i, rot_i) \mid i \in \mathbb{N}^\ast \}       \\ % chktex 35
    DYNA & \coloneq \{dyna_i \coloneq (mod_i, pos_i, rot_i, v_i) \mid i \in \mathbb{N}^\ast \}. \\ % chktex 35
\end{align*}%
Here, \(mod\) represents the type of the object, e.g., the model of the vehicle such as a motorcycle or a car, \(pos\) represents the position of the object in three-dimensional space, \(rot\) represents the direction the object is facing in three-dimensional space, and \(v\) represents the speed of the respective objects. % chktex 35
Each attribute (weather, brightness, object speed, etc.) is bounded by predefined ranges that reflect realistic conditions and conform to \textsc{Carla}'s valid parameters.
These ranges are also configurable, allowing testers to tailor the space of possible values to suit different testing requirements or fidelity levels.

A perturbation can be formally defined as a sequence of metamorphic transformations, denoted as \(q \coloneq \langle c_1, c_2, \ldots, c_k \rangle \), which are applied sequentially to a source scenario \(s\) to produce a follow-up scenario \(q(s)\).
Each transformation \(c_i\) in \(q\) corresponds to a unique input relation \(ir_i\) from the given set of \glspl{mr}.
Typically, a transformation \(c_i\) may represent a modification to the source scenario, such as adding a new object, removing an existing one, or replacing an existing object with a new one.
It may also represent no change at all, with no modification applied to that specific transformation.
To ensure realism, every transformation must remain within the predefined allowable ranges of the involved attributes.
This flexible structure allows perturbations to be constructed selectively, so that not every available transformation needs to be included.
However, we require that at least one actual transformation be applied to produce a follow-up scenario with meaningful variation.

% \subsubsection{Fitness}

% The fitness evaluation in \emph{CoCoMEGA} is based on a joint fitness function that quantifies the severity of \gls{mr} violations for each complete solution, together with individual fitness functions that attribute credit back to the participating scenario and perturbation individuals.

\subsubsection{Archive Strategy}\label{sec:archive-strategy}

To avoid premature convergence and encourage the exploration of diverse failures, \emph{CoCoMEGA} maintains archives of scenarios, perturbations, and complete solutions, and relies on a heterogeneous distance metric~\cite{wilson1997improved} over solutions to select archive members that are both high-fitness and well spread in the search space.

The heterogeneous distance used in \emph{CoCoMEGA} provides a unified way to measure how different two individuals (i.e., scenarios or perturbations) are, even though scenarios contain multiple types of elements and attributes.
At the attribute level, numerical values such as position or speed contribute to a normalized difference. In contrast, categorical values, such as object type, contribute a small fixed penalty when mismatched and zero when identical.
These attribute-level comparisons allow us to measure the similarity between two objects.
To compare sets of objects, such as static or dynamic objects in a scenario, the distance pairs objects across the two sets according to their attribute-level distances and aggregates the resulting best-match distances, where each best match corresponds to the minimum distance achievable by pairing an object with its most similar counterpart in the other set, capturing structural differences even when the two scenarios contain different numbers of objects.
Finally, the distances for static objects, dynamic objects, and global attributes are combined into a single value, yielding an overall measure of scenario similarity that accounts for variations in objects, attributes, and environmental conditions.
This unified metric enables consistent comparison across complex, heterogeneous scenarios commonly found in \gls{ads} testing.

The archive strategy is designed around the heterogeneous distance and maintains a balance between exploiting high-fitness solutions and exploring diverse regions of the search space.
At each generation, the algorithm begins by inserting the individual with the highest fitness into the archive, thereby preserving strong solutions.
It then incrementally fills the remaining archive slots by repeatedly selecting the individual that maximizes the increase in overall diversity when added to the current archive, encouraging the search to cover a broad range of behaviors and reducing the likelihood of premature convergence.
Diversity is quantified using the Pure Diversity metric~\cite{wang2017diversity}, which evaluates the spread of individuals to capture both primary dissimilarities and the overall extent of coverage in the search space, resulting in an archive that both retains high-fitness individuals and fosters exploration.

\subsubsection{Genetic Operators}\label{sec:genetic-operators}

Standard evolutionary operators (selection, crossover, mutation) are applied separately to the two populations, which are periodically merged with their corresponding archives, yielding an iterative co-evolutionary process that gradually accumulates a set of distinct, high-severity violating test cases.
Since the complexity of the scenario and perturbation representations precludes the use of simple genetic operators, \emph{CoCoMEGA} employs specialized operators tailored to each population, as described below.

\paragraph{Selection}
The selection operator for both populations uses standard tournament selection, the most prevalent method in evolutionary algorithms~\cite{whitley2018next}.
Candidate individuals are selected for reproduction based on their relative fitness rankings.

\paragraph{Crossover}
The crossover operator for scenarios creates new offspring by combining characteristics from two parent scenarios while preserving their essential structure.
The parameters of the ego vehicle \(ego\) and the trajectory \(WP\) remain unchanged, while global attributes \(ATTR\), static objects \(STAT\), and dynamic objects \(DYNA\) are exchanged between parents using a uniform crossover.
This approach allows offspring to inherit traits from both parents, such as weather conditions, vehicle positions, and obstacle configurations, while maintaining the overall integrity of the scenario.

For perturbations, the crossover operator works pairwise across sequences of metamorphic transformations from two parents.
Each corresponding transformation is combined using uniform crossover, which may swap parameters when both transformations affect the same type of element, such as adding vehicles or modifying global attributes.
If the paired transformations modify different types of elements, they are preserved as is.
This mechanism enables new perturbation sequences that inherit traits from both parents while maintaining valid and meaningful modifications to the scenario.

\paragraph{Mutation}
The mutation operator for scenarios introduces variations by modifying global attributes \(ATTR\) and the properties of static objects \(STAT\) and dynamic objects \(DYNA\) while keeping the ego vehicle \(ego\) and trajectory \(WP\) unchanged.
Numerical attributes are perturbed using polynomial mutation~\cite{deb1995simulated}, and categorical attributes are altered using integer randomization~\cite{luke2013metaheuristics}.
Additionally, the operator can incrementally add or remove objects from the scenario according to a probabilistic scheme, helping explore a broader range of conditions and preventing the scenario from growing excessively over generations.
These changes generate subtle but meaningful variations in scenarios, such as adjusting weather, brightness, vehicle positions, or pedestrian behaviors, while preserving the overall structure.

For perturbations, mutation is applied to each metamorphic transformation in a sequence, using the same numerical and categorical mutation strategies as for scenarios.
Depending on the type of transformation, the mutation may assign a new value to a global attribute, alter parameters of static or dynamic objects, or enable or disable a transformation entirely.
This approach ensures that offspring perturbations remain valid while introducing diversity, producing a rich set of variations that can explore different potential outcomes in the scenarios.

% Empirical results on CARLA \todo{[ref]} with the INTERFUSER \gls{ads} \todo{[ref]} showed that CoCoMEGA can efficiently generate severe and diverse \gls{mr} violations and outperforms baseline search strategies in terms of both the number and diversity of unsafe scenarios discovered.

\subsubsection{How \gls{ours} Extends CoCoMEGA for \gls{as} Testing}

\emph{CoCoMEGA} provides a first step towards co-evolutionary \gls{mt}. Still, it does not fully address several demands that arise in the development and testing of real-world \glspl{as}. \Gls{ours} builds directly on \emph{CoCoMEGA} and extends it along the following four key dimensions.

\paragraph{Differential Testing}
\emph{CoCoMEGA} focuses on testing a single system under test by maximizing the degree of \gls{mr} violation for that system. In practice, however, engineers frequently need to compare multiple versions or configurations of an \gls{as} (e.g., before and after a software update, or across alternative controller designs) and detect \deletedtext{regressions}\addedtext{degradations} or unexpected behavioral divergence.
\deletedtextsec{\emph{\Gls{ours}} explicitly incorporates \gls{dt} by defining objective functions over the difference in \gls{mr} violation between two \gls{as} variants and guides search towards scenarios where their behaviors diverge the most. This allows \emph{\gls{ours}} to uncover \deletedtext{regressions}\addedtext{degradations} and inconsistent behaviors that \emph{CoCoMEGA}, by design, cannot expose.}
\addedtextsec{A key motivation for \gls{ours} is that behavioral divergence is a paired property of two system versions under the same test condition. Running a single-version testing method independently on each version may find severe \gls{mr} violations, but it does not directly optimize for cases where the degree of violation changes across versions. \gls{ours} therefore evaluates each generated scenario--perturbation pair on both versions and uses the resulting difference in \gls{mr} violation as the search objective, enabling a controlled search for update-induced behavioral changes.
}

\paragraph{Constrained Search Space}
\emph{CoCoMEGA} performs evolutionary search without explicitly enforcing constraints that ensure the realism of generated scenarios, so many candidates can fall outside the system's operational environment or correspond to implausible traffic situations. This not only reduces the usefulness of the discovered failures but also wastes simulation budget on cases that engineers are unlikely to encounter in practice. In contrast, \emph{\gls{ours}} constrains the search to remain close to a diverse archive of realistic scenarios derived from previous \gls{as} executions. This steers the search towards realistic, operationally plausible scenarios that are more likely to arise in the operational environment.

\paragraph{Scenario Initialization}
\emph{CoCoMEGA} initializes its populations largely at random, which is appropriate for generic search but does not exploit information accumulated during iterative development and\deletedtext{ regression} testing.
\emph{\Gls{ours}} introduces \emph{\gls{ri}} strategies that bias the initial population towards regions of the scenario space that are close to previously detected failures. This improves search efficiency, accelerates the discovery of problematic scenarios, and better reflects how engineers iteratively test evolving \gls{as} versions.

\paragraph{Interpretability Layer}
The output of \emph{CoCoMEGA} is a set of high-dimensional solutions with associated fitness values, leaving it to engineers to manually infer why certain regions of the scenario space are particularly problematic.
\emph{\Gls{ours}} adds an interpretability component, which learns rule-based models that approximate the relationship between scenario attributes and \gls{mr} violation outcomes. These rules provide human-readable explanations of the form ``if a combination of conditions on scenario parameters holds, then the likelihood or magnitude of violation is high''. Such explanations help engineers understand and communicate why the \gls{as} fails, support debugging and design decisions, and make the testing outcomes more actionable in industrial settings.

\paragraph{Summary}
Overall, \emph{\gls{ours}} extends \emph{CoCoMEGA} from a general-purpose co-evolutionary \gls{mt} engine to a practice-oriented framework that supports \gls{dt}, respects domain constraints, leverages prior knowledge, and produces interpretable insights tailored to real-world \gls{as} development.

\section{Problem and Challenges}\label{sec:problem}

In this section, we first present a motivating example to illustrate the context of our research.
Next, we formally define the problem we aim to solve.
Finally, we outline the key challenges that arise in this context and the need for a systematic approach to address these challenges.

\subsection{Motivating Example}

Consider an \gls{ads} that has been fine-tuned to operate in a new urban environment with different road layouts and traffic patterns.
The original version of the system was primarily trained and tested in suburban settings, where traffic is relatively light and road structures are simpler.
However, the updated version has undergone fine-tuning to better handle the complexities of urban driving, including navigating dense traffic, frequent stops, and complex intersections.
After fine-tuning, it is essential to assess how the system's behavior has changed and whether these changes translate into improved safety and performance in urban conditions.

Evaluating such changes is challenging.
First, determining whether the updated \gls{ads} behaves correctly in complex urban scenarios is difficult, as there is often no clear test oracle specifying the expected outcome for every situation.
For example, when encountering multiple pedestrians crossing at different speeds in a crowded street, it is nontrivial to define precise ground-truth behaviors against which the system's decisions can be judged.
Second, the generated test scenarios must remain representative of the new urban environment.
Testing the system with unrealistic or overly synthetic scenarios may fail to reveal meaningful behavioral differences that arise in real city driving.
Moreover, when behavioral differences are detected between the original and updated systems, engineers need a way to understand why these differences occur.
Simply identifying failing or divergent scenarios is insufficient without interpretable explanations that reveal which environmental factors or interactions contribute to the observed changes.
Finally, this evaluation process should be efficient enough to be performed regularly, especially if the \gls{ads} undergoes frequent updates or fine-tuning.

\subsection{Problem Definition}

This motivating example highlights the need for an automated, effective, and time-efficient system-level testing methodology that systematically identifies scenarios in which the behavior of an updated \gls{as} has improved or declined relative to the original version, thereby supporting repeated, continuous testing throughout the system's development lifecycle.

Specifically, we re-express the problem as a search problem.
Let \(\mathcal{S}\) denote the space of all possible operating scenarios, and let \(\mathcal{Q}\) denote the space of all possible perturbations derived from a set of \glspl{mr} that share the same output relation \(or\).
We define a function \(\Delta E_{or}(s, q)\) that quantifies the difference in the extent of \gls{mr} violation of \(or\) between two versions of an \gls{as}, given a source scenario \(s \in \mathcal{S}\) and its follow-up scenario \(q(s)\), which results from applying perturbation \(q \in \mathcal{Q}\) to \(s\).
For illustration, consider a testing context where the goal is to evaluate how different versions of an \gls{ads} behave when encountering a pedestrian under various weather conditions.
The space \(\mathcal{S}\) includes all possible scenarios defined by factors such as road geometry, traffic density, and initial pedestrian placement.
The space \(\mathcal{Q}\) includes possible perturbations, such as changes in weather (e.g., fog or rain), derived from an \gls{mr} expecting the \gls{ads} to reduce speed when visibility decreases.
\annotate{Comment 3.8}\addedtext{The specific factors and their value ranges defining \(\mathcal{S}\) and \(\mathcal{Q}\) depend on the subject system and the simulation environment in which it operates.}
Now, given a source scenario \(s \in \mathcal{S}\) with clear visibility, applying a perturbation \(q \in \mathcal{Q}\) yields a follow-up scenario \(q(s)\) with reduced visibility.
The function \(\Delta E_{or}(s, q)\) then measures how inconsistently the two \gls{ads} versions respond to the same visibility change, specifically quantifying the difference in the extent to which they violate the expected speed reduction behavior defined by \(or\).
A larger value indicates greater inconsistency between system versions in meeting the expected \glspl{mr}.

\begin{definition}[Problem]
    The problem is to find a set \(\mathcal{SP}\) of scenario-perturbation pairs \((s, q) \in \mathcal{S} \times \mathcal{Q}\) such that the extent of violation of the expected output relation \(or\) differs between two versions of the \gls{as} when applied to the same perturbed scenario \(q(s)\):
    \[
        \mathcal{SP} = \{(s, q) \in \mathcal{S} \times \mathcal{Q} \mid \Delta E_{or}(s, q) > 0\}
    \]
\end{definition}

\subsection{Challenges}

However, assessing the impact of updates presents several challenges:

\begin{itemize}
    \item \textbf{Scalability and Coverage.}
          Thoroughly testing \glspl{as} in all conditions is infeasible due to the vast space of possible execution scenarios.
          Many factors influence \gls{as} behavior, such as road conditions, weather, traffic dynamics, and pedestrian interactions in the case of \glspl{ads}.
          Ensuring that the testing framework effectively captures the full range of behavioral changes between the original and updated versions is a significant challenge.
          Without comprehensive coverage, some critical failure cases may go undetected, posing safety risks.
    \item \textbf{Oracle Problem.}
          Determining whether a behavioral change in the \gls{as} constitutes an improvement or a decline is non-trivial.
          For example, tasks in \glspl{ads}, including perception, planning, and decision-making, are highly complex, and their correctness cannot usually be explicitly and precisely defined.
          The unpredictable nature of real-world traffic conditions further complicates this problem, making it difficult to establish ground-truth expectations for every possible scenario.
          The absence of well-defined test oracles can lead to ineffective evaluations, as it becomes challenging to distinguish between beneficial adaptations and unintended degradations.
    \item \textbf{Scenario Realism.}
          While artificially constructed edge cases can reveal vulnerabilities, failures found in realistic scenarios, particularly those closely resembling situations already encountered during system execution, are often of greater practical importance.
          Such scenarios are inherently more likely in deployment, making any detected failure pose a higher operational risk.
          Therefore, guiding the testing process toward realistic conditions enhances the relevance of test results, ensuring that identified violations reflect risks that are plausible in practice rather than purely theoretical.
    \item \textbf{Interpretability.}
          \Glspl{as} rely on \gls{ml} or \gls{dl}-based components with a vast number of parameters and intricate decision-making mechanisms.
          Fine-tuning involves modifying these parameters to optimize performance, but understanding how these modifications influence overall system behavior is not straightforward.
          Without systematic, well-designed testing methodologies, it is difficult to diagnose issues and ensure that system changes align with expected improvements.
    \item \textbf{Time Constraints.}
          In many scenarios, fine-tuning occurs automatically and frequently, particularly in self-adaptive systems that continuously refine their models.
          Given the rapid pace of such updates, an efficient testing process is essential.
          The challenge is to develop an automated, fast testing approach that provides insights into \gls{as} behavioral changes without significantly delaying deployment.
\end{itemize}

Addressing these challenges is critical to ensuring the safety of updated \glspl{as}. A robust testing methodology should efficiently identify behavioral changes, overcome the oracle problem, enable interpretability, and operate within realistic time constraints.

\section{Related Work}\label{sec:related_work}

This section reviews prior work in two main areas that intersect with our contributions, namely the Search-based Testing of \glspl{as} and the \acrlong{mt} of \glspl{as}.
We focus on prior work that targets system-level testing of \glspl{as}, including \glspl{ads}, drones, and robots, where tests are generated or prioritized using search-based optimization, fuzzing, or \gls{mt} in simulation. We exclude studies that address only isolated modules (such as perception components) or rely solely on random test generation without feedback from system behavior. Within this scope, we prioritize approaches that aim to uncover safety-critical behaviors and introduce design elements closely related to our method.

\subsection{Search-based Testing of Autonomous Systems}

Search-based testing has emerged as a key strategy for the system-level evaluation of \glspl{as} across a range of domains, including autonomous vehicles, aerial drones, robotic platforms, and other complex multi-input \gls{ai}-based systems. The central idea is to automate the exploration of complex, high-dimensional scenario spaces using evolutionary search to uncover failure-inducing behaviors.

Ben Abdessalem et al.~\cite{benabdessalem2016testing, benabdessalem2018testing} pioneered a multi-objective strategy that combines search-based testing with surrogate models based on neural networks to assess advanced driver assistance systems in simulation.
%In subsequent works, the authors extended their approach to detect conflicts between automated system functions~\cite{benabdessalem2018testing} and to test vision-based control systems~\cite{benabdessalem2018vision}, demonstrating that their methods uncovered substantially more critical scenarios than baseline techniques.
Dreossi et al.~\cite{dreossi2019compositional} proposed a compositional, search-based framework tailored for \gls{ml}-enabled \glspl{ads}, leveraging constraints from both the perception and system input spaces to efficiently isolate counterexamples.
Sartori~\cite{sartori2019simulation} proposed a simulation-based testing framework for autonomous robots that leverages search-based techniques to guide the generation and selection of virtual test worlds, incorporating test objectives, prior results, and dynamic agents. % Through two industrial case studies involving mobile robots, the approach demonstrated its capability to reveal safety-critical failures efficiently, even in low-fidelity simulations.
Haq et al.~\cite{haq2022efficient} introduced \emph{SAMOTA}, which employs many-objective optimization with surrogate models to efficiently surface safety violations by approximating expensive simulations.
Kolb et al.~\cite{kolb2021fitness} developed a set of fitness function templates tailored for search-based testing of \glspl{ads} in intersection scenarios. This work builds on earlier efforts in highway environments~\cite{hauer2019fitness} and demonstrates that tailored fitness definitions can significantly enhance the generation of diverse, safety-relevant tests compared to random testing baselines.
Luo et al.~\cite{luo2021targeting} presented \emph{EMOOD}, a search-based framework that leverages evolutionary strategies to systematically uncover test scenarios involving compound violations of system requirements. Birchler et al.~\cite{birchler2023cost, birchler2023single} proposed test case prioritization strategies aimed at optimizing the fault detection efficiency in regression testing. Their empirical results demonstrated that their approach can enhance early fault detection and testing efficiency relative to baseline methods.
% Khatiri et al.~\cite{khatiri2023simulation} proposed \emph{SURREALIST}, a search-based testing framework for Unmanned Aerial Vehicles that generates simulation test cases in the neighborhood of real flight logs by first replicating and then smoothly perturbing recorded trajectories. Their approach uncovered unstable and potentially unsafe system behaviors that did not occur in the original logs, improving the realism and fault-detection capability of simulation-based tests.
% Shimanuki et al.~\cite{shimanuki2025cortex} proposed \emph{CORTEX-AVD}, a framework that integrates \textsc{Carla} and \textsc{Scenic}~\cite{fremont2019scenic} with a genetic algorithm to automatically generate corner cases from textual scenario descriptions using a multi-factor fitness function. Their results showed that \emph{CORTEX-AVD} significantly increased the generation of high-risk events and improved simulation utility compared to random sampling.

Fuzzing techniques have also been employed to trigger faulty behaviors by systematically perturbing operating scenarios. Originally developed for software testing, fuzzing is the automated, repeated mutation of inputs to explore a wide range of system behaviors and increase the likelihood of exposing faults.
Li et al.~\cite{li2020av} introduced \emph{AV-Fuzzer}, which leverages genetic algorithms to perturb traffic participants in the \textsc{Apollo} platform~\cite{baiduapollo}, leading to broader discovery of critical safety violations compared to methods such as random fuzzing or adaptive stress testing.
Cheng et al.~\cite{cheng2023behav} proposed \emph{BehAVExplor}, a fuzzing approach that uses unsupervised clustering to model ego vehicle behavior and guide scenario generation based on behavior diversity and violation potential. Evaluated on \textsc{Apollo} and \textsc{LGSVL}~\cite{rong2020lgsvl}, it outperformed state-of-the-art baselines by detecting a significantly larger and more diverse set of violations.
Crespo-Rodriguez et al.~\cite{crespo2024pafot} proposed \emph{PAFOT}, a position-based framework that models the movements of surrounding traffic agents using a nine-position grid around the ego vehicle and applies a genetic algorithm to evolve collision-inducing scenarios. To improve local search, \emph{PAFOT} incorporates a local fuzzing strategy, which helps uncover additional safety violations near high-risk cases, leading to better performance than \emph{AV-Fuzzer} and random testing in \textsc{Carla}.
In a more recent study, Ji et al.~\cite{ji2025dofuzz} proposed \emph{DoFuzz}, a search-based framework that generates naturalistic driving scenarios from traffic patterns respecting basic vehicle dynamics and trajectories learned from real driving data, and prioritizes test executions based on a freshness score that favors behaviorally novel cases. Evaluated on multiple \textsc{Carla} \emph{Leaderboard}~\cite{carlaleaderboard} driving tasks, \emph{DoFuzz} discovered more unique failures and achieved faster convergence compared to \emph{BehAVExplor}.

Algorithmic innovations have further enriched the field. Gambi et al.~\cite{gambi2019automatically} integrated search-based testing with \emph{Procedural Content Generation}, a technique for algorithmically creating diverse virtual environments, to automatically construct virtual road scenarios for testing lane-keeping functionality.
Goss and Akba\c{s}~\cite{goss2022eagle} proposed a two-phase search method that first explores the scenario space broadly and then refines around failure-inducing cases to focus testing on critical regions. Zheng et al.~\cite{zheng2020rapid} proposed a quantum genetic algorithm to reduce the computational load associated with population size.
Humeniuk et al.~\cite{humeniuk2023ambiegen} introduced \emph{AmbieGen}, a modular, search-based test-generation framework that uses evolutionary algorithms to identify fault-revealing scenarios for \glspl{ads} and robots. % Through extensive evaluations on lane-keeping assist systems and obstacle-avoiding robots, their method revealed significantly more failures compared to the baseline methods.
In a subsequent study, they proposed \emph{RIGAA}~\cite{humeniuk2024reinforcement}, a reinforcement learning-informed genetic algorithm that integrates a pre-trained reinforcement learning agent into the evolutionary search process to improve the efficiency of simulator-based test generation. Evaluated on autonomous robot and vehicle systems, \emph{RIGAA} significantly outperformed prior methods, including \emph{AmbieGen}, revealing more failures in the vehicle domain and achieving faster convergence in both scenarios.

In summary, search-based testing has proven highly effective in uncovering critical failures in \glspl{as} by systematically exploring scenario spaces and guiding test generation toward high-risk behaviors and thereby surfacing edge cases. Techniques in this domain have evolved from early multi-objective formulations and surrogate-assisted search to more recent innovations involving fuzzing and behavioral clustering. Despite these advancements, search-based testing remains computationally expensive, particularly when dealing with high-dimensional scenario parameters and complex operating scenarios. This cost becomes a critical bottleneck in modern \gls{as} development pipelines, where systems are frequently retrained, fine-tuned, or updated to adapt to new data, hardware, or operating environments. These challenges highlight the need for scalable and adaptive testing frameworks that can efficiently identify critical behavioral changes across \gls{as} versions and minimize redundant simulation efforts, to keep pace with the rapid development cycles of modern \glspl{as}.

\subsection{Metamorphic Testing of Autonomous Systems}

\Acrfull{mt}~\cite{chen2018metamorphic,chen2020metamorphic} was introduced as a solution to the test oracle problem in domains where ground-truth output is hard to define.
Recent work highlights the effectiveness of \gls{mt} in many domains, including \glspl{as}, where it enables the detection of failures by checking the consistency of outputs across related inputs rather than relying on a known ground truth~\cite{segura2020metamorphic}.
% Recent applications of \gls{mt} to \glspl{ads} have shown their usefulness in uncovering context-dependent failures by defining \glspl{mr}.

% At the module level, \gls{mt} has been extensively applied to assess the dependability and robustness of perception systems. For instance, Yang et al.~\cite{yang2024metalidar} introduced \emph{MetaLiDAR}, an automated \gls{mt} framework that employs object-level manipulations, such as insertions, deletions, and affine transformations, to produce realistic variants of point cloud data for evaluating LiDAR-based object detection. They later proposed \emph{MetaSem}~\cite{yang2024metasem}, a framework based on semantic-level modifications to driving scenes (e.g., changes to traffic signs and road markings), revealing additional inconsistencies missed by prior methods.
% Zhou and Sun~\cite{zhou2019metamorphic} illustrated that injecting even minor perturbations outside the region of interest in LiDAR data could lead to critical perception errors, such as missed obstacles, underscoring the vulnerability of perception modules.

% Beyond the module level, \gls{mt} has also gained traction for system-level testing.
Lindvall et al.~\cite{lindvall2017metamorphic} developed a simulation-based testing framework that integrates \gls{mt} with model-based testing to automatically generate and evaluate test cases for autonomous drones under variations of geometric and environmental conditions.
Their approach revealed unstable behaviors and critical failure cases, including collisions and perception failures, demonstrating its effectiveness in uncovering subtle flaws that traditional methods often miss. Tian et al.~\cite{tian2018deeptest} introduced \emph{DeepTest}, which defines transformation-based \glspl{mr} involving lighting and weather variations to identify behavioral inconsistencies in Deep Neural Network-based \glspl{ads}. This idea was expanded in \emph{DeepRoad}~\cite{zhang2018deeproad}, which applied generative adversarial networks to generate high-fidelity weather conditions, enabling more realistic and effective scenario transformations.
% Pan et al.~\cite{pan2021metamorphic} proposed an \gls{mt} framework that focused on fog-related driving conditions, using variations in fog density and direction to evaluate performance.
Han and Zhou~\cite{han2020metamorphic} employed a metamorphic fuzzing approach, using \glspl{mr} to filter false positives by identifying genuine faults among automatically generated edge-case scenarios.
In a more recent effort, Cheng et al.~\cite{cheng2024evaluating} developed \emph{Decictor}, a scenario-based \gls{mt} framework aimed at assessing optimal decision-making in \glspl{ads}. By applying minor, carefully controlled changes to the driving environment, \emph{Decictor} identifies situations where the system diverges from the optimal trajectory without altering the ground-truth path.
Fredericks et al.~\cite{fredericks2024towards} proposed \emph{DroneMR}, an \gls{mt} framework for drone systems that combines evolutionary search with contextual \gls{mr} inference to detect failures under uncertain conditions. Applied to a micro-drone platform in simulation, \emph{DroneMR} generated diverse operating contexts and corresponding metamorphic test cases, showing its potential for both design-time and run-time validation in safety-critical settings.

Overall, \gls{mt} provides a practical means to detect failures in both module-level and system-level testing of \glspl{as}, especially in the absence of a test oracle. Its power lies in the definition of \glspl{mr} that express consistent behavioral expectations under controlled input modifications, enabling the detection of latent inconsistencies.
Nonetheless, one major limitation lies in the difficulty of formulating comprehensive and safety-relevant \glspl{mr}, especially in dynamic, real-world operating scenarios characterized by environmental variability and unpredictable agent interactions.
The task of identifying effective \glspl{mr} often requires significant domain expertise and manual effort, making automation a persistent open challenge~\cite{li2025metamorphic}.
However, because \glspl{mr} encode general behavioral properties rather than exact output specifications, they can remain effective even when only partially specified, as violations of these \glspl{mr} may still indicate anomalous or flawed behaviors.
% However, \glspl{mr} need not be fully complete or perfectly specified to be useful. Even partial or approximate \glspl{mr} have demonstrated value in uncovering flawed system behaviors.

\subsection{Discussion}

The current body of research identifies search-based testing and \gls{mt} as the two principal methodologies for system-level evaluation of \glspl{as}. Search-based testing has proven effective at systematically exploring large, complex scenario spaces to uncover rare but safety-critical failures, benefiting from advancements such as surrogate modeling, behavior-guided fuzzing, and other innovative approaches. On the other hand, \gls{mt} has emerged as a practical solution to the oracle problem, enabling validation in scenarios where ground-truth outputs are unavailable by checking the consistency of outputs under controlled input transformations defined by \glspl{mr}.

Despite the progress achieved by existing techniques, several limitations remain that hinder their practical deployment in real-world \gls{as} development cycles. Most notably, current search-based approaches are computationally expensive, requiring extensive simulation time and resources to explore high-dimensional scenario spaces. This limitation becomes especially problematic in modern development settings, where \glspl{as} are continuously updated to accommodate new environments, system configurations, or operating policies. In such contexts, the cost of re-running full-scale test campaigns after each system update is prohibitive. Furthermore, current approaches assess \glspl{as} in isolation, so they cannot reveal whether a newly discovered failure is a longstanding flaw or a degradation introduced by the latest software update. This inability to distinguish degradations from persistent issues makes it difficult for developers to prioritize fixes, allocate debugging resources effectively, and assess the impact of recent updates on system safety.
% , as well as to decide whether the updated system is ready for deployment.
Finally, existing methods typically lack mechanisms for fault interpretability, offering little insight into why failures occur or which scenario attributes most contribute to inducing unsafe behaviors. This lack of explanatory power limits their utility in iterative development, where actionable feedback is essential for effective debugging and refinement.

Building on these observations, our work investigates how to integrate search-based testing and \gls{mt} into a unified framework that addresses the practical limitations of current approaches. The central question we explore is how to design a testing methodology that remains effective and efficient in uncovering critical behavioral failures induced by system updates, while also providing insights into the key attributes of failure-inducing scenarios.
% The following sections describe our proposed approach in detail.

\section{Our Approach}\label{sec:approach}
\begin{figure*}[htbp]
    \centering
    \includegraphics[width=\linewidth]{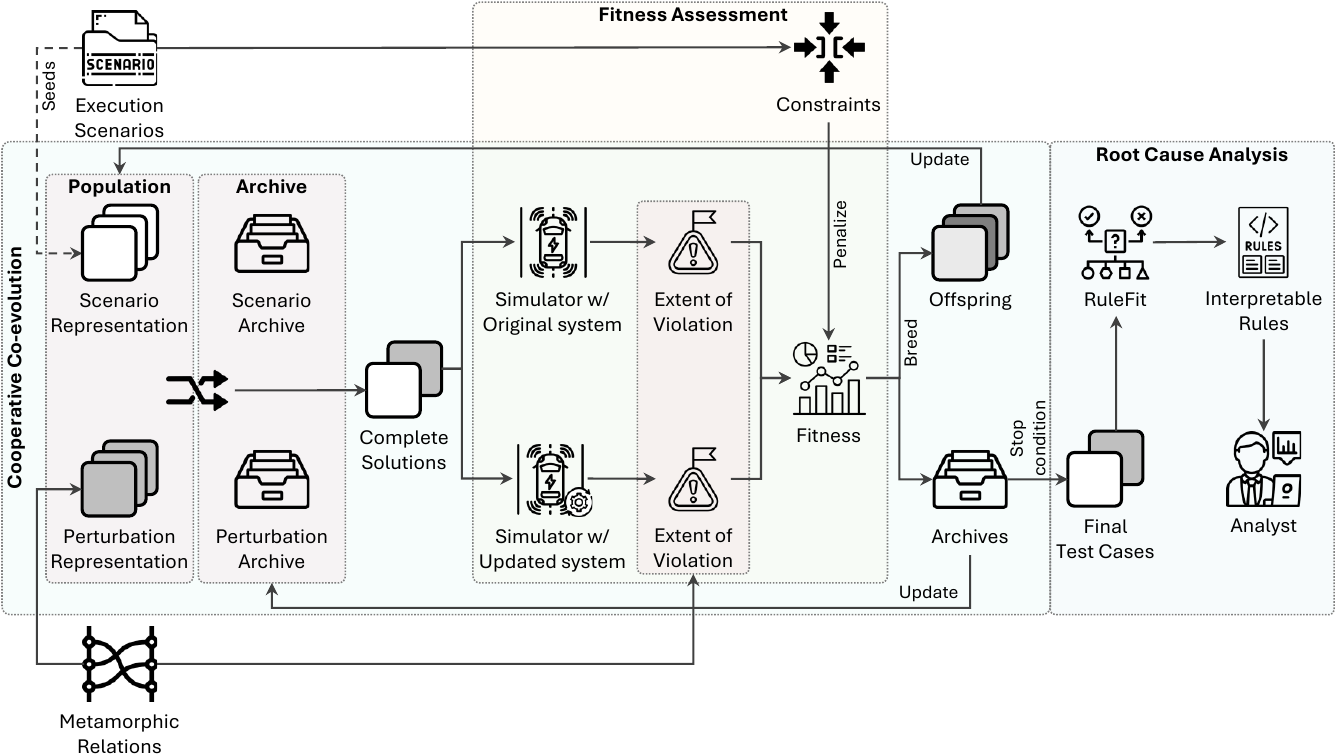}
    \caption{Overview of \emph{\gls{ours}}.}\label{fig:workflow}
    \Description{Overview of CoCoMagic.}
\end{figure*}

\Cref{fig:workflow} presents an overview of our proposed approach, \emph{\acrfull{ours}}.
Our approach integrates a constrained archive-based \gls{ccea} with \gls{dt} and \gls{mt} to effectively and efficiently generate test cases that expose differences in \gls{mr} violations between two versions of an \gls{as}.
\Gls{ccea} decomposes the original high-dimensional search problem into lower-dimensional subproblems, each handled by an independent population, not only enabling parallelism but also increasing the efficiency of the search process~\cite{ma2019survey}.
Specifically, we cast the problem as a two-population \gls{ccea}: the first population explores source scenarios, and the second consists of perturbations derived from the predefined \glspl{mr}.
Archive structures store elite individuals from both populations to guide the search toward promising regions of the space.
Through collaboration, individuals from both populations combine to form complete test cases, i.e., complete solutions, that reveal differences in \gls{mr} violations.
To further guide the search, we use a set of \glspl{mr} that share the same output relation, allowing the search to explore a broader, more diverse range of test cases while ensuring that each candidate test case directly targets relevant relation violations.
Additionally, constraints ensure that the generated test cases remain as relevant and realistic as possible in real-world conditions.
Finally, root cause analysis is performed on the generated complete solutions to interpret and explain the observed differences in \gls{mr} violations between the two \gls{as} versions.

In the following sections, we first introduce a fitness evaluation framework (\cref{sec:fitness}), which incorporates a joint fitness function~\cite{yang2008large,panait2006archive} for measuring the difference in \gls{mr} violations in complete solutions formed through collaboration between individuals from the two populations, as well as an individual fitness function~\cite{yang2008large,panait2006archive} for assessing each individual's contribution.
We then describe how constraints are incorporated into the search process to ensure that all generated solutions remain applicable to real-world conditions (\cref{sec:constraint}).
Next, we present our constrained \gls{ccea}-based search algorithm, which leverages the fitness functions and constraints to effectively guide the search (\cref{sec:algorithm}).
Other components of the approach follow our previous approach introduced in \cref{sec:cocomega} and are reused here without modification.
Finally, we describe the root cause analysis applied to the generated complete solutions, which helps interpret and explain the observed differences in \gls{mr} violations between the two \gls{as} versions (\cref{sec:rca}).

\subsection{Fitness Function}\label{sec:fitness}

We aim to define fitness functions that effectively guide the search toward solutions that maximize the difference in \gls{mr} violations between two versions of an \gls{as}.
To this end, we first introduce a method for quantifying the extent of \gls{mr} violation between a source and a follow-up scenario.
We then extend this definition to compute the difference in \gls{mr} violations between two versions of the \gls{as}.

\begin{definition}[Extent of \gls{mr} violation]
    Given a system under test \(V\), a source scenario \(s\) and a perturbation \(q\) derived from a set of \glspl{mr} that share the same output relation \(or\), and a function \(D_{or}\) which quantifies the deviation from \(or\), we define the extent of violation of \(or\) as follows:

    \begin{equation}\label{eq:violation}
        % E(s, q) \coloneq \frac{1}{|T|} \sum_{(t_s, t_q) \in T} diff_{or}(t_s, t_q)
        E(s, q, V, D_{or}) \coloneq D_{or}(V(s), V(q(s)))
    \end{equation}
    % where \(T\) represents the set of matched time points, which are pairs of time steps \((t_s, t_q)\) taken from the time series of outputs produced by the source scenario \(s\) and the follow-up scenario \(q(s)\), respectively.
    % These time points are aligned based on semantic equivalence, such as representing the same physical time or corresponding phases of a driving task.
    % In essence, each pair \((t_s, t_q)\) identifies moments where the outputs from both scenarios should be comparable under the expected behavior defined by the \gls{mr}.
    where \(V(s)\) and \(V(q(s))\) denote the outputs produced by the system \(V\) when executing the source scenario \(s\) and the follow-up scenario \(q(s)\), respectively.
    % \Cref{eq:violation} quantifies the difference between the expected output in \(s\) and the actual output in \(q(s)\).
\end{definition}

The exact formulation of \(D_{or}\) depends on the specific \(or\) of the \gls{mr} in use, and thus the extent of violation is computed accordingly.
For example, if \(or\) specifies that the output time series of the ego vehicle's speed should remain unchanged between the source and follow-up scenarios, \(D_{or}\) can be defined as the mean absolute difference between matched points in the two speed series.
The matching can be performed either pairwise by time index or using more advanced alignment algorithms, e.g., to handle temporal shifts or unequal lengths.
Accordingly, we can define \(D_{or}\) as follows:
\[
    D_{or}(T_s, T_q) \coloneq \frac{1}{|M|} \sum_{(t_s, t_q) \in M} \left( |t_s - t_q| - \phi \right)
\]
where \(M \subseteq T_s \times T_q\) denotes the set of matched point pairs between the two time series \(T_s\) and \(T_q\), and \({\phi}\) is a tolerance threshold that accounts for acceptable deviations due to noise or minor variations.

\Cref{eq:violation} captures how much the behavior in the follow-up scenario deviates from what is expected.
A higher value indicates a greater deviation, suggesting a more significant violation of the metamorphic relation.
In \cref{eq:violation}, the pair \((s, q)\) constitutes a complete solution.

\begin{definition}[Difference in \gls{mr} violations]
    Given a source scenario \(s\) and a perturbation \(q\) derived from a set of \glspl{mr} with the same output relation \(or\), and two versions of an \gls{as}, denoted \(V_{ref}\) and \(V_{test}\), the difference in \gls{mr} violation between the two versions is defined as follows:

    \begin{equation}\label{eq:violation-diff}
        Diff(s, q) \coloneq \left|\max(E(s, q, V_{test}, D_{or}), 0) - \max(E(s, q, V_{ref}, D_{or}), 0)\right|
    \end{equation}
    % where \(E_{ref}(s, q)\) and \(E_{test}(s, q)\) represent the extent of \gls{mr} violations for \(V_{ref}\) and \(V_{test}\), respectively, as defined in \cref{eq:violation}.
\end{definition}

We use the maximum of \(E\) and \(0\) to ignore negative values of \(E\), which indicate no violation of the given \glspl{mr}.
In such cases, the difference is not meaningful and should not \annotate{Comment 3.9}\deletedtext{influence the}\addedtext{affect} fitness\addedtext{, as our primary objective is to identify complete solutions that reveal behavioral divergences in \gls{mr} violations.}
\Cref{eq:violation-diff} quantifies how differently the two system versions respond to the same set of \glspl{mr}, highlighting changes that strengthen or alleviate \gls{mr} violations.
In our framework, \(Diff\) serves as the joint fitness function, which we aim to maximize during the search process to identify complete solutions that best expose behavioral discrepancies between the two versions.

To guide the optimization of each population independently, we define an individual-level fitness function that captures how well a given scenario or perturbation contributes to the discovery of differences in \gls{mr} violations.
This function assesses each individual based on its best-performing collaboration with members of the opposite population.
For a source scenario, its fitness is computed as the highest violation difference it produces when paired with any perturbation from the current population.
Conversely, for a perturbation, its fitness corresponds to the maximum violation difference observed across all source scenarios with which it is combined.

\begin{definition}[Individual Fitness Function]
    Formally, given an individual \(i\), its fitness is defined as follows:

    \begin{equation}\label{eq:fitness}
        fitness(i) \coloneq
        \begin{cases}
            \max_j \left(Diff(i, q_j)\right) & \text{if } i \text{ is a scenario}     \\
            \max_j \left(Diff(s_j, i)\right) & \text{if } i \text{ is a perturbation}
        \end{cases}
    \end{equation}
    where each \(q_j\) or \(s_j\) is drawn from the opposite population.
\end{definition}

This approach encourages individuals to specialize in combinations that are most likely to reveal behavioral differences between system versions.
Prior work has shown that relying on the maximum fitness obtained with collaborators, rather than the average or minimum, leads to more effective co-evolutionary outcomes~\cite{panait2006archive,bull1997evolutionary,bull1998evolutionary,wiegand2001empirical}.

\subsection{Constraints}\label{sec:constraint}

\annotate{Comment 2.4}\deletedtext{The updated version of an \gls{as} is typically retrained to adapt to changes in its operating environment.
As a result, it may be exposed to new environmental conditions that were absent or insufficiently tested in the previous version.
To ensure the system's reliability and safety under these new conditions, it is essential to evaluate its behavior accordingly.
To this end, we impose constraints on the search process to guide test case generation toward scenarios that reflect these newly introduced conditions.
Specifically, the search is constrained to generate complete solutions that are not only effective at revealing differences in \gls{mr} violations but also aligned with the new operating conditions observed during execution time, referred to as execution scenarios.}
\addedtext{During deployment, an \gls{as} continuously operates in an evolving environment and is therefore exposed to operating conditions that reflect the current state of the environment.
When the system is updated (e.g., retrained) to adapt to such changes, these recently observed conditions become particularly important for evaluation, as they are up to date and grounded in real-world execution.
To capture these conditions, we run simulations of the system under test, monitor environmental conditions as they arise, and collect a representative set of execution scenarios that characterizes operating conditions.
This collected set is then used to constrain the subsequent search for the updated version, ensuring that generated test cases remain aligned with realistic execution-time scenarios to an extent configurable by engineers.
}
%These execution scenarios represent specific, real-world situations newly encountered by \glspl{as} during actual operation. They are collected at runtime through monitoring mechanisms integrated into \glspl{as} and encompass a wide range of situations, providing a realistic and representative basis for testing.

\deletedtext{These execution scenarios are generated online during simulation runs of the system under test.}\addedtext{During the scenario collection phase,} the framework continuously monitors system behavior and environmental conditions, encoding each encountered situation in terms of its salient features (e.g., environmental context and agent behavior). As new situations arise, their feature representations are compared against those of previously recorded scenarios, and only \addedtext{sufficiently distinct or representative scenarios are retained in order to avoid redundancy.}
\deletedtext{scenarios that are sufficiently different, i.e., novel or particularly representative, are retained.}
Over time, this novelty-driven filtering process yields a compact yet diverse set of execution scenarios that captures a wide range of operational conditions \deletedtext{to constrain and evaluate}\addedtext{and serves as a basis for constraining} the search.

To enforce relevance, we quantify how similar a generated complete solution is to these real-world execution scenarios by measuring its dissimilarity to a set of execution scenarios \(\mathbb{S}\) collected \deletedtext{beforehand}\addedtext{in execution time}.
\addedtext{This constraint biases the search toward solutions that are realistic and operationally grounded, ensuring that generated test cases reflect conditions the system has recently encountered and is likely to encounter again in practice.}\deletedtext{This helps focus the search on solutions that are both meaningful and realistic, ensuring that the generated test cases reflect conditions the system is likely to encounter in practice.}

\begin{definition}[Dissimilarity to execution scenarios]
    Given a source scenario \(s\), a perturbation \(q\) derived from a set of \glspl{mr} sharing the same output relation, and a set of execution scenarios \(\mathbb{S}\), the dissimilarity of the complete solution formed by \(s\) and \(q\) to \(\mathbb{S}\) is defined as:

    \begin{equation}\label{eq:dissimilarity}
        Dissim(s, q, \mathbb{S}) \coloneq \min_{i \in \{s, q(s)\}} \left(\min_{j \in \mathbb{S}} \left(dist(i, j)\right)\right)
    \end{equation}
    where \(q(s)\) denotes the follow-up scenario and \(dist(i, j)\) represents a distance measure between scenarios \(i\) and \(j\).
\end{definition}

A typical choice for \(dist\), which accommodates both continuous and categorical variables, is the heterogeneous distance metric as introduced in \cref{sec:archive-strategy}.
\Cref{eq:dissimilarity} measures how dissimilar a complete solution is from the set \(\mathbb{S}\) by considering the minimum distance between either the source or follow-up scenario and its nearest neighbor in \(\mathbb{S}\).
This approach ensures that at least one component of each complete solution closely resembles a real execution scenario from \(\mathbb{S}\), preserving relevance.
Using the minimum rather than the average distance avoids inflating dissimilarity when a complete solution is very close to one scenario but far from others, which can result in a misleadingly moderate average that suggests the solution is farther from the set than it actually is.
This constraint helps maintain the plausibility and practical relevance of the generated test cases, focusing the search on behavior differences that are not only theoretically interesting but also applicable in real-world \gls{as} deployments.

\subsection{Algorithm}\label{sec:algorithm}

Building on the fitness functions defined in \cref{sec:fitness}, the constraints described in \cref{sec:constraint}, and other components reused from our previous approach introduced in \cref{sec:cocomega}, this section presents our proposed algorithm.

\begin{algorithm}
    \small
    \caption{Cooperative co-evolutionary algorithm.}\label{alg:alg}
    \LinesNumbered{}
    \KwIn{
        Population size \(n\)\newline
        % Maximum size of archive population \(l\)\newline
        Dissimilarity threshold \({\theta}\)\newline
        Execution scenarios \(\mathbb{S}\)
    }
    \KwOut{Archive of complete solutions \(X\)}

    \BlankLine{}
    Population of scenarios \(P_s \gets initPopulation(n)\){\;}
    Population of perturbations \(P_q \gets initPopulation(n)\){\;}
    Archive of scenarios \(X_s \gets P_s\){\;}
    Archive of perturbations \(X_q \gets P_q\){\;}
    Archive of complete solutions \(X \gets \emptyset \){\;}

    \BlankLine{}
    \While{not{(stopping condition)}}{
        \(P_s, P_q, X \gets assessFitness(P_s, P_q, X_s, X_q, X, \theta, \mathbb{S})\){\;}
        \(X_s \gets updatePopulationArchive(P_s)\){\;}
        \(X_q \gets updatePopulationArchive(P_q)\){\;}
        \(P_s \gets breed(P_s) \cup X_s\){\;}
        \(P_q \gets breed(P_q) \cup X_q\){\;}
        % \(P_s \gets selection(P_s, n)\){\;}
        % \(P_q \gets selection(P_q, n)\){\;}
        % Offspring of scenarios \(O_s \gets breed(P_s)\){\;}
        % Offspring of perturbations \(O_q \gets breed(P_q)\){\;}
        % \(P_s \gets P_s \cup O_s\){\;}
        % \(P_q \gets P_q \cup O_q\){\;}
    }

    \BlankLine{}
    \KwRet{\(X\)}\;
\end{algorithm}

\Cref{alg:alg} presents the pseudocode of our algorithm.
It takes as input a population size \(n\) and a set of execution scenarios \(\mathbb{S}\), and it outputs an archive \(X\) containing distinct complete solutions.
The algorithm begins by initializing two populations: a scenario population \(P_s\), which is either randomly generated or sampled from \(\mathbb{S}\) (line 1), and a perturbation population \(P_q\) (line 2).
Sampling the scenario population from execution scenarios enhances both the diversity of the initial population and its relevance to real-world conditions.
The choice of initialization strategy influences the similarity of the resulting complete solutions to real-world scenarios, as further discussed in \cref{sec:rq3}.
Corresponding population archives \(X_s\) and \(X_q\) are created at lines 3 and 4.
An empty archive of complete solutions \(X\) is also initialized at line 5.
The core of the algorithm is a co-evolutionary process where \(P_s\) and \(P_q\) co-evolve, guided by \(X_s\) and \(X_q\), until a predefined \emph{stopping condition} is met (line 6).
This process consists of the following iterative steps:
\begin{enumerate*}
    \item Evaluating the fitness of individuals in both \(P_s\) and \(P_q\), updating the archive \(X\) with complete solutions and their joint fitness values as determined by the two system versions (line 7);
    \item Updating the population archives \(X_s\) and \(X_q\) based on individual fitness values (lines 8--9); and
    \item Breeding new individuals for \(P_s\) and \(P_q\), and merging them with \(X_s\) and \(X_q\) to form the next generation of both populations (lines 10--11).
          % \item selecting individuals from \(P_s\) and \(P_q\) for breeding (lines 6--7);
          % \item breeding the selected individuals from \(P_s\) and \(P_q\) to produce offspring (lines 8--9); and
          % \item merging the offspring populations \(O_s\) and \(O_q\) with the original populations \(P_s\) and \(P_q\) to form their next generation  (lines 10--11).
\end{enumerate*}
After the \emph{stopping condition} is satisfied, the algorithm returns the final archive \(X\) (line 13).

The function \(assessFitness\) initially computes joint fitness values for complete solutions formed by combining individuals from \(P_s\), \(P_q\), \(X_s\), and \(X_q\).
To improve computational efficiency, complete solutions that already exist in \(X\) are excluded from re-simulation, avoiding redundant evaluations.
After updating the archive \(X\), each individual's fitness value is assigned based on the joint fitness obtained from the complete solutions it contributes to.
Finally, fitness clearing~\cite{petrowski1996clearing} is applied to each population to maintain variety within the populations, reducing the risk of premature convergence~\cite{petrowski1996clearing} and the impact of genetic drift~\cite{sareni1998fitness}, which causes the population to converge towards individuals that resemble those with high fitness~\cite{eigen1971selforganization,goldberg1987genetic,horn1998timing,mahfoud1995population}.

\begin{algorithm}
    \small
    \caption{assessFitness}\label{alg:fitness}
    \LinesNumbered{}
    \KwIn{
        Population of scenarios \(P_s\)\newline
        Population of perturbations \(P_q\)\newline
        Archive of scenarios \(X_s\)\newline
        Archive of perturbations \(X_q\)\newline
        Archive of complete solutions \(X\)\newline
        % Niche capacity \({\kappa}\)\newline
        Dissimilarity threshold \({\theta}\)\newline
        Execution scenarios \(\mathbb{S}\)
    }
    \KwOut{
        Updated population of scenarios \(P_s\)\newline
        Updated population of perturbations \(P_q\)\newline
        Updated archive of complete solutions \(X\)
    }

    \BlankLine{}
    Set of complete solutions \(CS \gets collaborate(P_s, P_q, X_s, X_q)\){\;}

    \BlankLine{}
    \ForEach{Complete solution \(cs \in CS\)}{
        \If{\(cs \notin X\)}{
            \(execute(cs)\){\;}
            \(cs.fitness \gets Diff(cs.s, cs.q)\){\;}
            \(X \gets X \cup \{cs\} \){\;}
        }
    }

    \BlankLine{}
    \ForEach{Population \(P \in \{P_s, P_q\} \)}{
        \ForEach{Individual \(i \in P\)}{
            \(i.fitness \gets assessIndividualFitness(i, X)\){\;}
            \If{\(Dissim(i.s, i.q, \mathbb{S}) > \theta \)}{
                \(i.fitness \gets \frac{i.fitness}{e^{c(Dissim(i.s, i.q, \mathbb{S}) - \theta)}}\){\;}
            }
        }
        \(fitnessClearing(P)\){\;}
    }

    \BlankLine{}
    \KwRet{\(P_s, P_q, X\)}\;
\end{algorithm}

The pseudocode for \(assessFitness\) is provided in \cref{alg:fitness}, with the following inputs: the population of scenarios \(P_s\), the population of perturbations \(P_q\), the population archive of scenarios \(X_s\), the population archive of perturbations \(X_q\), the archive of complete solutions \(X\), the dissimilarity threshold \({\theta}\), and the set of execution scenarios \(\mathbb{S}\).
The function returns updated populations \(P_s\) and \(P_q\) with assigned individual fitness values, along with an updated archive \(X\) incorporating newly created complete solutions and their corresponding fitness values.
The algorithm begins by generating an initial set of complete solutions by combining individuals from both populations and population archives (line 1).
Specifically, each individual in \(P_s\) collaborates with every individual in \(X_q\), and vice versa for \(P_q\) and \(X_s\).
For each generated complete solution \(cs\) (line 2), if \(cs \notin X\), i.e., \(cs\) has not been previously evaluated (line 3), it is evaluated by executing both versions of the \gls{as} (line 4).
The difference in \gls{mr} violations between the two versions is then quantified using \cref{eq:violation-diff} (line 5), and \(cs\), along with its evaluated result, is added to \(X\) (line 6).
Upon updating \(X\), the algorithm assigns fitness values to all individuals in both populations.
For each population \(P \in \{P_s, P_q\} \) (line 9) and for each individual \(i \in P\) (line 10), the individual fitness of \(i\) is determined as the maximum joint fitness obtained from all complete solutions involving \(i\), calculated according to \cref{eq:fitness} (line 11).
This strategy of assessing individual fitness using an elitist approach (i.e., selecting the individual with the highest fitness value) aligns with findings from previous experimental research~\cite{luke2013metaheuristics,ma2019survey,yousefizadeh2025using}.
Next, the algorithm applies a penalty to the individual fitness if the dissimilarity between the individual \(i\) and the execution scenario set \(\mathbb{S}\), as defined in \cref{eq:dissimilarity}, exceeds the threshold \({\theta}\) (line 12--13).
The penalized fitness is computed as:

\begin{equation}\label{eq:penalization}
    fitness \coloneq \frac{fitness_{raw}}{e^{c \left(Dissim(i.s, i.q, \mathbb{S}) - \theta\right)}}
\end{equation}
where \(fitness_{raw}\) denotes the individual fitness value before penalization and \(c\) is a constant that controls penalty severity.
This exponential-decay penalty function~\cite{mandrioli2025testing} ensures that individuals with greater dissimilarity to execution scenarios receive fitness values that decrease exponentially.
Finally, the algorithm performs fitness clearing on each population \(P\) (line 16).
The algorithm concludes by returning the updated \(P_s\), \(P_q\), and \(X\) (line 18).

The function \(updatePopulationArchive\) updates the population archives \(X_s\) and \(X_q\), which play a key role in guiding the search process.
Each individual collaborates with archive members to form complete solutions, which are evaluated for fitness.
Several strategies can be employed to update these archives, such as selecting individuals with the highest fitness values or choosing individuals at random~\cite{ma2019survey}.
To balance exploitation and exploration, a hybrid approach may be used, i.e., selecting the top-performing individual along with randomly chosen individuals~\cite{ma2019survey,sharifi2023identifying}.
In our case study, we adopt a hybrid approach that selects the individual with the highest fitness and fills the remaining with individuals that maximize the overall diversity of the archive, as described in \cref{sec:archive-strategy}.
This ensures that high-fitness solutions are retained while preserving diversity within the archive, promoting broader exploration of the search space, and helping to prevent premature convergence on suboptimal solutions.

The function \(breed\) generates new individuals by combining selected parents from the current populations.
This process drives exploration of the search space and supports the discovery of better solutions over successive generations.
It relies on selection, crossover, and mutation operators to produce offspring that inherit characteristics from their parents while introducing variation to maintain diversity.
% All offspring generated through this process should be valid, ensuring they remain compatible with the structure and constraints of the search space.
Our algorithm is designed to be flexible, allowing users to define and configure selection, crossover, and mutation operators tailored to the specific representations of scenarios and perturbations.
This user-defined setup ensures adaptability to various \glspl{as} and problem contexts, while preserving compatibility with the structure and constraints of the search space.
In the domain of \gls{ads} testing, the selection, crossover, and mutation operators described in \cref{sec:genetic-operators} can be applied to both scenarios and perturbations.

\subsection{Root Cause Analysis}\label{sec:rca}

After generating \gls{mr}-violating solutions that elicit behavioral differences between two \glspl{as}, we introduce an interpretability approach designed to analyze the root causes of these differences in \gls{mr} violations.
Our goal is to uncover the underlying reasons behind the observed behavioral differences, enabling engineers to understand under which conditions the two versions behave differently and how significantly those conditions contribute to the differences.
To this end, we extract interpretable rules that describe the behavioral differences between the two versions of an \gls{as}.
These rules specify, in a human-readable format, the conditions under which differences occur and quantify the extent to which each rule contributes to the observed differences in \gls{mr} violations.
This analysis helps identify potential degradations in the updated \gls{as} and enables targeted improvements to enhance system safety, especially by focusing on high-impact conditions.

To extract interpretable rules that explain behavioral differences between two versions of an \gls{as}, we employ \emph{RuleFit}~\cite{friedman2008predictive}, a method that generates human-readable rules derived from decision trees trained on input data.
According to Margot et al.~\cite{margot2021new}, \emph{RuleFit} is the most accurate rule-based algorithm in their evaluation of predictive performance.
% In addition, it is one of the few rule-based algorithms that support regression tasks, making it well-suited for analyzing differences in \gls{mr} violations.
\emph{RuleFit} operates by first training an ensemble of decision trees (we use \glspl{gbdt} in \emph{\gls{ours}}) and converts each tree's decision paths into binary rules, where each rule represents a specific combination of feature conditions (e.g., ego vehicle speed \(> 70 km/h\) and weather \(=\) foggy).
These derived binary rules are then combined with the original features in a sparse linear regression model using Lasso regularization~\cite{tibshirani1996regression}.
Lasso assigns weights to both raw features and derived rules, automatically filtering out less important ones through coefficient shrinkage, thereby improving interpretability and focusing attention on the most relevant patterns.

In our context, we apply \emph{RuleFit} to analyze the differences in \gls{mr} violations between two versions of an \gls{as} by training a model on the generated complete solutions.
Each input vector represents a scenario-perturbation pair, i.e., a complete solution incorporating features from both the source scenario and the follow-up scenario resulting from applying perturbations.
These features typically include continuous and categorical variables, such as ego vehicle initial speed, object positions, and environmental conditions relevant to testing \glspl{ads}.
To ensure compatibility with \emph{RuleFit}, preprocessing such as one-hot encoding is often required for categorical features, converting them into binary indicators that the model can process alongside continuous variables.
The target variable is the observed difference in \gls{mr} violations between the two system versions for each complete solution.
In the end, the \emph{RuleFit} model outputs a set of human-readable IF-clause rules along with coefficients and support values for the original and derived features.
Each rule's coefficient quantifies its contribution to predicting the difference in \gls{mr} violations, while the support value indicates the proportion of complete solutions where the rule's conditions hold.

By examining these rules, we can identify scenario patterns that most strongly influence behavioral differences between the two versions.
For example, a rule such as ``IF ego vehicle speed \(> 70 km/h\) AND foggy weather \(=\) true, THEN the difference in \gls{mr} violation increases by 2.5'' suggests that under high-speed, foggy conditions, the two system versions behave more inconsistently, with the updated system likely exhibiting a higher \gls{mr} violation.
The increased value of 2.5 indicates the magnitude of this effect on the difference in \gls{mr} violations.
If the average difference in \gls{mr} violations across all complete solutions is, for instance, 1.0, this rule highlights a significant deviation from the norm.
Such insights help developers and testers focus their efforts by highlighting specific scenario conditions where updates may have introduced unintended behaviors or safety risks.
These interpretable rules directly support both root cause analysis and actionable decision-making, such as prioritizing which scenario types to investigate further or monitor more closely in future system updates.

\section{Empirical Evaluation}\label{sec:evaluation}

In this section, we empirically evaluate the effectiveness and efficiency of \emph{\gls{ours}} for generating test cases that reveal differences in \gls{mr} violations between two versions of the \gls{as} under test, comparing it against two baseline methods.
We further investigate how the use of constraints and population initialization strategies affects the similarity between the identified test cases and the execution scenarios that represent real-world operating conditions.
Finally, we assess how our interpretability approach helps capture the underlying root causes of the observed differences in \gls{mr} violations between the two \glspl{as} under test.
Specifically, we answer the following \glspl{rq}.

\begin{enumerate}[label={RQ\arabic*}, font=\bfseries, labelsep=0pt, wide=0pt, listparindent=\parindent]
    \item: How effectively can \emph{\gls{ours}} find test cases revealing differences in \gls{mr} violations compared to baseline methods?\label{rq1}

          To answer \labelcref{rq1}, we evaluate the performance of different search methods in terms of the number of distinct complete solutions identified that reveal the differences in violations of predefined \glspl{mr} within a given search budget.
    \item: How efficiently can \emph{\gls{ours}} find test cases revealing differences in \gls{mr} violations compared to baseline methods?\label{rq2}

          To answer \labelcref{rq2}, we examine the speed at which different search methods identify distinct complete solutions that reveal the differences in violations of predefined \glspl{mr}.
    \item: To what extent do the applied constraints and population initialization strategies affect the similarity of identified test cases to execution scenarios?\label{rq3}

          To answer \labelcref{rq3}, we analyze the impact of constraints and population initialization strategies.
          First, we evaluate how these factors affect the similarity between the identified complete solutions and the execution scenarios.
          Next, we assess their effect on the overall effectiveness of the search process.
          Understanding these aspects is essential to determine whether our approach effectively guides the search toward more realistic solutions and to identify any potential negative side effects.

    \item: To what extent does our interpretability approach contribute to identifying the root causes of the differences in \gls{mr} violations?\label{rq4}

          To answer \labelcref{rq4}, we examine the accuracy and support of interpretable rules, extracted from the identified complete solutions, in explaining the observed differences in \gls{mr} violations. Further, we ask four industrial experts to provide feedback on the rules via a questionnaire study.  
\end{enumerate}

\subsection{Subject Systems and Simulation Environment}

Our experiments were conducted using \textsc{Carla}~\cite{dosovitskiy2017carla}, a widely used open-source simulator tailored for autonomous driving research. \textsc{Carla} enables the creation of complex, interactive driving scenarios featuring configurable weather, traffic, and environmental elements. To ensure realistic perception and behavior, we incorporated \textsc{InterFuser}~\cite{shao2022safety}, a state-of-the-art multi-modal end-to-end \gls{ads} that ranks among the top performers on the official \textsc{Carla} \emph{Leaderboard}~\cite{carlaleaderboard}. \textsc{InterFuser} fuses data from LiDAR and Camera sensors to deliver rich scene information.

To compare two realistic iterations of the same \gls{ads}, we considered two versions of \textsc{InterFuser}, namely a reference version (\(V_{\text{ref}}\)) and an updated test version (\(V_{\text{test}}\)). \(V_{\text{ref}}\) corresponds to the original \textsc{InterFuser} model released by its authors. To obtain \(V_{\text{test}}\), we followed the official retraining workflow provided with the framework. Specifically, we generated an additional training dataset using the provided data-collection pipeline. We then retrained the original \textsc{InterFuser} model on this newly collected dataset, thereby yielding an updated version. This construction of \(V_{\text{ref}}\) and \(V_{\text{test}}\) mirrors a common iteration cycle in practical \gls{as} development, where an existing model or component is updated by retraining on newly acquired operational data.

Experiments were executed on a dedicated high-performance server equipped with 56 CPU cores and 2 GPUs (each with 48GB of memory). To increase throughput, we deployed 4 parallel \textsc{Carla} instances in separate Docker containers, leveraging CUDA 12.4 for GPU-accelerated computation. This setup enabled concurrent scenario evaluations at scale using \textsc{Carla} version 0.9.10.1.

\subsection{Baseline Methods}

Since our work is the first to systematically generate test cases aimed at revealing differences in \gls{mr} violations between two versions of an \gls{as}, there are no existing methods tailored for this exact problem. To enable meaningful comparison, we evaluate \emph{\gls{ours}} against two baseline strategies that reflect commonly used search paradigms in test generation.

\begin{itemize}
    \item \textbf{\emph{\Acrfull{rs}:}}\glsunset{rs} This baseline involves generating test cases through uniform random sampling without any guidance from fitness feedback. Although simple, \emph{\gls{rs}} provides a useful reference point to assess the inherent difficulty of the search and the value of guided optimization.

    \item \textbf{\emph{\Acrfull{sga}:}}\glsunset{sga} This variant applies a standard genetic algorithm~\cite{holland1992adaptation} using a single population, treating the entire solution space as a uniform search domain. Unlike our method, it does not differentiate between scenario structure and perturbation behaviors. Comparing with \emph{\gls{sga}} helps highlight the benefits of our cooperative co-evolutionary setup.
\end{itemize}

By including these two baselines, we aim to illustrate both the relative performance gains achieved through guided search and the contribution of our modular co-evolutionary design to the quality and diversity of the generated test cases.

To investigate the impact of constraining mechanisms and population initialization strategies on the effectiveness of \emph{\gls{ours}}, we designed different variants of \emph{\gls{ours}} with specific configurations.
The first configuration represents the standard constrained version, where the search is forced to remain within a specified similarity to recorded execution scenarios.
In the second configuration, denoted as \emph{\gls{ours}\textbackslash{c}}, the constraint was removed to assess its influence on the quantity and quality of generated test cases. To evaluate the effect of the initialization strategy, we introduced an alternative method, termed \emph{\acrfull{ri}}, in which the initial population is sampled directly from previously collected execution scenarios rather than generated randomly. We applied this initialization strategy to both the constrained and unconstrained variants, yielding 4 experimental configurations in total.

To assess the interpretability of the rules extracted by our approach, we compare \emph{RuleFit} used in \emph{\gls{ours}} against two baseline methods, i.e., \emph{\acrfull{rf}}\glsunset{rf} and \emph{\acrfull{gbdt}}\glsunset{gbdt}, both commonly used for regression tasks.

\subsection{Execution Scenarios}

We collected a diverse set of execution scenarios by running \textsc{InterFuser} in \textsc{Carla} \emph{Leaderboard} environments.
\textsc{Carla} \emph{Leaderboard} is a standardized evaluation framework that enables running an \gls{ads} through a variety of driving tasks in simulated urban environments. We collected these scenarios across four towns and eight distinct weather conditions, including clear skies, light rain, and heavy rain, under varying lighting conditions, e.g., noon and sunset.

To ensure temporal and spatial diversity, we sampled execution scenarios at fixed time intervals (every 5 seconds), with a random offset from the start of each route execution.
We also limited the number of sampled scenarios per route-weather configuration to encourage balanced coverage, resulting in approximately 800 distinct execution scenarios used for our experiments.
These execution scenarios are used for the constraining mechanism and population initialization strategy in \emph{\gls{ours}} and its variants.

\subsection{Metamorphic Relations}

\begin{table}[htbp]
    \small
    \caption{\glspl{mr} involved in the experiments.}\label{tab:mrs}
    \begin{threeparttable}
        \begin{tabularx}{\linewidth}{llX}
            \toprule
            \textbf{Category} & \textbf{\gls{mr}}                 & \textbf{Description}                                                                                                                                                                                                                        \\
            \midrule
            \multirow{5}{*}{\rotatebox[origin=c]{90}{\makecell{\(GP_1\)                                                                                                                                                                                                                                         \\ {\footnotesize Speed Reduction}}}} & \(MR_1\)\tnote{*}                       & If a pedestrian appears on the roadside, then the ego vehicle should slow down.                                                                                                                                                                                       \\
                              & \(MR_2\)\tnote{*,\({\dagger}\)}   & If the driving time changes into night, then the ego vehicle should slow down.                                                                                                                                                              \\
                              & \(MR_3\)\tnote{\({\dagger}\)}     & Adding a vehicle in the front of the ego vehicle, the speed of the ego vehicle should decrease.                                                                                                                                             \\
                              & \(MR_4\)\tnote{\({\dagger}\)}     & Adding a pedestrian in the front of the ego vehicle, the speed of the ego vehicle should decrease.                                                                                                                                          \\
                              & \(MR_5\)\tnote{\({\dagger}\)}     & Changing from sunny to rainy, the speed of the ego vehicle should decrease.                                                                                                                                                                 \\
            \midrule
            \multirow{12}{*}{\rotatebox[origin=c]{90}{\makecell{\(GP_2\)                                                                                                                                                                                                                                        \\Actor Invariance}}}                        & \(MR_6\)\tnote{\({\ddagger}\)}    & In a given driving scenario where the ego vehicle detects a target obstacle and attempts to avoid a collision, the ego vehicle should keep the steering unchanged when the speed of the ego vehicle changes (increased or decreased by a certain factor). \\
                              & \(MR_7\)\tnote{\({\ddagger}\)}    & In a given driving scenario where the ego vehicle detects a target obstacle and attempts to avoid a collision, the ego vehicle should keep the steering unchanged when adjusting (i.e., scaled down or up) the size of the ego vehicle.     \\
                              & \(MR_8\)\tnote{\({\ddagger}\)}    & In a given driving scenario where the ego vehicle detects a target obstacle and attempts to avoid a collision, the ego vehicle should keep the steering unchanged when adjusting (i.e., scaled down or up) the size of the target obstacle. \\
                              & \(MR_9\)\tnote{\({\ddagger}\)}    & In a given driving scenario where the ego vehicle detects a target obstacle and attempts to avoid a collision, the ego vehicle should keep the steering unchanged when changing the position of the ego vehicle.                            \\
                              & \(MR_{10}\)\tnote{\({\ddagger}\)} & In a given driving scenario where the ego vehicle detects a target obstacle and attempts to avoid a collision, the ego vehicle should keep the steering unchanged when changing the speed of the target obstacle.                           \\
                              & \(MR_{11}\)\tnote{\({\ddagger}\)} & In a given driving scenario where the ego vehicle detects a target obstacle and attempts to avoid a collision, the ego vehicle should keep the steering unchanged when adding additional actors.                                            \\
            \bottomrule
        \end{tabularx}
        \begin{tablenotes}
            \footnotesize
            \item[*] These \glspl{mr} are derived from study~\cite{deng2023declarative}.
            \item[\({\dagger}\)] These \glspl{mr} are derived from study~\cite{deng2021bmt}.
            \item[\({\ddagger}\)] These \glspl{mr} are derived from study~\cite{iqbal2024metamorphic}.
        \end{tablenotes}
    \end{threeparttable}
\end{table}

We used a set of 11 \glspl{mr} from \cref{tab:mrs}, drawn from established research~\cite{deng2023declarative,deng2021bmt,iqbal2024metamorphic}, selected \annotate{Comment 1.3}\deletedtext{based on their}\addedtext{for} compatibility with both the \gls{ads} under test and the simulation platform. Some \glspl{mr} from prior literature were excluded due to incompatibility with the constraints of our simulator.
For instance, an \gls{mr} \deletedtext{requiring}\addedtext{that requires replacing} background buildings\deletedtext{ to be replaced} with trees could not be used, as it would require modifying the static map provided by \textsc{Carla}, which is not feasible within our experimental setup.
\addedtext{
When applying these \glspl{mr}, we ensured that the generated source scenarios satisfied their preconditions.
For example, to apply \(MR_6\), the source scenario must contain a target obstacle that the ego vehicle can detect and attempt to avoid.
Consequently, the selected \glspl{mr} are both valid and implementable within the \textsc{Carla} simulation environment.
Although these \glspl{mr} are specified at a high level of abstraction, they can be concretely instantiated through appropriate parameterization and scenario configuration.
For instance, changing the position of the ego vehicle can be realized by adjusting its spawn point from one lane to another in the \textsc{Carla} simulator.
}

To facilitate evaluation, we grouped the \glspl{mr} into two categories, i.e., \(GP_1\) and \(GP_2\), based on their output behavior.
\(GP_1\) includes \glspl{mr} with expected speed reductions in response to environmental or traffic changes.
\(GP_2\) captures invariance-based relations, i.e., changes in actors should not significantly alter the vehicle's steering behavior.
% These invariance constraints were defined using a 1-degree steering deviation threshold.
% Thresholds used in the output relations were derived from prior work and validated through pilot studies.
% These preliminary runs ensured that selected thresholds aligned with practical driving expectations and simulator fidelity, even though our goal was not to optimize them.

In the following subsections, we report the results for \(GP_1\).
The outcomes for \(GP_2\) are consistent with those of \(GP_1\), thereby demonstrating the robustness of \emph{\gls{ours}} across different \gls{mr} groups.
To avoid redundancy, the detailed results for \(GP_2\), including all supporting figures, are presented in the appendix.

\subsection{Parameter Settings}\label{sec:setting}

All methods were executed with a simulation budget of \deletedtext{200}\addedtext{500} scenario evaluations, chosen based on pilot experiments that demonstrated sufficient convergence behavior under this limit. To ensure fair comparisons, each method was allowed to use up to \deletedtext{200}\addedtext{500} scenario evaluations. \annotate{Comment 3.14}\addedtext{Additionally, to mitigate the impact of simulator nondeterminism, we applied a re-evaluation procedure in the simulator-based violation assessment. Each test case was first simulated once, and if its resulting \gls{mr} violation exceeded a minimum threshold of \(0.2\), it was re-executed twice. The final violation value was then obtained by aggregating the three violation values using their median. This strategy was adopted to limit the effect of simulation noise on promising candidates while keeping the computational cost manageable.}

We configured each method to use a population of 5 individuals, balancing between computational cost and search effectiveness. Given the high cost of each fitness evaluation (approximately two minutes per test case), this smaller population size aligns with common practice in search-based testing literature~\cite{benabdessalem2020automated,chugh2017survey}.
For both \emph{\gls{ours}} and \emph{\gls{sga}}, we adopted tournament selection with a size of 3, which strikes a good balance between exploration and exploitation~\cite{whitley2018next,lavinas2018experimental,kaelo2007integrated}. This choice also helps avoid premature convergence, especially in small populations.
\annotate{Comment 2.5}\addedtext{
The \emph{\gls{sga}} baseline employs the same fitness function as the joint fitness function used in \emph{\gls{ours}}, but applied to a single population whose individuals have the same structure, consisting of both source scenarios and perturbations.
Accordingly,}
we employed the same mutation and crossover operators for both \emph{\gls{ours}} and \emph{\gls{sga}}, with the mutation and crossover rates set to 0.2 and 0.8, respectively, consistent with recommendations in the evolutionary computation literature~\cite{mirjalili2018genetic,sette2001genetic}.
As for \emph{\gls{ours}}, which relies on cooperative co-evolution, we tuned its specific hyperparameters via lightweight exploratory runs. The final settings included an archive size of 3 per population\deletedtext{, with crossover and mutation rates mirroring those used in \emph{\gls{sga}}}.
The penalty severity constant \(c\) in \cref{eq:penalization} was set to 6.66 based on pilot experiments that balanced constraint satisfaction and search exploration.

Due to the nature of population-based algorithms, some methods may slightly exceed the \deletedtext{200-simulation}\addedtext{500-simulation} budget per run, since evaluations occur in batches.
In such cases, the actual number of simulations consumed per generation can vary between methods, leading to misalignment in budget usage.
To ensure consistent comparisons, we applied linear interpolation on all evaluation metrics at points where the actual simulations deviated from the predefined budget.
This process estimates metric values precisely at the intended search budgets by interpolating between the closest available points, enabling a fair assessment under an aligned computational budget.

\subsection{Evaluation Metrics}

To evaluate the effectiveness of different methods in identifying test cases that reveal differences in \gls{mr} violations between two \glspl{ads}, we introduce the following metrics:

\begin{itemize}
    \item \textbf{\emph{\Gls{ds}}} denotes the number of distinct complete solutions that expose significant behavioral differences between the \gls{ads} versions. We require such solutions to satisfy the following conditions:
          \begin{enumerate*}
              \item its fitness value, i.e., the measured difference in \gls{mr} violations across \gls{ads} versions, exceeds a predefined fitness threshold \(\theta_f\), and
              \item it is sufficiently different from other selected solutions based on a minimum distance threshold \(\theta_d\).
          \end{enumerate*}
          Formally, let \(CS_V\) be the set of complete solutions identified by method \(V\) that satisfy the above criteria. The number of distinct solutions is defined as \(DS(V) \coloneq |CS_V|\).

    \item \textbf{\emph{Fitness}} quantifies the degree to which a complete solution exposes behavioral differences in \gls{mr} violations between the two \glspl{ads}, with larger absolute values indicating more pronounced differences.
\end{itemize}

To evaluate the efficiency of different methods, we analyze the \emph{\gls{ds}} metric as a function of the search budget and employ the following additional metrics:

\begin{itemize}
    \item \textbf{\emph{\Gls{auc}}} measures the area under the \emph{\gls{ds}}-budget curve, providing an aggregate measure of how quickly a method accumulates distinct complete solutions over the course of the search, where higher values indicate that a method finds more distinct solutions earlier during the search.
          This metric is particularly suitable for our setting, as it captures both the speed and volume of distinct solutions discovered, two central factors in assessing test generation efficiency.
          It allows us to compare algorithms not only at final budget levels but throughout the entire simulation horizon.

    \item \textbf{\emph{Execution Time}} records the wall-clock time taken by each method to complete the search within the allocated budget.
          This metric provides complementary insights into the algorithms' computational efficiency, highlighting differences in their computational costs and offering a practical perspective on their suitability for cost-sensitive testing scenarios.
\end{itemize}

To evaluate the realism of the generated test cases, we employ the following metric:

\begin{itemize}
    \item \textbf{\emph{\Gls{aed}}} measures how closely the generated complete solutions resemble real-world execution scenarios.
          For each test case, we compute its distance to the nearest execution scenario (see \cref{eq:dissimilarity}) and then aggregate these values by averaging over all test cases in a given run.
          This metric provides insights into how well the generated scenarios align with observed real-world scenarios.
\end{itemize}

To evaluate the interpretability of the rules extracted from the identified complete solutions, we utilize the following metrics:

\begin{itemize}
    % \emph{\Gls{mae}} is a metric used to evaluate a model's prediction accuracy in a scale-independent manner, which enables fair comparisons of predictive performance across different methods.
    \item \textbf{\emph{\Gls{mae}}} is a widely used metric for evaluating the accuracy of predictive models, measuring the average magnitude of errors between predicted and actual values~\cite{willmott2005advantages,hyndman2006another}.
          In our context, \emph{\gls{mae}} quantifies how accurately the interpretable rules characterize the differences in \gls{mr} violations between two versions of the \gls{ads} under test.
          A lower \emph{\gls{mae}} value indicates greater model accuracy in capturing behavioral differences.
          % Specifically, \emph{\gls{mase}} is calculated by dividing the \emph{\gls{mae}} of a model's predictions for differences in \gls{mr} violations by the \emph{\gls{mae}} of a naive predictor, which we define as using the median of the observed behavioral differences~\cite{friedman2008predictive}:
          Specifically, \emph{\gls{mae}} calculates the mean of the absolute differences between each pair of predicted and true values:

          % \begin{equation}
          %     MASE \coloneq \frac{\sum_{i \in \mathcal{SP}^\prime} \left|y_i - \hat{y}_i\right|}{\sum_{i \in \mathcal{SP}^\prime} \left|y_i - median(y)\right|}
          % \end{equation}
          \[
              MAE \coloneq \frac{1}{\left| \mathcal{SP}^\prime \right|} \sum_{i \in \mathcal{SP}^\prime} \left|y_i - \hat{y}_i\right|
          \]
          where \(\mathcal{SP}^\prime \) denotes the set of complete solutions, \(y_i\) is the observed difference in \gls{mr} violations for a given complete solution, and \(\hat{y}_i\) is the predicted value.
          % , and \(median(y)\) represents the median of all observed differences.

    \item \textbf{\emph{Support}} evaluates the extent to which the extracted interpretable rules cover the identified complete solutions, which is essential for their practical usefulness.
          A complete solution is considered supported if there exists at least one rule whose conditions are satisfied by the solution.
          A higher support indicates that the rules effectively capture the characteristics of a larger portion of the identified complete solutions.
\end{itemize}

\subsection{Experiment Design}

\subsubsection{Design for RQ1}

To answer \labelcref{rq1}, we evaluate the effectiveness of \emph{\gls{ours}} in identifying test cases, i.e., complete solutions, that amplify \addedtext{safety-related} behavioral differences between two versions of \textsc{InterFuser}.
Unlike existing \gls{mr} approaches that target outright violations of predefined properties, our objective is to identify solutions that exhibit large discrepancies in their effects on the two \gls{ads} versions. We compare \emph{\gls{ours}} against the two baseline methods under the same search budget, using the \emph{\gls{ds}} metric as the primary indicator of effectiveness.

To ensure a comprehensive evaluation, we assess each method under multiple configurations of the fitness threshold \(\theta_f\) and distance threshold \(\theta_d\).
The fitness threshold controls the required degree of behavioral difference between the two \glspl{ads}, reflecting the severity of the identified complete solutions; higher values correspond to greater differences.
The distance threshold, on the other hand, governs the minimal dissimilarity between retained solutions, thus controlling the diversity of the complete solutions.
By exploring a wide range of threshold values, we can compare the performance of different approaches under varying demands for the severity and diversity of the generated complete solutions.
Specifically, we use \(\theta_f \in \{1.0, 1.3, 1.5, 1.8, 2.0, 2.3\} \) to cover a meaningful spectrum of \deletedtext{behavioral differences observed across all methods}\addedtext{severity constraints applied on the generated test cases}.
This range was chosen to approximately span the 80th percentile of the fitness values observed in our experiments (around 2.3), ensuring that the selected thresholds capture realistic and informative levels of severity.
Solutions with fitness below 1.0 are excluded due to their limited discriminative value and minimal impact\deletedtext{ on differentiating between the two \glspl{ads}}.
For \(\theta_d\), we consider values from 0.0 to 4.0 in increments of 0.4, covering a range of diversity requirements from lenient to stringent.
The upper bound was set to the 95th percentile of pairwise distances across all generated solutions, ensuring the threshold range meaningfully reflects the observed distribution of solution diversity.

To complement the \emph{\gls{ds}} metric, we also assess the quality of solutions generated by each method based on their fitness values.
Specifically, we analyze the distribution of fitness values across all complete solutions produced by each approach, where a higher fitness indicates greater differences in \gls{mr} violations.
This provides insight into each method's capacity to uncover critical test cases\deletedtext{ that induce pronounced behavioral differences between the two \glspl{ads}}.

\subsubsection{Design for RQ2}

To address \labelcref{rq2}, we analyze the rate at which different methods discover complete solutions over the course of their search. To this end, we track the number of distinct solutions identified by each method as a function of the search budget, expressed as a percentage of the total number of simulations allowed.
The methodology follows that of \labelcref{rq1}, employing the same hyperparameters and repeating each method 10 times.
\deletedtext{The only difference is that the search budget is increased from 200 to 500 simulations. This larger budget provides a much stronger basis for evaluating the algorithms' efficiency over extended runs, yielding more reliable insights into their efficiency over time. The drawback, however, is that it significantly increases execution time, which can limit practicality in fast-paced development settings where quick feedback is essential.}

In addition to analyzing \emph{\gls{ds}} as a function of the search budget, we compute the \emph{\gls{auc}} of each \emph{\gls{ds}}-budget curve and measure the execution time for each method, providing a comprehensive assessment of their efficiency in generating distinct complete solutions.

\subsubsection{Design for RQ3}

To address \labelcref{rq3}, we analyze how different configurations of \emph{\gls{ours}} affect both the similarity of the identified test cases to \deletedtext{real-world} execution scenarios and the overall effectiveness of the search process.
Each variant was executed independently 10 times to account for stochastic variation in the search process.
For all runs, we employed three metrics:
\begin{enumerate*}
    \item the \emph{\gls{ds}} metric, capturing the number of generated distinct solutions,
    \item the average fitness of the test cases found by each variant, quantifying the capacity of uncovering critical test cases with severe behavioral discrepancies, and
    \item the \emph{\gls{aed}} metric, measuring how closely the generated test cases resemble real-world execution scenarios.
\end{enumerate*}

\subsubsection{Design for RQ4}\label{subsec:rq4}

To answer \labelcref{rq4}, we begin by assessing the accuracy of the interpretable rules derived from the complete solutions by evaluating the \emph{\gls{mae}} metric.
\emph{RuleFit}, used in \emph{\gls{ours}}, as well as the two baseline methods, \emph{\gls{rf}} and \emph{\gls{gbdt}}, are trained on the same dataset of complete solutions paired with their corresponding differences in \gls{mr} violations.
This dataset combines results from all search methods used in \labelcref{rq1} and \labelcref{rq3} and contains two types of complete solutions.
The first type includes solutions where the two systems exhibit observable behavioral differences.
Each of these solutions is assigned a fitness value that reflects both the direction and severity of the difference, with positive values indicating better performance by the original system and negative values indicating better performance by the updated system.
The second type consists of solutions where no behavioral difference is observed between the systems.
These solutions are assigned a fitness value of zero.
This combination allows the prediction methods to learn not only what conditions lead to discrepancies, but also which solutions result in consistent behavior across both versions.
By training on this combined dataset, \emph{RuleFit} and baseline methods can identify the conditions that cause the systems to diverge, while also recognizing when their behaviors remain the same.
% learn both the conditions that lead to system divergence and those that do not, allowing them to distinguish behavioral discrepancies while grounding them against baseline cases with no behavioral differences.
The combined dataset consists of 2,573 complete solutions, with 1,967 solutions exhibiting behavioral differences and 606 solutions showing no differences, split into 80\% for training and 20\% for validation.
To account for randomness during training, each method is evaluated across 500 independent runs with different random seeds to ensure robust and generalizable results.
To facilitate interpretability, \emph{RuleFit} is constrained to extract at most 30 rules per run.
Since our goal is explanation rather than maximal predictive accuracy, a smaller, high-importance rule set is easier to audit and apply in practice.
This threshold was chosen based on both empirical observations and considerations of human interpretability.
In preliminary experiments, we varied the maximum number of rules while keeping the dataset of complete solutions fixed and evaluated the resulting prediction accuracy.
The model achieved its highest accuracy when the limit was set to approximately 30 rules.
From a cognitive perspective, prior work on model interpretability suggests that human analysts can effectively process only a limited number of rules before comprehension and decision quality degrade~\cite{lakkaraju2016interpretable}.
Therefore, limiting the output to 30 rules preserves most of the explanatory power while keeping the results concise and practical for human review.
For a fair comparison, the hyperparameters of the baseline methods, including the maximum tree depth, the maximum number of leaf nodes, the number of trees, and other relevant parameters, were tuned using randomized search~\cite{bergstra2012random}, a hyperparameter optimization technique that samples parameter configurations from predefined ranges, aiming to approximate the best-performing settings for each method on the dataset.

To understand how thoroughly the rules support the identified complete solutions with behavioral differences (1,967 solutions), we analyze the distribution of the rules' \emph{support} across 500 independent runs of \emph{RuleFit}.
Specifically, in each run, we compute the proportion of complete solutions that are supported by the full set of extracted rules.
This analysis reveals the extent to which the rules capture the behavioral discrepancies reflected in the identified complete solutions.
A higher proportion indicates stronger support or broader coverage, suggesting that the rules provide a more comprehensive explanation of the behavioral differences between the two system versions.

\annotate{Comments 1.2, 2.3, and 3.15}\addedtext{
To complement the quantitative evaluation of the interpretability framework, we conducted a questionnaire study with four industrial domain experts to evaluate the practical value of the generated rules. The four participants are professionals in the autonomous driving domain with experience ranging from 2 to 9 years. Their expertise primarily covers perception, decision-making, simulation and testing, and safety-related aspects of autonomous systems. All participants have prior involvement in safety-critical systems or risk assessment, most with extensive experience.
Informed consent was obtained from all participants prior to the study for the use of their feedback. \Cref{fig:questionnaire_excerpt} presents a sample item from the questionnaire presented to the experts in this study. The full questionnaire is available in the replication package.

\begin{figure}[t]
\centering
\begin{tcolorbox}[
    width=\linewidth,
    colback=gray!3,
    colframe=black!60,
    boxrule=0.5pt,
    arc=2mm,
    left=2mm,
    right=2mm,
    top=1.5mm,
    bottom=1mm,
    fonttitle=\small\bfseries,
    colbacktitle=gray!15,
    coltitle=black
]
\small
% \centering
% {\fontsize{18}{22}\selectfont \bfseries Rule 1\par}
% \vspace{1.0mm}
\begin{tcolorbox}[
    colback=white,
    colframe=black!20,
    boxrule=0.3pt,
    arc=1mm,
    title=\textbf{Conditions Under Which the Two ADS Versions Differ},
    fonttitle=\bfseries\small,
    colbacktitle=blue!5,
    coltitle=black
]
Vehicles to the left of the ego vehicle are traveling at fairly consistent speeds. Static obstacles ahead are directly in front of or to the right of the ego vehicle.
\end{tcolorbox}

\vspace{0.2mm}

\begin{tcolorbox}[
    colback=white,
    colframe=black!20,
    boxrule=0.3pt,
    arc=1mm,
    title=\textbf{Observed Behavioral Discrepancy},
    fonttitle=\bfseries\small,
    colbacktitle=blue!5,
    coltitle=black
]
On average, the updated system performs moderately better under these conditions than the original system.
\end{tcolorbox}

\vspace{0.2mm}

\begin{tcolorbox}[
    colback=white,
    colframe=black!20,
    boxrule=0.3pt,
    arc=1mm,
    title=\textbf{Occurrence Frequency},
    fonttitle=\bfseries\small,
    colbacktitle=blue!5,
    coltitle=black
]
This rule applies to 5.74\% of all detected behavioral discrepancies.
\end{tcolorbox}

\vspace{0.5mm}

\begin{tcolorbox}[
    colback=white,
    colframe=black!20,
    boxrule=0.3pt,
    arc=1mm,
    title=\textbf{An Example Scenario Where This Rule Applies},
    fonttitle=\bfseries\small,
    colbacktitle=blue!5,
    coltitle=black,
    height=2.8cm,
    valign=center
]
\centering
\emph{Side-by-side recordings of the two ADS versions executing the corresponding follow-up scenario.}
\end{tcolorbox}

\vspace{1.5mm}

\begin{tcolorbox}[
    colback=white,
    colframe=black!20,
    boxrule=0.3pt,
    arc=1mm,
    title=\textbf{Questions},
    fonttitle=\bfseries\small,
    colbacktitle=green!5,
    coltitle=black
]
\textbf{Q1.} How easy is it to understand the conditions described by this rule?\\ \\
1 = Very difficult \\
2 = Difficult \\
3 = Moderately easy \\
4 = Easy \\
5 = Very easy \\

\vspace{1mm}

\textbf{Q2.} Does this rule help you narrow down which types of conditions to prioritize for testing the updated system? \\ \\
1 = Not at all \\
2 = Slightly \\
3 = Moderately \\
4 = Considerably \\
5 = Very much \\
\end{tcolorbox}

\end{tcolorbox}
\caption{Example of one questionnaire item used in the expert study. For each rule, experts were shown a natural-language description of the conditions, the observed behavioral discrepancy, the occurrence frequency, an illustrative example scenario, and the Likert-scale questions used for evaluation.}
\label{fig:questionnaire_excerpt}
\end{figure}

Each expert was provided with the top 10 interpretable rules, ranked by importance, that characterize the behavioral discrepancies between two \gls{ads} versions.
The rules were first extracted from raw \emph{RuleFit} output and translated into natural-language descriptions that are easier for practitioners to understand. This translation was performed through a systematic post-processing step described below. 

Each rule consists of one or more conditions connected by logical conjunctions. To interpret a rule, we first analyzed each condition by decomposing the corresponding feature name into its semantic components. 
For example, the feature named \(follow\_up\_walker\_left\_yaw\_max\_sin\) is decomposed into: follow-up scenario, pedestrian, front-left region, heading direction, maximum value, and sine component. Accordingly, this feature represents the maximum sine value of the heading direction of pedestrians located in the front-left region of the ego vehicle in the follow-up scenario.

Next, threshold values were interpreted in accordance with the attribute's meaning and range. For directional attributes such as \emph{yaw} and \emph{angle}, values were interpreted relative to the ego vehicle's heading using \textsc{Carla}'s left-hand coordinate system. For instance, the condition \emph{\(follow\_up\_walker\_left\_yaw\_max\_sin > 0.4\)} indicates that, in the follow-up scenario, at least one pedestrian ahead on the left of the ego vehicle has a heading direction with a positive sine value of significant magnitude. Since positive sine values indicate rightward movement, and negative values indicate leftward movement, this condition can be interpreted as indicating that at least one such pedestrian is moving rightward.
For distance and speed attributes, threshold values are interpreted relative to the scaled range of each feature, where values closer to zero indicate closer proximity or lower speed, and higher values indicate greater distance or higher speed.

After each condition in a rule was interpreted, the resulting condition-level interpretations were combined to derive the overall meaning of the rule. Finally, the resulting rule interpretation was reformulated into a coherent scenario description through minor manual rephrasing to improve clarity and readability. This final description preserves the rule's essential content while presenting it in a form easier for engineers to read and understand.

Each translated rule was presented together with its expected relative effect, its occurrence frequency across the generated test cases, and an illustrative example scenario in which the rule applies.
The experts were then asked to assess the understandability and actionability of each rule by answering two questions on a five-point Likert scale~\cite{likert1932technique}:
\begin{enumerate*}[label = (Q\arabic*)]
    \item ``How easy is it to understand the conditions described by this rule?'' (Rate from 1\,=\,Very difficult to understand to 5\,=\,Very easy to understand), and % chktex 38
    \item ``Does this rule help you narrow down which types of conditions to prioritize for testing the updated system?'' (Rate from 1\,=\,Not at all to 5\,=\,Very much). % chktex 38
\end{enumerate*}
}

Finally, we present \deletedtext{an}\addedtext{several} example rule\addedtext{s} extracted from one of the \emph{RuleFit} runs to illustrate the interpretability and practical value of the generated rules.
It showcases the interpretability approach's effectiveness in gaining a deeper understanding of the underlying factors contributing to the observed behavioral discrepancies.

\section{Results and Analysis}\label{sec:results}

In this section, we present the results and analysis of our empirical evaluation of \emph{\gls{ours}}.

% \todo{
% - Report the absolute numbers rather than the percentages when the absolute numbers are so low (just one or two).

% - Unsubstantiated claim regarding GP1 and GP2: The statement ``The outcomes for GP2 are consistent with those of GP1'' must be referenced (as it's in the appendix) or clarified/removed.

% - I wondered why SGA performs worse than random? Are you able to offer any reasoning behind this? Typically, anything worse than random is considered an unsuitable baseline. Along similar lines, can you explain why CoCoMagic is also worse than random for smaller budgets?
% }

\subsection{RQ1: Effectiveness of \texorpdfstring{\gls{ours}}{CoCoMagic}}

\begin{figure*}[htbp]
    \centering
    \includegraphics[width=\linewidth]{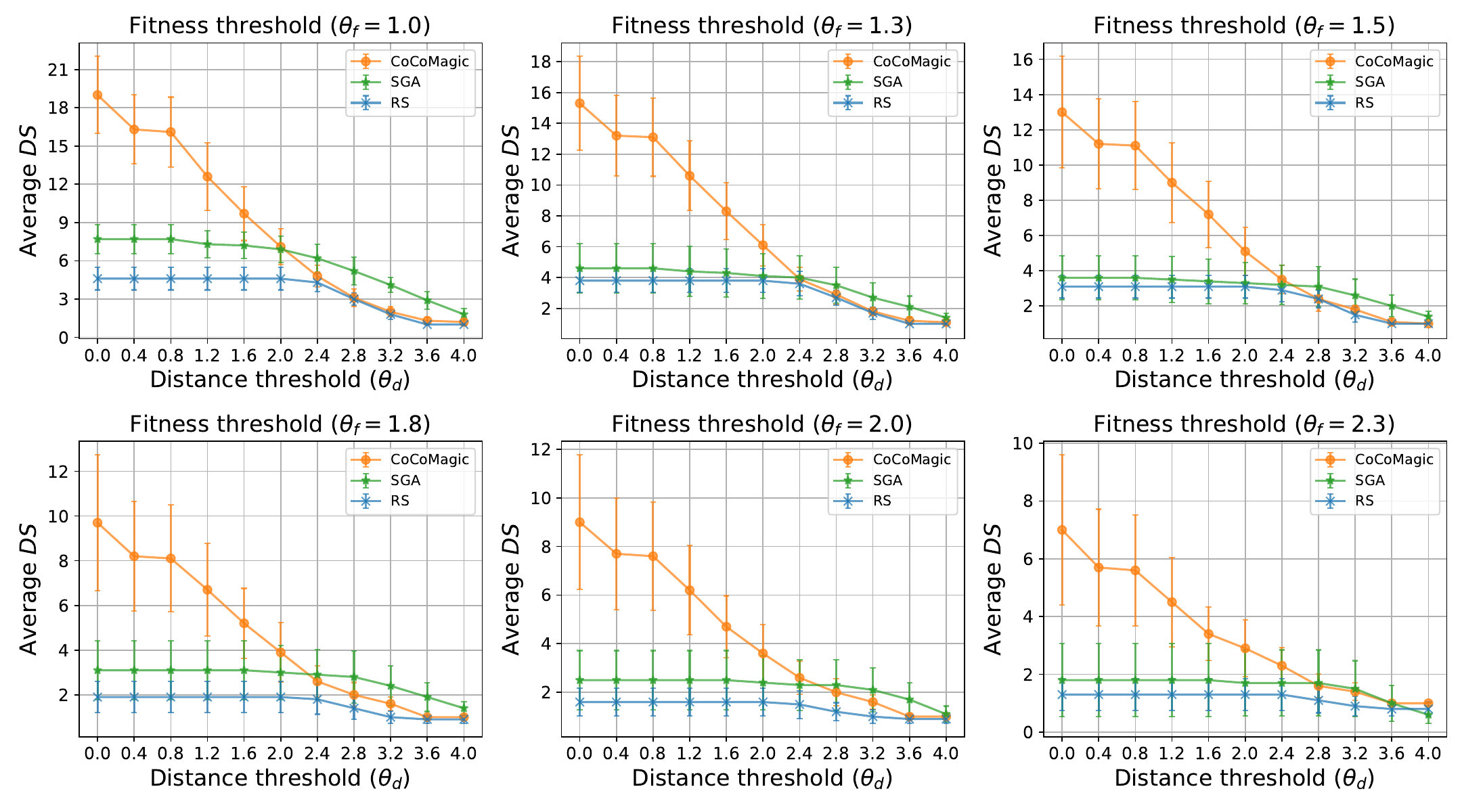}
    \caption{
        \emph{\Acrfull{ds}} vs.\ distance threshold (\(\theta_d\)) across different fitness thresholds (\(\theta_f\)).
        A higher \emph{\gls{ds}} value indicates more distinct test cases discovered by the method.
        Each curve plots the average \emph{\gls{ds}} value at different \(\theta_d\) settings, under a specific fitness threshold \(\theta_f\).
        This figure reveals how each method balances the quantity and diversity of test cases as \(\theta_f\) varies.
    }\label{fig:ds}
    \Description{Distinct solutions vs.\ distance threshold across different fitness thresholds.}
\end{figure*}

\Cref{fig:ds} reports the average \emph{\gls{ds}} values across 10 distinct experiments, achieved by \emph{\gls{ours}}, \emph{\gls{sga}}, and \emph{\gls{rs}}, at varying \(\theta_d\) and \(\theta_f\).
Each subplot shows the average \emph{\gls{ds}} values over executions, as a function of \(\theta_d\), with 95\% confidence intervals represented by error bars. Different subplots correspond to different fitness thresholds \(\theta_f\).

\addedtext{The performance advantage of \emph{\gls{ours}} is most evident at low and moderate distance thresholds. For every fitness threshold
\(\theta_f\), \emph{\gls{ours}} achieves the highest average \emph{\gls{ds}} values when
\(\theta_d\)
is relatively small, often by a large margin over both \emph{\gls{sga}} and \emph{\gls{rs}}. This trend remains consistent as \(\theta_f\) is increased, indicating that \emph{\gls{ours}} is substantially more effective at discovering a larger set of complete solutions across all severity constraints.

As \(\theta_d\) increases, the average \emph{\gls{ds}} value decreases for all methods, which is expected since stricter diversity requirements lead to the exclusion of more solutions, and thereby reduce the number of retained complete solutions. However, the rate and implications of this decrease differ across methods. \emph{\gls{ours}} shows a sharper decline as it begins with a much larger pool of archived solutions, and progressively fewer of them satisfy increasingly strict distinctness requirements. In contrast, \emph{\gls{sga}} and \emph{\gls{rs}} remain relatively flat over a wider range of \(\theta_d\). Specifically, \emph{\gls{sga}} begins to outperform \emph{\gls{ours}} under strict diversity requirements and relatively permissive fitness thresholds (\(\theta_d \geq 2.4\) and \(\theta_f \leq 1.8\)). This indicates that \emph{\gls{sga}} may produce a solution set with slightly greater diversity, which only becomes advantageous under sufficiently strict distance thresholds.
%although \emph{\gls{ours}} is more effective at finding a larger number of distinct solutions overall, \emph{\gls{sga}} tends to retain a slightly more dispersed subset. 
In other words, the main advantage of \emph{\gls{ours}} lies in the low-to-moderate \(\theta_d\) region, where it consistently produces more archived distinct solutions, whereas \emph{\gls{sga}} becomes competitive only after aggressive diversity filtering removes much of that advantage. \emph{\gls{rs}} remains generally inferior throughout, and only approaches the other methods when all techniques yield very few remaining solutions.
}

\deletedtext{
Across almost all settings, \emph{\gls{ours}} consistently outperforms both \emph{\gls{sga}} and \emph{\gls{rs}} in terms of discovering a greater number of distinct complete solutions that highlight behavioral differences between \gls{ads} versions.
Notably, under stricter fitness thresholds (\(\theta_f \geq 1.0\)), \emph{\gls{ours}} surpasses the baseline methods across all distance thresholds, demonstrating its strong capability in uncovering solutions that reveal large behavioral differences between the two \glspl{ads}.
For moderate fitness and distance thresholds (\(\theta_f = 1.0, \theta_d = 2.0\)), \emph{\gls{ours}} identifies an average of 2.2 distinct solutions within the specified budget, outperforming \emph{\gls{sga}} and \emph{\gls{rs}}, which yield only 1.0 and 0.9 distinct solutions, respectively.

% On average, \emph{\gls{ours}} discovers 108\% (\(p\)-value \(< 10^{-5}\)) more distinct solutions than \emph{\gls{sga}} and 287\% (\(p\)-value \(< 10^{-13}\)) more than \emph{\gls{rs}}, underscoring its effectiveness in uncovering behavioral differences between the two \gls{ads} versions.
% To assess the statistical significance of these differences, we conducted non-parametric Mann-Whitney U tests with Fisher's method for \(p\)-value correction~\cite{mann1947test}.

Varying the fitness threshold \(\theta_f\) reveals how each method adapts to increasing demands for the severity of behavioral differences in the generated solutions.
As \(\theta_f\) increases from 0.5 to 2.0, \emph{\gls{ds}} naturally decreases across all methods, since fewer solutions exhibit large enough behavioral differences to surpass stricter fitness criteria. Nonetheless, \emph{\gls{ours}} consistently maintains a clear lead over both \emph{\gls{sga}} and \emph{\gls{rs}} at every fitness threshold, indicating its robustness in discovering complete solutions even under stringent thresholds.
This advantage is particularly evident at higher fitness thresholds (\(\theta_f \in \{1.5, 2.0\} \)), where \emph{\gls{ours}} continues to discover more solutions than both \emph{\gls{sga}} and \emph{\gls{rs}}, who struggle to identify hardly any high-fitness solutions.
As the distance threshold \(\theta_d\) increases, all methods show a reduction in \emph{\gls{ds}} values, which is expected since stricter diversity requirements lead to the exclusion of more solutions, and thereby reduce the number of retained complete solutions.
Nevertheless, \emph{\gls{ours}} retains a significant advantage across nearly all \(\theta_d\) settings.
These results demonstrate that our approach is more effective at navigating the search space and detecting solutions with high behavioral discrepancies across system versions.}

\begin{figure}[htbp]
    \centering
    \includegraphics[width=.7\linewidth]{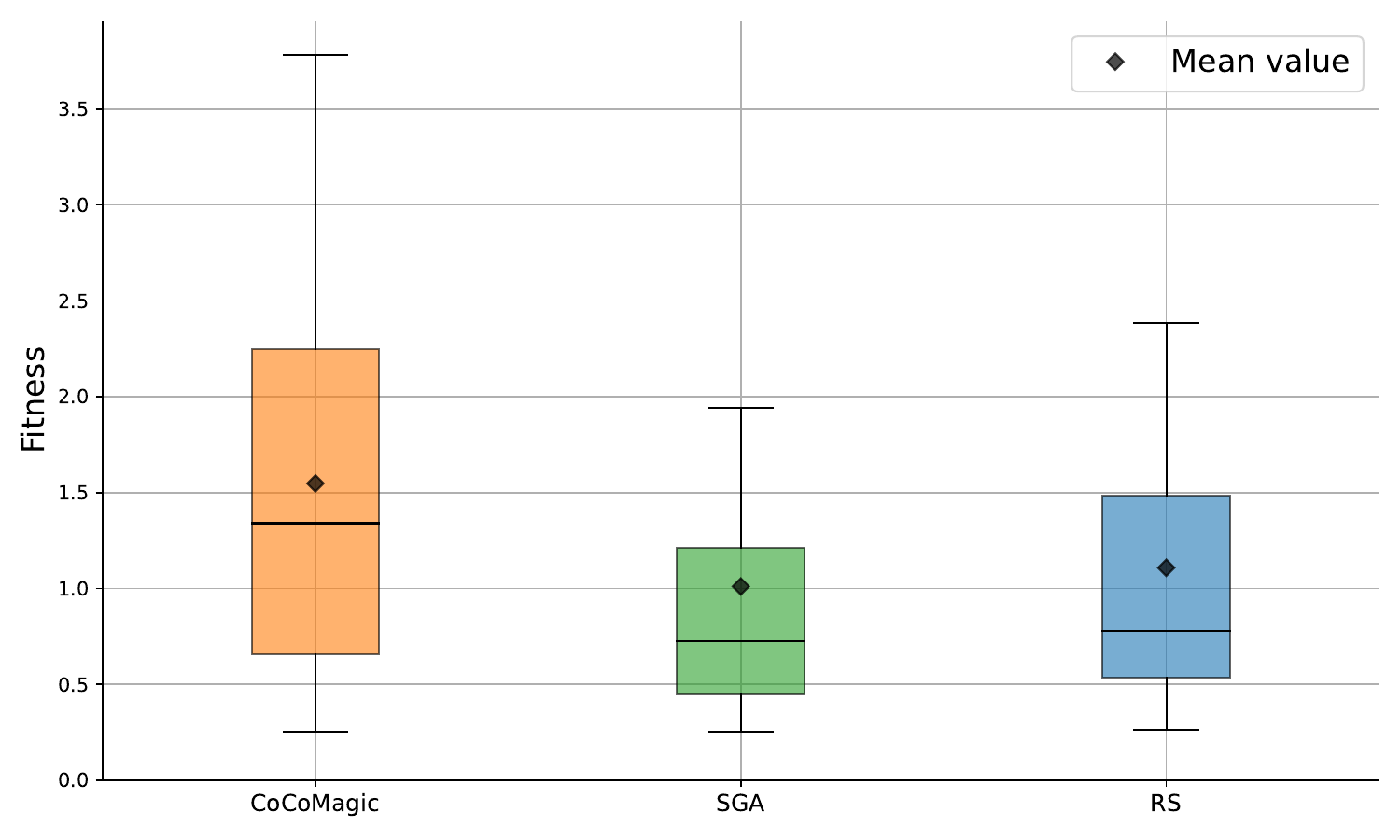}
    \caption{
        Fitness distribution of test cases generated by \emph{\gls{ours}}, \emph{\gls{sga}}, and \emph{\gls{rs}}.
        Higher fitness indicates that the discovered test cases induce greater differences in the violations exhibited by the two \glspl{ads}.
        Each box shows the fitness values for the test cases across 10 executions, providing insight into each method's peak effectiveness.
        This figure highlights the relative strength of each method in uncovering critical test cases with severe behavioral differences.
    }\label{fig:avg_fitness}
    \Description{Fitness distribution of test cases generated by CoCoMagic, SGA, and RS.}
\end{figure}

Beyond the \emph{\gls{ds}} metric, we evaluate the quality of complete solutions produced by each method, excluding those with insignificant fitness values (\(< 0.25\)). This thresholding serves as a post-processing step to filter out solutions with near-zero fitness, which typically arise from randomness in the process and do not reflect genuine differences in violations. \deletedtext{The filtering ensures that the comparison focuses only on meaningful, practical solutions.}
\Cref{fig:avg_fitness} presents the distribution of average fitness of complete solutions generated by \emph{\gls{ours}}, \emph{\gls{sga}}, and \emph{\gls{rs}} across 10 independent executions.
\emph{\Gls{ours}} achieves both higher median and mean fitness values compared to the baselines, indicating a greater capacity to uncover solutions that induce stronger behavioral discrepancies between the two system versions. On average, solutions generated by \emph{\gls{ours}} reach a fitness value of 1.55, representing a 53\% (\(p\)-value \(< 10^{-10}\)) increase over \emph{\gls{sga}} (1.01) and a 40\% (\(p\)-value \(< 10^{-4}\)) increase over \emph{\gls{rs}} (1.11).
To assess the statistical significance of these differences, we conducted non-parametric Mann-Whitney U tests with Fisher's method for \(p\)-value correction~\cite{mann1947test}.

In addition to the higher median and mean fitness, \cref{fig:avg_fitness} highlights that \emph{\gls{ours}} also attains a substantially larger range of fitness values compared to the baselines. While \emph{\gls{sga}} and \emph{\gls{rs}} peak at fitness levels around 1.95 and 2.35, respectively, \emph{\gls{ours}} reaches up to 3.80. This indicates that \emph{\gls{ours}} not only consistently yields stronger solutions on average but also uncovers more severe cases that trigger significantly larger behavioral differences between the two system versions.

\annotate{Comment 3.12}\addedtext{

\begin{figure*}[htbp]
    \centering
    \includegraphics[width=\linewidth]{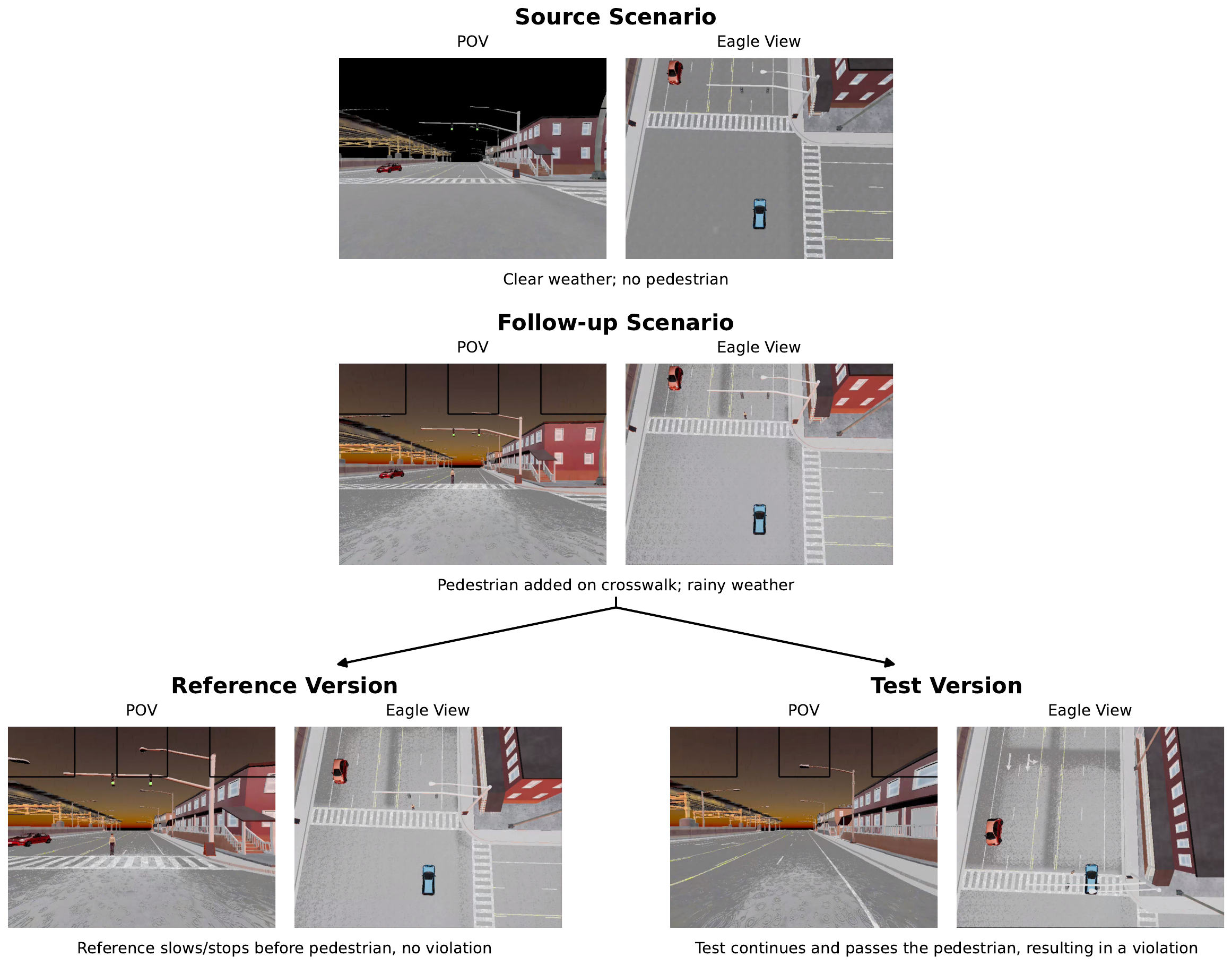}
    \caption{
        Illustrative discrepancy revealed by \emph{\gls{ours}}. The source scenario (top) depicts an ego vehicle traversing an intersection under clear weather conditions with no interfering agents. A follow-up scenario (middle) is generated by applying a metamorphic transformation that introduces a pedestrian in the crosswalk ahead of the ego vehicle and changes the weather to rainy. The follow-up scenario is then executed on two versions of the \gls{ads}: the reference (bottom-left) and the test (bottom-right). In the reference version, the vehicle slows down and stops at a safe distance from the pedestrian, satisfying the \gls{mr}. In contrast, the test version continues without sufficient deceleration and passes the pedestrian at close proximity, resulting in a violation.
    }\label{fig:example}
    \Description{Distinct solutions vs.\ distance threshold across different fitness thresholds for different configurations of CoCoMagic.}
\end{figure*}

To provide a concrete illustration of the types of discrepancies captured by \emph{\gls{ours}}, consider the scenario depicted in \cref{fig:example}. The source scenario demonstrates the ego vehicle approaching and traversing an intersection under clear weather conditions. The perturbation generated by \emph{\gls{ours}} includes two controlled changes: (i) placing a pedestrian on the crosswalk ahead of the ego vehicle, and (ii) modifying the weather from clear to rainy. The associated \gls{mr} specifies that, under these changes, the ego vehicle should exhibit more cautious behavior, particularly by reducing speed when approaching the pedestrian.

Both the reference (original) and test (retrained) versions of \emph{InterFuser} are evaluated on this test case. The reference system responds as expected by slowing down and stopping at a safe distance from the pedestrian, thereby satisfying the \gls{mr}. In contrast, the test system maintains its speed and passes the pedestrian at close proximity, failing to exhibit the required behavioral adjustment. This divergence results in a violation of the \gls{mr} in the test version (with a violation extent of 1.14), but not in the reference version, indicating a degradation with respect to the specified \gls{mr}.
}

In summary, the results clearly demonstrate the superior effectiveness of \emph{\gls{ours}} compared to both baseline methods in identifying a greater number of distinct test cases that not only satisfy diversity requirements but also induce more severe behavioral discrepancies between the two \gls{ads} versions. Its ability to generate both critical and diverse test cases highlights the strength of its co-evolutionary search strategy in navigating complex search spaces and uncovering meaningful discrepancies, making it an effective framework for rigorous differential testing of \glspl{as}.

\Finding{
    \emph{\Gls{ours}} outperforms both \emph{\gls{sga}} and \emph{\gls{rs}} in detecting severe differences in \gls{mr} violations, as measured by the average \emph{\gls{ds}} metric. In addition, it achieves an average fitness that is 53\% higher than \emph{\gls{sga}} and 40\% higher than \emph{\gls{rs}}, underscoring its effectiveness in uncovering more pronounced behavioral discrepancies.
}

\subsection{RQ2: Efficiency of \texorpdfstring{\gls{ours}}{CoCoMagic}}
\begin{figure*}[htbp]
    \centering
    \includegraphics[width=.9\linewidth]{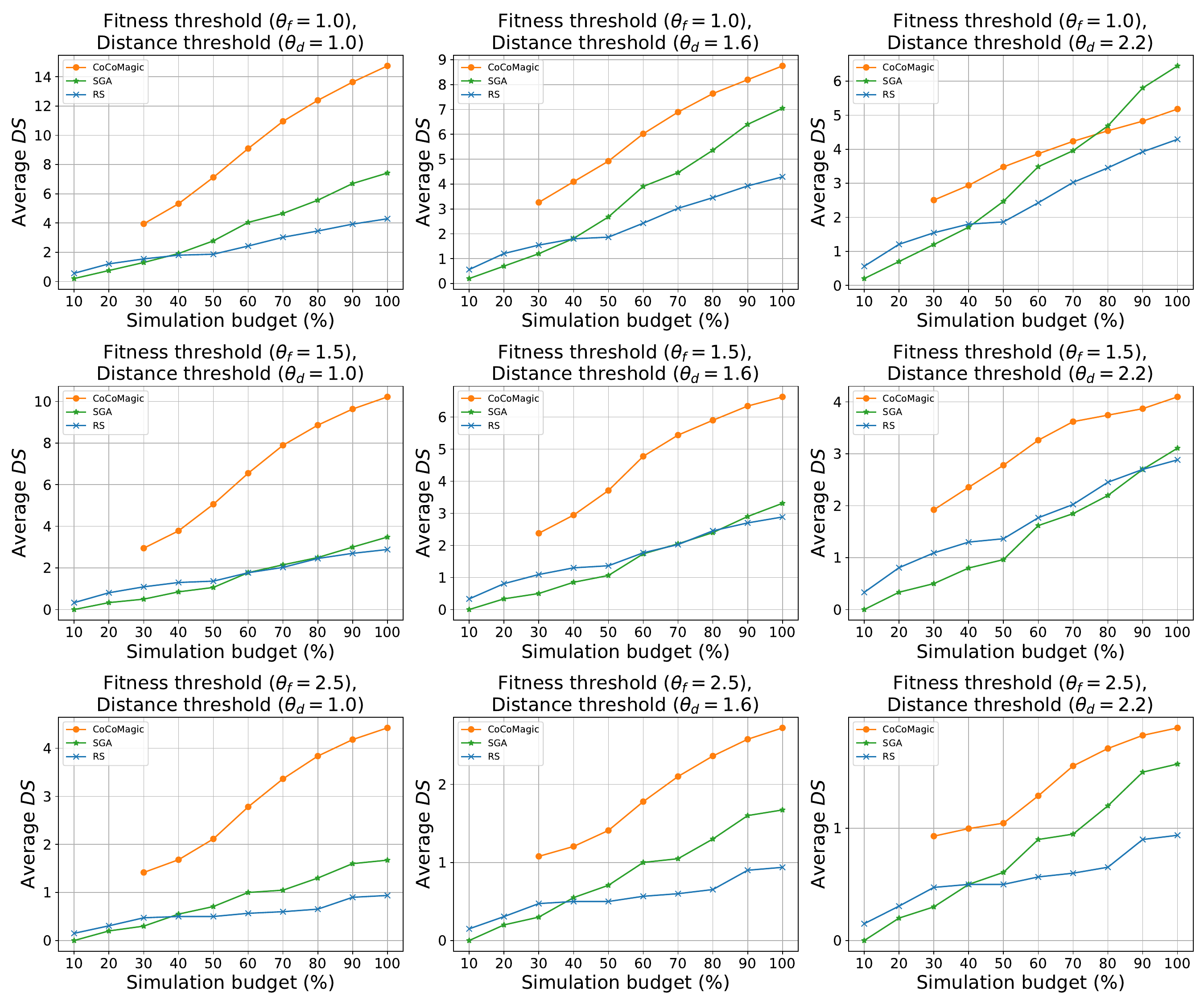}
    \caption{
        \emph{\Acrfull{ds}} progression across simulation budget levels, normalized to a total of 500 simulations. Each subplot illustrates how the \emph{\gls{ds}} value grows as more simulation resources are consumed, under specific combinations of fitness thresholds (\(\theta_f\)) and distance thresholds (\(\theta_d\)). The curves represent different methods, allowing a direct comparison of how quickly and effectively each approach uncovers severe and diverse behavioral discrepancies between \gls{ads} versions. Sharper increases and higher endpoints indicate greater efficiency in identifying test cases early in the search process.
    }\label{fig:ds_over_sim}
    \Description{Distinct solutions progression across simulation budget levels.}
\end{figure*}

\Cref{fig:ds_over_sim} presents the average \emph{\gls{ds}} values over simulation budget for each method, under nine different configurations of fitness thresholds \(\theta_f\) and distance thresholds \(\theta_d\). Each subplot corresponds to a specific threshold setting, capturing lenient (\(\theta_f = 1.0\)), moderate (\(\theta_f = 1.5\)), and high (\(\theta_f = 2.5\)) severity requirements, each paired with increasing levels of diversity (\(\theta_d = 1.0, 1.6, 2.2\)). The distance thresholds were selected based on their filtering effect on the generated solutions. On average, \(\theta_d = 1.0, 1.6\), and \(2.2\) remove about 10\%, 30\%, and 50\% of the solutions produced by all methods. These values, therefore, correspond to low, moderate, and high diversity constraints.

Across nearly all configurations, \emph{\gls{ours}} consistently outperforms the baseline methods in the total number of distinct solutions discovered \deletedtext{over time}\addedtext{as the simulation budget increases}. Moreover,
\emph{\gls{ours}} \deletedtext{maintains a steady and}\addedtext{exhibits a} steeper growth trajectory compared to the baselines, which reflects the advantage of the co-evolutionary setting in promoting the discovery of additional distinct solutions as the budget increases.
\addedtext{The advantage of \emph{\gls{ours}} is particularly noticeable during the early stages of the search. In most configurations, the method begins discovering distinct solutions earlier than the baselines and accumulates them faster. This indicates higher search efficiency, as a larger number of useful discrepancies can be identified using a smaller portion of the available simulation budget.}

As the fitness threshold becomes stricter (\(\theta_f \in \{1.5, 2.5\} \)), the number of retained solutions decreases for all methods, leading to lower overall \emph{\gls{ds}} values. \addedtext{This behavior is expected because fewer test cases satisfy the more demanding severity requirements.} Nevertheless, \emph{\gls{ours}} maintains a clear advantage over the others.
For example, under \addedtext{moderate} thresholds \(\theta_f = 1.5\) and \(\theta_d = 1.6\), \emph{\gls{ours}} nearly doubles the number of discovered distinct solutions relative to \deletedtext{\emph{\gls{rs}} and four times the \emph{\gls{ds}} values compared to \emph{\gls{sga}}}\addedtext{both \emph{\gls{sga}} and \emph{\gls{rs}}} by the 100\% budget mark.

\deletedtext{Under higher diversity restrictions \(\theta_d = 2.3\), \emph{\gls{rs}} starts to perform competitively, specifically at early stages of the search process (budget \(< 60\% \)). This demonstrates the advantage of random exploration in generating a diverse set of solutions when constraints are tight.
However, as the budget grows (budget \(\geq 70\% \)), \emph{\gls{ours}} surpasses \emph{\gls{rs}} in \emph{\gls{ds}} values, illustrating the advantage of systematic search and diversity optimization over pure exploration, in sustaining progress and uncovering more solutions over extended runs.
Notably, \emph{\gls{sga}} is consistently outperformed by \emph{\gls{rs}} under these settings.
This can be explained by premature convergence: in the absence of explicit diversity optimization mechanisms, \emph{\gls{sga}} tends to concentrate its search in a few suboptimal regions of the search space.
Consequently, while \emph{\gls{sga}} may generate a larger number of solutions compared to \emph{\gls{rs}}, many are either too similar to pass the \(\theta_d\) filtering or insufficiently severe to satisfy \(\theta_f\), resulting in fewer retained distinct solutions.}

\addedtext{When the diversity requirement becomes more restrictive (\(\theta_d = 2.2\)), the gap between the methods narrows in some configurations. In particular, under mild fitness constraints (\(\theta_f = 1.0\)), \emph{\gls{sga}} becomes competitive in later stages of the search (budget \(\geq 80\% \)). However, this advantage of \emph{\gls{sga}} is limited and largely disappears as the fitness requirement becomes stricter (\(\theta_f = 1.5\) and \(\theta_f = 2.5\)). This indicates that while \emph{\gls{sga}} may identify slightly more diverse but mild discrepancies, \emph{\gls{ours}} is more effective at discovering severe and diverse discrepancies simultaneously, which is the primary goal of the testing process.}

To quantitatively assess the overall efficiency of each method, we compute the \emph{\gls{auc}} for the \emph{\gls{ds}}-budget curves shown in \cref{fig:ds_over_sim}.
This metric aggregates the progression of distinct solution discovery over the entire simulation budget, offering a single representative value for each configuration. A higher \emph{\gls{auc}} indicates that a method finds more distinct solutions earlier and more consistently throughout the search process.
On average, \emph{\gls{ours}} outperforms \emph{\gls{sga}} by 126\% and \emph{\gls{rs}} by 169\% in terms of \emph{\gls{auc}}, demonstrating more efficient test case generation under budget constraints. Statistical analysis using the Wilcoxon signed-rank test~\cite{wilcoxon1945individual} confirms the significance of these improvements, with \(p\)-value \(< 10^{-2}\) against both baselines.

\begin{figure}[htbp]
    \centering
    \includegraphics[width=.5\linewidth]{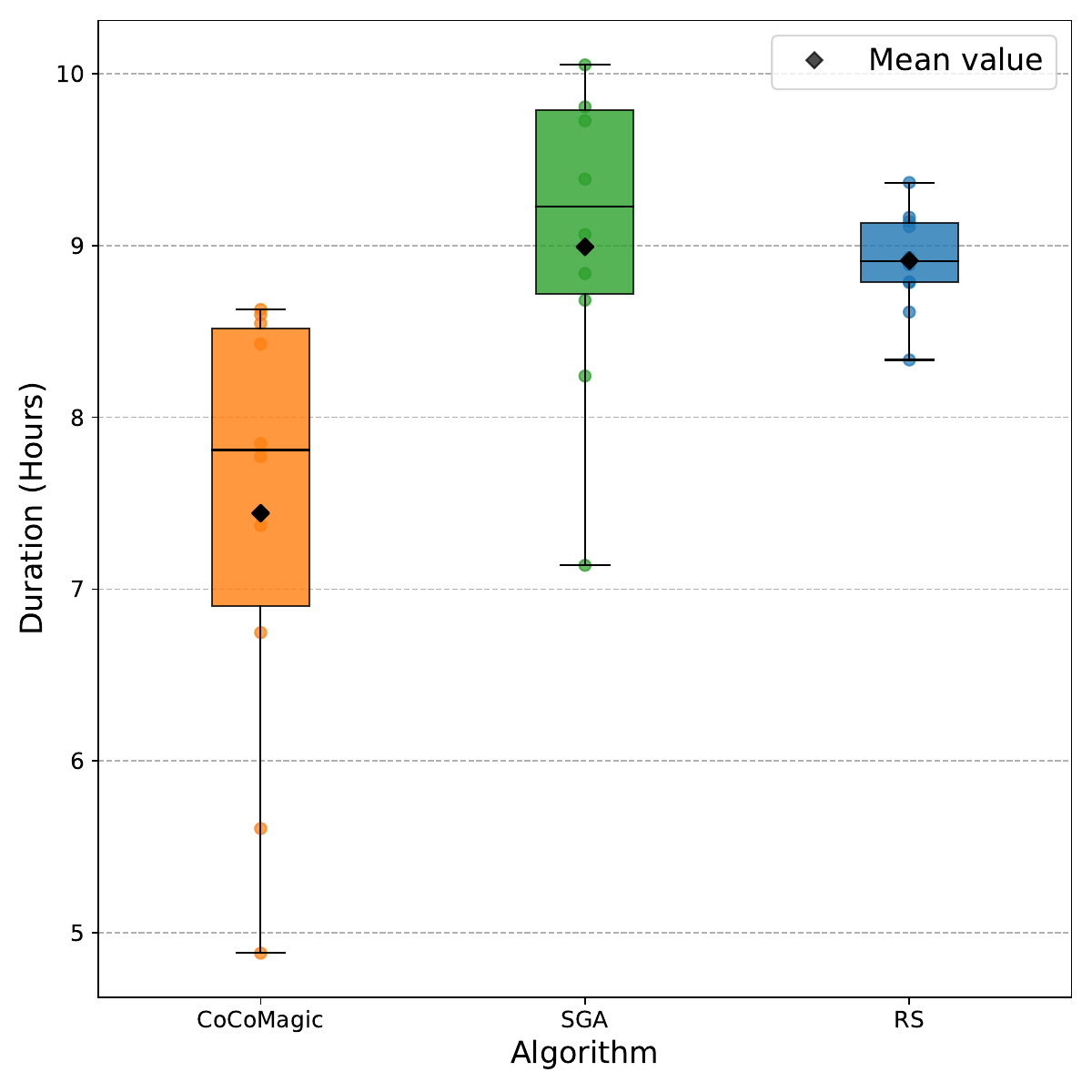}
    \caption{
        Distribution of execution times (in hours) for \emph{\gls{ours}}, \emph{\gls{sga}}, and \emph{\gls{rs}} under a fixed search budget. Lower execution time indicates greater computational efficiency in completing the search. Each box summarizes the distribution of execution times across 10 runs.
    }\label{fig:exec_time}
    \Description{Distribution of execution times (in hours) for CoCoMagic, SGA, and RS under a fixed search budget.}
\end{figure}

In addition to evaluating the discovery rate of distinct solutions over budget, we also examined the computational efficiency of the algorithms. \Cref{fig:exec_time} presents the distribution of execution times (in hours) across 10 runs for \emph{\gls{ours}}, \emph{\gls{sga}}, and \emph{\gls{rs}}. The results show that \emph{\gls{ours}} consistently requires less computational time, with an average of 7.44 hours, compared to 8.99 hours for \emph{\gls{sga}} and 8.91 hours for \emph{\gls{rs}}. This indicates that \deletedtext{our approach}\addedtext{\emph{\gls{ours}}} is not only more effective at utilizing the search budget but also faster to execute. Specifically, \emph{\gls{ours}} improves average execution time by 17.24\% over \emph{\gls{sga}} (\(p\)-value \(< 10^{-2}\)) and by 16.49\% over \emph{\gls{rs}} (\(p\)-value \(< 10^{-3}\)).
The longer execution time observed for \emph{\gls{sga}} and \emph{\gls{rs}} can be attributed to their tendency to generate more invalid complete solutions during the search.
These invalid solutions violate simulation constraints, for example, by placing objects in overlapping locations, which causes the simulator to terminate immediately at the start of execution.
Although such solutions are excluded from the simulation budget and discarded from the evolutionary process, the effort spent attempting to simulate them still incurs additional computational overhead, resulting in longer overall execution times.

Overall, these results demonstrate that \emph{\gls{ours}} is significantly more budget-efficient at discovering critical and diverse test cases, while also achieving faster execution than the baselines. These advantages make \emph{\gls{ours}} better suited for cost-sensitive testing scenarios.

\Finding{
    \emph{\Gls{ours}} outperforms \emph{\gls{sga}} by 126\% and \emph{\gls{rs}} by 169\% in terms of overall efficiency, as measured by the \emph{\gls{auc}} of the \emph{\gls{ds}}-budget curves, demonstrating its superior ability to discover critical and diverse test cases within a limited search budget. In addition, it achieves 17\% faster execution time than \emph{\gls{sga}} and 16.5\% faster than \emph{\gls{rs}}, confirming its advantage in computational efficiency.
}

\subsection{RQ3: Impact of Constraints and Population Initialization}\label{sec:rq3}

\begin{figure}[htbp]
    \centering
    \includegraphics[width=.7\linewidth]{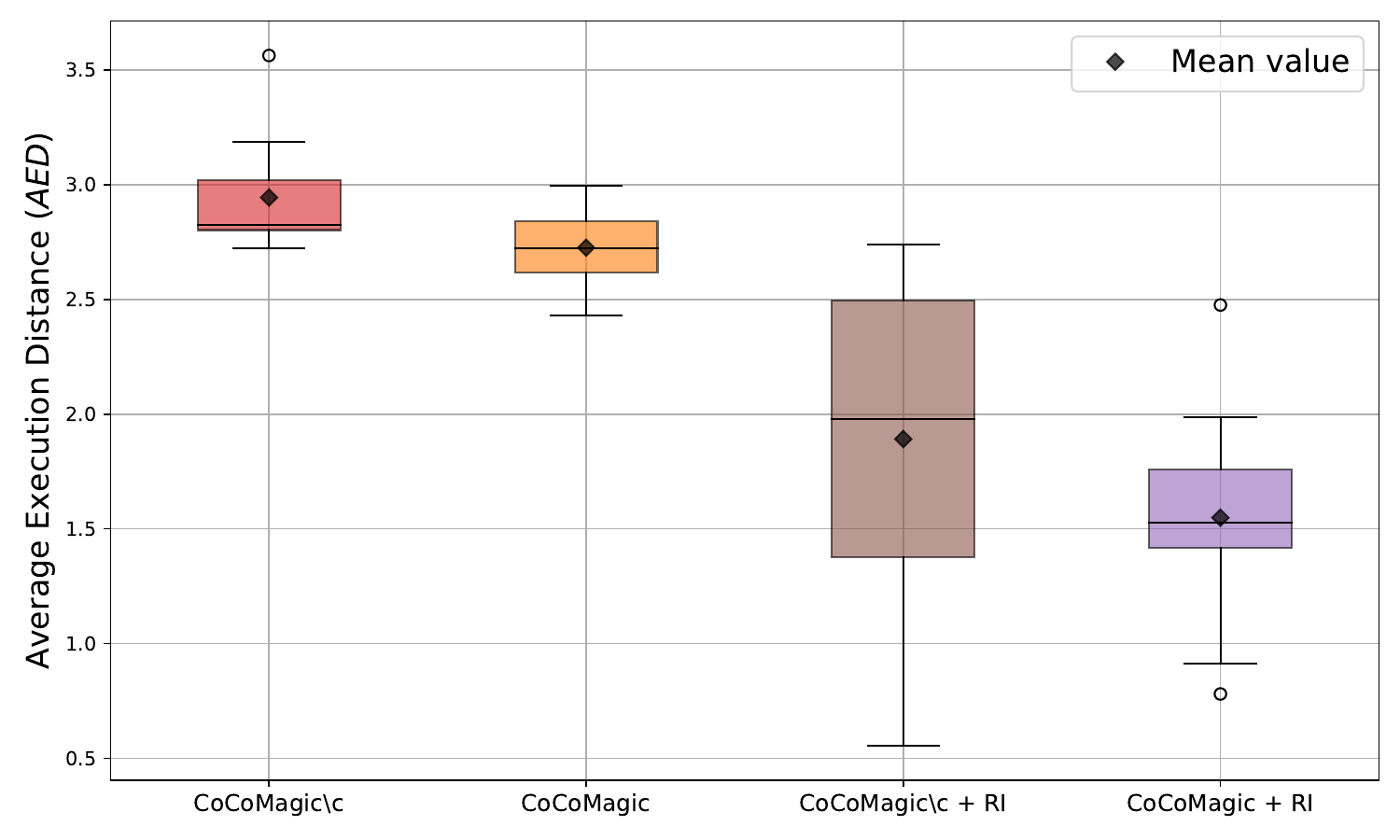}
    \caption{
        \emph{\Acrfull{aed}} between identified test cases and execution scenarios under different configurations of \emph{\gls{ours}} and baseline methods.
        Higher \emph{\gls{aed}} values indicate that the generated test cases are more distinct from the execution scenarios.
        Each box summarizes the distribution of \emph{\gls{aed}} values for a given method, highlighting how closely the identified test cases resemble previously observed real-world scenarios.
    }\label{fig:dist_to_scenario}
    \Description{Average Execution Distance between identified test cases and execution scenarios under different configurations of CoCoMagic and baseline methods.}
\end{figure}

\Cref{fig:dist_to_scenario} shows the \emph{\gls{aed}} between the generated complete solutions and the set of previously observed execution scenarios.
This metric reflects how closely the identified solutions resemble behaviors observed during actual system executions.
Lower \emph{\gls{aed}} values indicate that the generated complete solutions are behaviorally closer to known execution scenarios, whereas higher values suggest the discovery of more dissimilar solutions.

The most noticeable trend in \cref{fig:dist_to_scenario} is the substantial reduction in \emph{\gls{aed}} values achieved by both \emph{\gls{ri}} configurations.
\emph{\Gls{ours} with RI} produces the lowest \emph{\gls{aed}} values overall, followed by its unconstrained counterpart. This outcome confirms the effectiveness of \emph{\gls{ri}} in guiding the search toward solutions that remain close to observed ones.
By seeding the population with scenarios observed during actual executions, \emph{\gls{ri}} biases the search toward regions of the search space that are more contextually grounded.
In contrast, the unconstrained and constrained configurations without \emph{\gls{ri}} (i.e., \emph{\gls{ours}\textbackslash{c}} and \emph{\gls{ours}}) yield significantly higher \emph{\gls{aed}} values, with distributions clustered around a much higher median.
This confirms that, in the absence of \emph{\gls{ri}}, the search is more likely to generate complete solutions that deviate from what is commonly observed, resulting in less realistic but potentially more severe solutions.

The impact of the constraining mechanism is particularly evident when comparing constrained and unconstrained variants within each initialization strategy. In both cases, applying the constraint results in lower \emph{\gls{aed}} values, indicating that it effectively guides the search toward solutions closer to the observed execution scenarios.
Quantitatively, applying the constraint alone without \emph{\gls{ri}} yields a modest improvement of 7.43\% in average \emph{\gls{aed}} value (2.73) compared to its unconstrained counterpart (2.94). The benefit becomes more pronounced when \emph{\gls{ri}} is introduced.
The unconstrained variant with \emph{\gls{ri}} reduces the average \emph{\gls{aed}} value by 35.76\% (1.89) relative to the non-\emph{\gls{ri}} variant.
When combining both \emph{\gls{ri}} and constraining, the search is tightly focused on execution scenarios, resulting in the most significant improvement: a 47.41\% (1.55) reduction in \emph{\gls{aed}}.
This highlights the complementary role of constraints in enhancing behavioral realism, especially when the initial population is already grounded in the observed execution scenarios.
Overall, these results illustrate the effectiveness of both \emph{\gls{ri}} and the constraining mechanism in promoting the realism of generated test cases.

% \Cref{fig:dist_to_scenario} presents the distribution of distances between identified complete solutions and the original execution scenarios, measured using the same distance metric as in \cref{sec:constraint}.

\begin{figure*}[htbp]
    \centering
    \includegraphics[width=\linewidth]{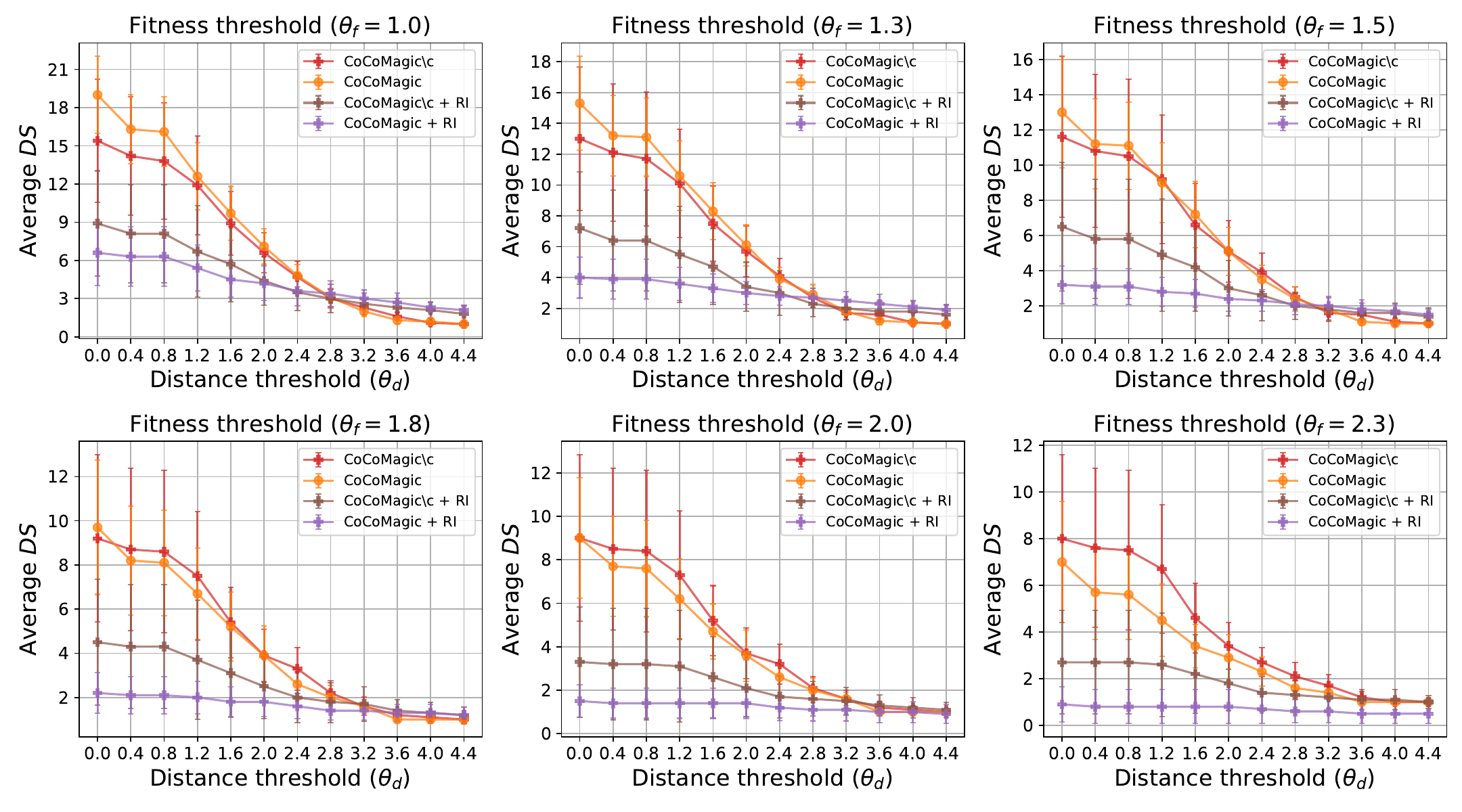}
    \caption{
        \emph{\Acrfull{ds}} vs.\ distance threshold (\(\theta_d\)) across different fitness thresholds (\(\theta_f\)) for different configurations of \emph{\gls{ours}}.
        A higher \emph{\gls{ds}} value indicates more distinct test cases discovered by the method.
        Each curve plots the average \emph{\gls{ds}} value at different \(\theta_d\) settings, under a specific fitness threshold \(\theta_f\).
        This figure reveals how each method balances the quantity and diversity of test cases as \(\theta_f\) varies.
    }\label{fig:conf_comp}
    \Description{Distinct solutions vs.\ distance threshold across different fitness thresholds for different configurations of CoCoMagic.}
\end{figure*}

\Cref{fig:conf_comp} compares the average number of distinct solutions discovered by each configuration of \emph{\gls{ours}}, across varying distance thresholds (\(\theta_d\)) and fitness thresholds (\(\theta_f\)).
\deletedtext{Across all settings, it is apparent that removing the constraint (\emph{\gls{ours}\textbackslash{c}}) consistently yields a higher average \emph{\gls{ds}} value compared to the constrained version (\emph{\gls{ours}}). This suggests that unconstrained search allows the algorithm to explore a broader portion of the search space, improving its ability to uncover diverse solutions.}
\addedtext{
Across the six fitness-threshold settings, \emph{\gls{ours}} and \emph{\gls{ours}\textbackslash{c}} exhibit similar overall behavior, with only modest differences observed across combinations of \(\theta_f\) and \(\theta_d\).
For lower fitness thresholds (\(\theta_f=1.0,1.3,1.5\)), \emph{\gls{ours}} tends to achieve slightly higher average \emph{\gls{ds}} values than \emph{\gls{ours}\textbackslash{c}} at smaller distance thresholds. However, this gap is relatively small and diminishes as \(\theta_d\) increases, with both configurations converging to comparable performance at larger distance thresholds.
For stricter fitness thresholds (\(\theta_f=1.8,2.0,2.3\)), a mild shift can be observed, where \emph{\gls{ours}\textbackslash{c}} slightly exceeds \emph{\gls{ours}}. This suggests that, under stricter severity requirements, removing the constraint can help retain more solutions.
}

Introducing \emph{\gls{ri}} leads to a noticeable reduction in \emph{\gls{ds}} values, for both constrained and unconstrained variants.
This reduction can be attributed to the tighter focus imposed by sampling initial populations from execution scenarios, which biases the search toward more realistic, but potentially more restricted regions of the search space.
However, \emph{\gls{ri}} exhibits reduced sensitivity to the distance threshold \(\theta_d\); both \emph{\gls{ri}} variants maintain a more stable \emph{\gls{ds}} value compared to their counterparts, as \(\theta_d\) increases. This robustness is likely due to the higher diversity inherent in the sampled execution scenarios used for initialization, which enables the generation of complete solutions that are already dispersed across different behavioral regions.

These trends are also reflected quantitatively in the overall reduction of \emph{\gls{ds}} values compared to the unconstrained, non-\emph{\gls{ri}} baseline (\emph{\gls{ours}\textbackslash{c}}).
Applying the constraint alone results in a 3.97\% (\(p\)-value \(= 0.028\)) decrease in the \emph{\gls{ds}} metric, highlighting the limiting effect of realism enforcement on exploratory diversity.
Introducing \emph{\gls{ri}} to the unconstrained variant leads to a slightly larger reduction of 21.22\% (\(p\)-value \(< 10^{-34}\)), while combining both constraints and \emph{\gls{ri}} produces the most substantial drop, with a 33.83\% (\(p\)-value \(< 10^{-78}\)) decrease in \emph{\gls{ds}}.
These results confirm the previously discussed trade-off: while constraints and \emph{\gls{ri}} improve behavioral realism, they reduce the algorithm's ability to discover a wide range of distinct solutions.

\deletedtext{Interestingly, under lower fitness thresholds, the \emph{\gls{ri}} variants begin to outperform their non-\emph{\gls{ri}} counterparts at higher values of \(\theta_d\). For example, when \(\theta_f \in \{0.5, 0.8\} \) and \(\theta_d \geq 3.2\), both \emph{\gls{ri}} configurations yield more solutions than their counterparts.
This is likely explained by the inherent diversity of the execution scenarios, which is carried over to the initial population through direct sampling. Such initialization helps sustain diversity when the diversity requirement (via \(\theta_d\)) is high and fitness constraints are relatively loose.
However, as fitness thresholds increase (\(\theta_f \geq 1.0\)), the \emph{\gls{ri}} variants fall behind, indicating that although \emph{\gls{ri}} supports early diversity, it may hinder the discovery of highly critical solutions, as it limits the algorithm's ability to detect severe edge-cases that are not represented in the selected execution scenarios.}

\begin{figure}[htbp]
    \centering
    \includegraphics[width=.7\linewidth]{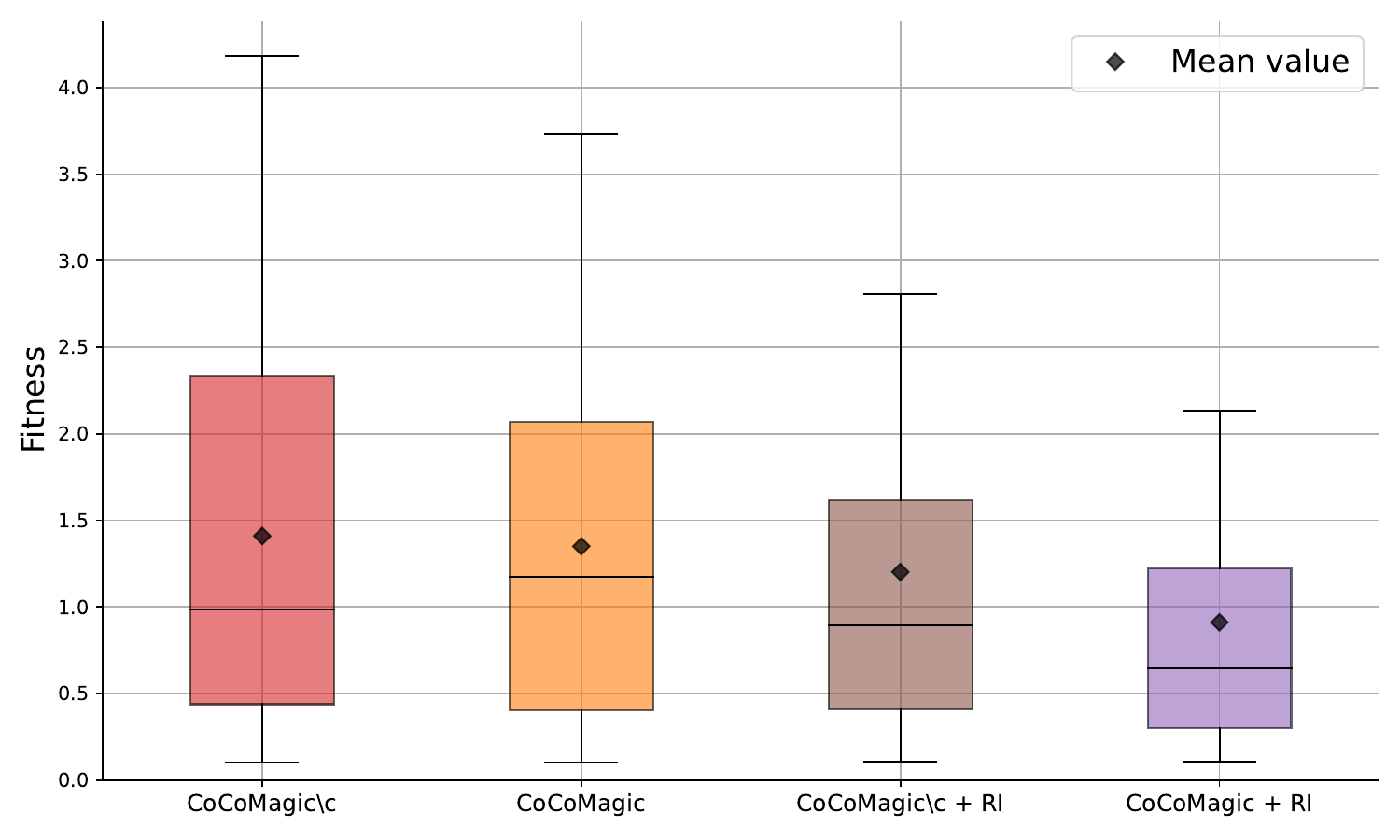}
    \caption{
        Average fitness of test cases generated by different configurations of \emph{\gls{ours}}.
        Higher average fitness indicates that the discovered test cases induce greater differences in the violations exhibited by the two \glspl{ads}.
        Each box shows the average fitness values of the generated test cases across 10 executions.
        This figure highlights the relative strength of each method in uncovering test cases with severe behavioral differences.
    }\label{fig:avg_fitness_conf_comparison}
    \Description{Average fitness of test cases generated by different configurations of CoCoMagic.}
\end{figure}

\Cref{fig:avg_fitness_conf_comparison} presents the fitness distribution of solutions identified by each configuration, excluding those with insignificant fitness values (\(< 0.1\)), across 10 executions.
The metric reflects the severity of the discovered solutions, with higher values indicating more critical discrepancies between the two \gls{ads} versions.

As shown in \cref{fig:avg_fitness_conf_comparison}, the unconstrained variant (\emph{\gls{ours}\textbackslash{c}}) achieves the highest average fitness across runs, followed closely by the constrained variant. This observation may suggest that removing the constraint allows the search to better exploit regions of the search space that lead to more severe discrepancies. However, the small reduction in average fitness observed for \emph{\gls{ours}} and \emph{\gls{ours}\textbackslash{c}+RI} relative to \emph{\gls{ours}\textbackslash{c}} is not statistically significant, suggesting that there is no solid evidence that introducing constraints or initialization with execution scenarios diminishes the ability to find severe discrepancies. In contrast, the reduction observed for \emph{\gls{ours}+RI} compared to \emph{\gls{ours}\textbackslash{c}} is statistically significant. This indicates that when both constraints and initialization are applied together, the search is more strongly biased toward typical or nominal driving patterns, restricting access to rare edge-case behaviors that produce higher-fitness solutions.

Quantitative comparisons provide insights into the effect of constraints and \emph{\gls{ri}} on the severity of discovered solutions.
Applying the constraint alone leads to a reduction of 4.20\% in average fitness values, with the constrained configuration (\emph{\gls{ours}}) achieving an average fitness of 1.35 compared to 1.41 for the unconstrained baseline (\emph{\gls{ours}\textbackslash{c}}).
Similarly, incorporating \emph{\gls{ri}} into the unconstrained variant yields a 14.72\% decrease, with an average fitness of 1.20. However, neither of these reductions is statistically significant (\(p \approx 0.42\) and \(p \approx 0.07\), respectively).
In contrast, combining both constraints and \emph{\gls{ri}} produces a substantial and statistically significant reduction, with an average fitness of 0.91, corresponding to a 35.37\% drop from the baseline (\(p < 10^{-5}\)).

Overall, these findings indicate that applying either constraints or \emph{\gls{ri}} alone can slightly reduce the severity of the generated solutions by limiting exploration. Yet, this effect is negligible and statistically insignificant. In contrast, combining the two results in a significant reduction in severity, demonstrating that their joint influence can substantially limit the search from reaching highly critical discrepancies.

In summary, the results of \labelcref{rq3} reveal a clear trade-off between behavioral realism and exploratory power when applying constraints and \emph{\gls{ri}}.
While both mechanisms independently and jointly reduce the number and severity of generated test cases, they significantly enhance alignment with real-world scenarios, as evidenced by lower \emph{\gls{aed}} values.
\emph{\Gls{ri}} proves especially effective in grounding the search, but it also significantly restricts access to edge-case solutions, particularly under stricter fitness requirements. Conversely, unconstrained and randomly initialized search enables broader exploration and uncovers more critical solutions, though at the cost of lower realism. These findings emphasize the importance of selecting the appropriate configuration based on the testing objective and of prioritizing realism, severity, or diversity in the discovered solutions.

\Finding{
    Both constraints and \emph{\gls{ri}} improve the behavioral realism of generated test cases, as measured by \emph{\gls{aed}}, with their combination yielding around 50\% improvement over the unconstrained baseline.
    However, these mechanisms also tend to reduce the number of discovered solutions.
    The constraint alone leads to a 4\% reduction in \emph{\gls{ds}} values.
    Incorporating \emph{\gls{ri}} further decreases \emph{\gls{ds}} values by up to 34\%.
    This reveals a clear trade-off: constraints and \emph{\gls{ri}} enhance realism but may reduce the quantity and severity of the identified test cases.
}

\subsection{RQ4: Interpretability Approach}\label{sec:rq4}

\subsubsection{Quantitative Evaluation of Interpretability}
\begin{figure}[htbp]
    \centering
    \includegraphics[width=.5\linewidth]{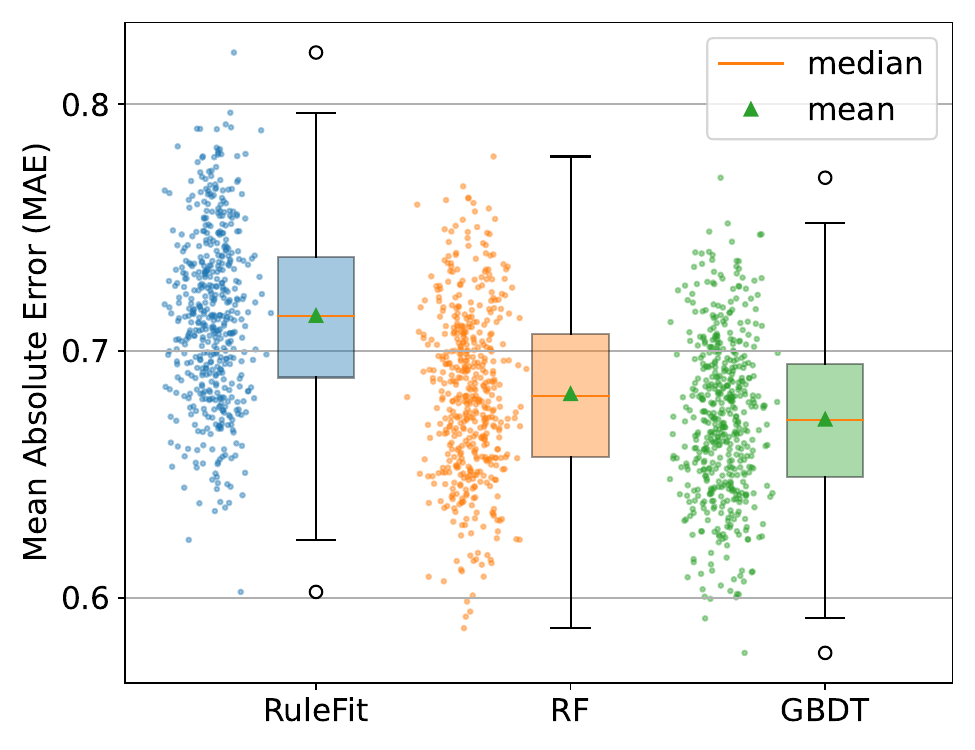}
    \caption{
        Distribution of \emph{\acrfull{mae}} for \emph{RuleFit} in \emph{\gls{ours}} and baseline methods.
        A lower \emph{\gls{mae}} value indicates greater model accuracy in capturing behavioral differences between the two system versions.
        Individual points beside each box represent the \emph{\gls{mae}} values for each identified test case.
    }\label{fig:mae}
    \Description{Distribution of Mean Absolute Error for RuleFit in CoCoMagic and baseline methods.}
\end{figure}

\Cref{fig:mae} presents the distribution of \emph{\gls{mae}} values for predictions made by \emph{RuleFit} in \emph{\gls{ours}} and the baseline methods, \emph{\gls{rf}} and \emph{\gls{gbdt}}.
The results indicate that \emph{RuleFit} achieves an average \emph{\gls{mae}} value of 0.71, which is higher than that of \emph{\gls{rf}} (0.68) and \emph{\gls{gbdt}} (0.67).
Although \emph{\gls{rf}} and \emph{\gls{gbdt}} yield slightly lower average \emph{\gls{mae}} value, indicating higher accuracy in modeling behavioral differences in \gls{mr} violations, they lack interpretability as they tend to generate a large number of complex rules.
In decision tree-based models like \emph{\gls{rf}} and \emph{\gls{gbdt}}, these rules correspond to decision paths, which are sequences of feature-based conditions followed from the root to a leaf node in each tree.
On average, \emph{\gls{rf}} produces 1009.95 such paths and \emph{\gls{gbdt}} produces 2010.00, making it difficult to extract clear, human-understandable insights from their predictions.
This limitation reduces their practical usefulness for understanding the behavioral discrepancies between system versions.
In contrast, \emph{RuleFit} strikes a balance between accuracy and interpretability, producing a concise set of human-readable rules (28.72 rules on average) that facilitate understanding while still achieving accuracy close to that of \emph{\gls{gbdt}} and \emph{\gls{rf}}.
Furthermore, the rules generated by \emph{\gls{rf}} and \emph{\gls{gbdt}} are more complex, averaging 3.91 and 4.10 conditions per rule, respectively, whereas \emph{RuleFit} rules average only 2.09 conditions.
This simplicity enhances the interpretability of \emph{RuleFit} rules, making them more accessible for human analysts.

% Furthermore, the \emph{\gls{mase}} distribution of \emph{RuleFit} shows less variance, indicating more consistent and reliable predictions in capturing differences in \gls{mr} violations across solutions.

% \begin{figure}[htbp]
%     \centering
%     \includegraphics[width=.85\linewidth]{figs/support.pdf}
%     \caption{
%         Distribution of the \emph{support} of the rules extracted by \emph{RuleFit} in \emph{\gls{ours}}.
%         Higher \emph{support} values indicate that the rules cover more complete solutions.
%         The left subplot shows the distribution across all 500 runs, with horizontal jitter added to avoid overlapping points and improve visibility.
%         The right subplot shows the same data as a histogram with a logarithmic scale for better visualization.
%     }\label{fig:support}
% \end{figure}

% \Cref{fig:support} shows the distribution of the \emph{support} of the rules extracted by \emph{RuleFit}, where each dot in the left subplot represents the \emph{support} value of the full set of rules extracted in a single run.
% Horizontal jitter is applied to the dots in the left subplot to reduce overlap and improve clarity.
% The right subplot presents the same data as a histogram with a logarithmic scale on the x-axis to better visualize the distribution.
The \emph{support} distribution demonstrates that the rules extracted by \emph{RuleFit} consistently cover\deletedtext{ nearly} all of the identified complete solutions.
\deletedtext{In 498 out of 500 independent runs, t}\addedtext{T}he generated rules achieved 100\% \emph{support}, meaning every identified complete solution was covered by at least one rule.
\deletedtext{The remaining two runs exhibited slightly lower \emph{support} values of 99.8\% and 99.6\%, respectively.}
% The remaining few runs show only minor deviations, with \emph{support} values still exceeding 99.5\%.
Given the maximum of 30 rules allowed per run, achieving \deletedtext{near-}complete coverage of all identified solutions is particularly noteworthy.
This high level of \emph{support} underscores the effectiveness and consistency of \emph{RuleFit} in capturing a compact yet comprehensive set of interpretable rules or patterns that explain the observed behavioral differences across diverse runs.

\annotate{Comments 1.2, 2.3, and 3.15}\addedtext{
\subsubsection{Expert Assessment of Interpretability Rules}
\begin{figure*}[htbp]
    \centering
    \includegraphics[width=\linewidth]{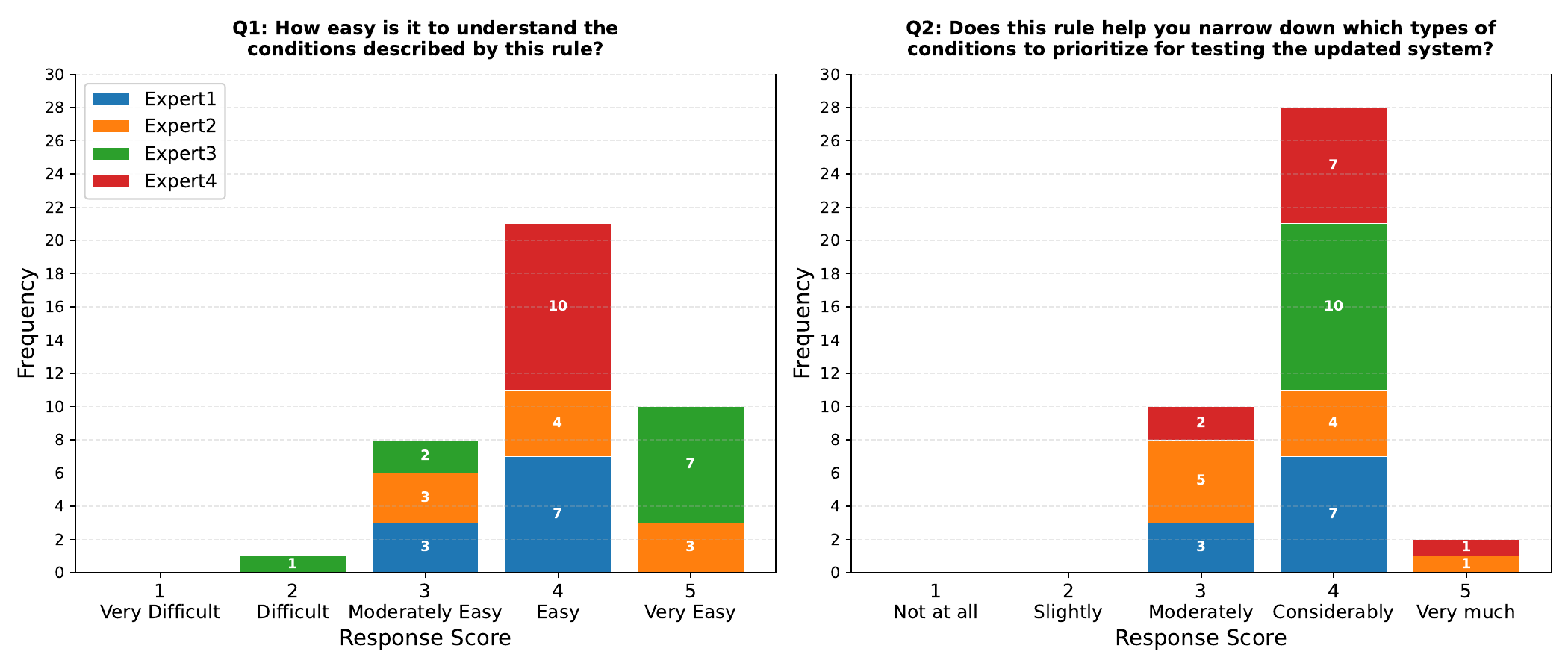}
    \caption{
        Distribution of expert ratings for the generated rules across two evaluation criteria: ease of understanding (\(Q1\)), and usefulness for prioritizing testing conditions (\(Q2\)). Ratings are based on a five-point Likert scale.        
    }\label{fig:expert_assessment}
    \Description{Expert assessment Results}
\end{figure*}

\Cref{fig:expert_assessment} shows the distribution of expert ratings for the generated rules across the two evaluation questions. The results indicate that the rules are generally understandable and practically useful. % The average scores are 4.0 for \textbf{Q1} (ease of understanding), and 3.8 for \textbf{Q2} (usefulness for prioritizing testing).

The responses to \(Q1\) show a clearly positive pattern. Across the 40 ratings collected for this question (10 rules \(\times\) 4 experts), the average score is 4.0, which corresponds to the \emph{Easy} response. In total, 31 out of 40 ratings (77.5\%) are either \emph{Easy} or \emph{Very Easy}, while only a single rating falls below \emph{Moderately Easy}. This indicates that the conditions described by the rules are largely understandable to practitioners, and experts rarely found the rules confusing or difficult to interpret. The results are also relatively consistent across experts. In most cases, the ratings assigned to a given rule differ by at most 1 point. Larger disagreements are rare, with only two rule assessments showing a difference of 2 points between experts.

A similar trend can be observed for \(Q2\).  Out of the 40 ratings, 30 (75.0\%) are \emph{Considerably} or \emph{Very much}, and none of the responses fall in the two lowest categories. 
The mean score is 3.8, lying between \emph{Moderately} and \emph{Considerably} useful and closer to the latter. This suggests that, on average, the experts found the rules useful for narrowing down which conditions to prioritize when testing the updated system.

Overall, the questionnaire results indicate that the generated rules are understandable to domain experts and useful for narrowing the testing focus to relevant driving conditions. These findings support the claim that the interpretability component can help engineers move from raw test cases toward more actionable testing insights.
% However, to further increase benefits, practitioners should likely receive professional training on how to optimally use the generated rules, which was limited in our case due to time constraints.
}

\deletedtext{
To further illustrate the practical value of this approach, we present a representative example from one of the runs.
% , accompanied by the corresponding feature contribution analysis.
The representative rule is ranked as the most important among all extracted rules.
% and contains the most influential features identified through the feature contribution analysis.
The rule states ``IF the relative position of the ego vehicle during the third quartile of the trajectory is less than 23.78 meters to the left of its starting position in the source scenario, there are at most 16 static objects in the source scenario, and the relative position of the ego vehicle in the middle of the trajectory is less than 68.73 meters to the left of its starting position in the follow-up scenario, THEN the updated \gls{ads} tends to exhibit more \gls{mr} violations than the original version, with a predicted difference of 0.492, supported by 71.03\% of the identified complete solutions''.
This rule captures a set of interacting spatial and contextual conditions under which the updated \gls{ads} is more likely to generate violations compared to the original.
Each condition contributes an interpretable insight into when behavioral differences between the two versions are most likely to occur.
An example scenario satisfying this rule is illustrated in Fig.~11.
The first condition requires that, during the third quartile of the trajectory in the source scenario, the ego vehicle's lateral displacement remains less than 23.78 meters to the left of its starting position.
This suggests that the ego vehicle does not deviate substantially from its initial lane or execute a sharp left maneuver at this stage.
The second condition constrains the source scenario itself, requiring that the number of static objects not exceed 16, limiting the level of environmental complexity to a moderately challenging but not excessively congested scene.
The third condition concerns the follow-up scenario, requiring that the ego vehicle's lateral displacement at the midpoint of its trajectory is less than 68.73 meters to the left of its starting position.
This indicates that the vehicle may have executed a left turn or a lane change in the middle of the trajectory, which could trigger different responses across the two system versions, possibly due to updated planning or lane-keeping mechanisms.
Taken together, these spatial and contextual constraints highlight a systematic relationship between vehicle positioning and environmental structure, revealing conditions under which the updated \gls{ads} is more prone to generating \gls{mr} violations than the original.
}

\annotate{Comment 2.6 and 3.12}\addedtext{
\begin{table}[htbp]
    \small
    \caption{Automatically generated rules describing scenarios that induce behavioral differences between \gls{ads} versions.}\label{tab:rules_summary}
    \begin{threeparttable}
        \begin{tabularx}{\linewidth}{Xc}
            \toprule
            \textbf{Driving Condition Description} & \textbf{Performance Change} \\
            % & \textbf{Explanation of Behavior} \\
            \midrule

            The ego vehicle is moving at a slow speed, and a pedestrian on the left is extremely close.
            & Degradation \\
            % & The extreme proximity of the pedestrian requires precise and fast detection and handling, which may not be adequately achieved under such tight spatial constraints. \\
            
            Pedestrians ahead of the ego vehicle are very close and all are moving.
            & Degradation \\
            % & Multiple moving pedestrians create competing deceleration demands, increasing the likelihood of insufficient or mistimed braking. \\
            
            Pedestrians ahead move along the road direction while a left-side pedestrian is close.
            & Degradation \\
            % & Forward pedestrians appear non-threatening, while the nearby left pedestrian may not trigger sufficient deceleration. \\

            % % Potential conflict to rules 8 and 9.
            % Vehicles on the left move consistently, and there is an obstacle ahead or to the right.
            % & improvement
            % & The forward or right-side obstacle may not be correctly identified as requiring deceleration when the left-side vehicles provide a dominant but non-threatening motion pattern that diverts attention. \\

            % % Potential conflict to rule 1.
            % Pedestrians are mostly in front or slightly to the right of the ego vehicle.
            % & improvement
            % & Pedestrians concentrated in the forward and slightly rightward area create a dense cluster of potential hazards, making it difficult to assess the correct deceleration priority and timing for each individual pedestrian. \\

            A left-side pedestrian is moving toward the right.
            & Improvement \\
            % & This is an unusual crossing pattern, as pedestrians commonly move from right to left; this atypical motion can lead to misinterpretation and delayed or incorrect driving responses. \\
            % & Failure to detect the pedestrian or misjudging the pedestrian's speed or trajectory in this situation can lead to delayed or insufficient braking. \\

            % Vehicles ahead are centered or to the right of the ego vehicle.
            % & Degradation \\
            % & Vehicles positioned centrally or to the right may partially occlude other road users or obstacles, leading to delayed perception and insufficient deceleration. \\

            Ego vehicle is moving slowly under very dark lighting conditions.
            & Improvement \\
            % & Severely reduced visibility under dark lighting conditions can impair the detection of surrounding road users and obstacles, leading to insufficient or delayed deceleration despite the slow speed. \\

            There are slow-moving vehicles on the left, and the front vehicles are oriented to the right.
            & Degradation \\
            % & Slow-moving vehicles on the left may divert attention from rightward front vehicles, and their conflicting directions can complicate motion assessment and deceleration decisions. \\

            The ego vehicle is moving while a static obstacle ahead is extremely close.
            & Degradation \\
            % & Close obstacles can obstruct the field of view and leave inadequate reaction time, increasing the risk of delayed or insufficient braking. \\

            \bottomrule
        \end{tabularx}
    \end{threeparttable}
\end{table}

To further illustrate the practical value of the interpretability approach, \Cref{tab:rules_summary} presents several representative rules from one run, together with their associated performance changes. Each rule characterizes a specific driving condition under which the two \gls{ads} versions exhibit different \gls{mr} violations. The \emph{Driving Condition Description} column reports natural-language descriptions derived from \emph{RuleFit} rules via the post-processing step described in \Cref{subsec:rq4}, while the \emph{Performance Change} column indicates whether the observed divergence corresponds to a degradation or an improvement with respect to \(GP_1\). This is inferred from the sign of the coefficient associated with the rule in \emph{RuleFit}, where a positive coefficient indicates a degradation, and a negative coefficient indicates an improvement.
For instance, the first rule indicates that when the ego vehicle is moving at a low speed while a pedestrian on the left is in very close proximity, the updated version tends to exhibit more severe \gls{mr} violations.
These rules are not meant to be interpreted as definitive diagnostic conclusions. Instead, they provide actionable starting points for identifying driving conditions that warrant further investigation when analyzing behavioral differences between system versions.
}

% The feature contribution analysis further supports the importance of these conditions.
% Among all features, trajectory-related features, such as the relative position of the ego vehicle at different stages of the trajectory, emerge as the most influential in distinguishing behavioral differences between the two versions.
% These features act as contextual descriptors that shape how perturbations take effect, since their impact depends on whether the vehicle is, for instance, driving straight, turning, or changing lanes.
% This interpretation aligns with the conditions encoded in the representative rule, thereby strengthening its validity.
% Additionally, the \emph{score} (see \cref{eq:score}) that quantifies the overall shift in feature contributions between the two versions is a positive value, indicating that the updated system tends to exhibit more \gls{mr} violations than the original.
% This insight can help engineers prioritize improvements in the updated system's handling of specific trajectory patterns, particularly in terms of spatial perception and interactions with static objects.
% Together, the example rule and the feature contribution analysis demonstrate how \emph{RuleFit} and \emph{\gls{shap}}-based analysis can jointly produce actionable insights.
% They provide a deeper understanding of the specific, contextual, and feature-driven conditions underlying behavioral differences between system versions.

In summary, the results demonstrate that \emph{RuleFit}, as used in \emph{\gls{ours}}, effectively balances accuracy and interpretability in modeling behavioral differences in \gls{mr} violations.
It produces a compact set of human-readable rules with competitive accuracy, indicating a reliable ability to capture behavioral discrepancies.
% The feature contribution analysis complements this by identifying the most influential features and shows strong alignment with the patterns captured by the extracted rules.
It also provides practical insights into whether and under which conditions the updated system exhibits more violations than the original, supporting informed decisions about real-world deployment.
This approach provides interpretable, actionable explanations of the factors driving behavioral discrepancies, thereby facilitating a more comprehensive understanding of system updates.

\Finding{
    In our interpretability approach, \emph{RuleFit} effectively balances accuracy and interpretability when modeling differences in \gls{mr} violations, producing a compact set of human-readable rules with accuracy comparable to \emph{\gls{gbdt}} and \emph{\gls{rf}}. \addedtext{Additionally, the expert assessment confirms that the generated rules are generally understandable and practically useful, particularly for guiding testing efforts and identifying relevant conditions for further investigation.}
    % The feature contribution analysis complements this by highlighting the most influential features and offering a holistic view of how system updates affect behavior.
}

\section{Discussion}\label{sec:discussion}

In this section, we discuss several important aspects of our proposed framework, including the trade-offs involved in using constraints and population initialization strategies, the applicability of our approach to other types of \glspl{as}, practical considerations\deletedtext{ regarding scalability and the significance of distinct solutions}, and potential threats to the validity of our results.

\subsection{Trade-offs in Using Constraints and the Population Initialization Strategy}

In search-based generation of realistic and effective test cases, the use of constraints and the choice of population initialization strategy are critical factors shaping the realism, severity, and quantity of the resulting scenarios.
These two factors influence the balance between producing test cases that closely reflect real-world conditions and exploring a wide range of potentially risky or severe behavioral discrepancies.
Achieving this balance inevitably involves trade-offs.
Tighter controls or heavy reliance on observed execution scenarios may improve realism but can limit the discovery of critical edge cases, whereas looser controls may enhance quantity but may reduce practical relevance.

Constraints guide the search toward regions of the solution space that are more relevant to real-world conditions.
By penalizing unrealistic or infeasible conditions, they ensure that the generated test cases remain grounded in scenarios that could occur in practice.
This targeted focus enhances the practical value of identified test cases by reducing time spent analyzing irrelevant or impossible cases.
However, collecting a diverse and representative set of execution scenarios is important to capture the breadth of conditions under which failures may occur. Excessive similarity among collected execution scenarios can cause the testing process to miss distinct failure modes that may arise under different yet equally plausible conditions.
Moreover, overly restrictive constraints can narrow the search space excessively, omitting rare but potentially critical situations that might expose severe behavioral discrepancies.
Conversely, relaxing the constraints expands exploration and increases the diversity of generated test cases, but at the cost of admitting unrealistic or invalid scenarios that require additional computational filtering.
Based on our empirical findings, a constraint level that penalizes approximately 60--70\% of candidate test cases per generation tends to strike an effective balance, maintaining both solution realism and sufficient severity.

Similarly, the population initialization strategy introduces its own set of trade-offs.
Initializing the search with execution scenarios sampled from real-world data (i.e., the \emph{\gls{ri}} strategy) accelerates convergence by starting from scenarios already representative of operational contexts.
This often yields test cases that are immediately applicable and realistic, making the approach well-suited when computational budgets or testing time are constrained.
However, the drawback is that reliance on known scenarios can bias the search toward familiar behaviors, limiting the discovery of novel or unexpected situations that have not been previously observed but may reveal critical behavioral discrepancies.
In contrast, a purely random initialization promotes broader exploration, improving the likelihood of uncovering rare, high-severity cases that challenge system safety or reliability.
The trade-off is that such cases may be less realistic and, without a sufficiently diverse and complete set of representative execution scenarios, the search may overlook critical, realistic conditions that could pose safety risks.

Selecting the right balance between constraints and initialization strategies depends on the specific objectives of the testing.
Our experimental results (see \cref{sec:rq3}) show that combining constraints with \emph{\gls{ri}} produces test cases with the highest realism. However, this approach tends to limit the discovery of the most severe behavioral discrepancies.
Therefore, practitioners must weigh the trade-offs carefully, tailoring their approach to ensure both targeted realism and adequate exploration to effectively identify potential behavioral discrepancies and safety issues.

\subsection{Applicability to Other Autonomous Systems}

Although our evaluation focuses on autonomous driving, the proposed differential, search-based testing workflow is generic and transferable to a wide range of \glspl{as}. The core components of the framework, including the search algorithms, differential analysis logic, and constraining mechanisms, operate independently of any specific \gls{as} domain. All domain assumptions are encapsulated in a small set of plug-in modules, allowing practitioners to adapt the framework by implementing only domain-specific interfaces without modifying the generic core.

In general, adapting the framework to a new domain requires the practitioners to define three domain-specific components that interface the generic core with the semantics of the target \gls{as}. As a hypothetical illustrative example, consider a robotic assembly system in which a controller coordinates a robotic arm to assemble components in an assembly line. In this setting, first, the practitioners specify a set of domain-relevant \glspl{mr} that encode expected behavioral consistency under controlled changes to the environment. For instance, an \gls{mr} \(or\) may state that increasing the assembly-line speed by \(x\) should not increase the component failure rate by more than \(y\).

Based on the selected \glspl{mr}, practitioners define a scenario representation that captures the characteristic parameters of a scenario, including those subject to change by the selected \glspl{mr}. A simple representation can be a vector whose dimensions capture key operating conditions (e.g., the speed of the assembly line and the position of items on the line). Second, the practitioners implement the scenario execution module, which runs a given scenario, typically using a simulator, and returns the metrics required by the \glspl{mr} (e.g., the failure rate).
Finally, for each specified \gls{mr}, the practitioners define how a follow-up scenario is derived from a source scenario through perturbations (e.g., increasing the speed parameter by \(x\)), and specify the corresponding output relation that maps the observed outcomes to a continuous violation severity score, i.e., extent of \gls{mr} violation (see \cref{eq:violation}).
In this example, the function \(D_{or}\) that quantifies the deviation from \(or\) could be defined based on the difference in failure rates between the source and follow-up scenarios.
For instance, such a function can be defined as \(D_{or} \coloneq r_{\text{follow-up}} - \left(1 + y\right) \times r_{\text{source}}\), where \(r_{\text{source}}\) and \(r_{\text{follow-up}}\) denote the observed failure rates in the source and follow-up scenarios, respectively, and the result of \(D_{or} > 0\) indicates an \gls{mr} violation.
Once these domain-specific modules are in place, they can be integrated into the framework, allowing the core workflow to operate directly on the specified domain without further modification.

\subsection{Practical Considerations}

In this section, we discuss \deletedtext{two}\addedtext{three} practical considerations regarding the scalability of our approach\deletedtext{ and}\addedtext{,} the significance of distinct solutions in the context of \gls{as} safety\addedtext{, and the interpretation of differences in \gls{mr} violations}.

\subsubsection{Scalability and Computational Cost}

In terms of scalability, the dominant cost in our workflow is scenario execution rather than the generic search and differential analysis logic. The overhead of population management, constraint handling, differential evaluation, and diversity management is comparatively small as these steps operate on compact scenario representations and scalar violation scores. As a result, the scalability of the overall approach is primarily driven by the throughput of the underlying simulator or testbed, the available computational resources, and the extent to which scenario executions can be distributed across parallel simulation instances within a given simulation budget. This reflects a common characteristic of search-based testing for \glspl{as}, where each additional candidate solution requires an expensive end-to-end execution to obtain safety-relevant outcomes.

Given this cost structure, search efficiency translates directly into scalability and cost-effectiveness. Since our framework uncovers severe differential violations with fewer executed scenarios than the baselines, it reduces the number of expensive simulator runs required to reach a fixed testing objective.
In addition, the framework explicitly supports parallel execution of scenarios by dispatching independent evaluations to multiple simulation agents. Since each candidate scenario can be executed and scored independently, increasing the number of concurrent simulator instances directly improves throughput and can substantially reduce wall-clock testing time within the same simulation budget.
Together, these properties make the framework well-suited to iterative development settings, where timely feedback on update-induced behavioral \deletedtext{regressions}\addedtext{degradations} is essential, and where testing throughput is constrained primarily by simulation capacity rather than by the optimization logic itself.

\subsubsection{Practical Significance of Distinct Solutions}

As we use \emph{\gls{ds}} as a primary measure of effectiveness and efficiency, it is important to clarify its practical significance in the context of \gls{as} safety.
Each distinct solution corresponds to a pair of source and follow-up scenarios that yield different outcomes in terms of \gls{mr} violations, indicating behavioral divergence between system versions.
In practice, such divergences may correspond to safety-critical situations, depending on the application domain.
For example, in autonomous driving, these differences may indicate collisions, near-misses, or traffic rule violations.
In other \glspl{as}, they may lead to mission failures, unsafe interactions, or violations of operational constraints.

Distinct solutions indicate that the testing approach is exploring different regions of the operational space, including varied environmental conditions and interactions with external agents.
A higher number of distinct solutions, particularly when they are sufficiently dissimilar, suggests that the testing process is uncovering a broader and more diverse set of critical scenarios and edge cases that could compromise system safety.
An \gls{as} that exhibits fewer unexpected behavioral differences across such a wide range of scenarios provides stronger evidence of robustness and reliability.
Conversely, the discovery of many distinct behavioral differences highlights areas where the system may require additional analysis, targeted testing, or refinement.

Overall, the number of distinct solutions exposing behavioral differences serves as a practical proxy for how thoroughly an \gls{as} has been challenged under safety-relevant conditions.
The more distinct solutions identified, the greater the insight gained into potential safety risks and system vulnerabilities.

\subsubsection{Interpretation of Differences in \gls{mr} Violations}

\annotate{Comments 1.1, 2.2,\\and 3.6}\addedtext{
The behavioral differences identified by \emph{\gls{ours}} are expressed through differences in \gls{mr} violations across system versions.
These differences are not intended as a measure of absolute system correctness, but rather as a confirmation of improvement or degradation with respect to the particular safety property specified by the \gls{mr}.
When the original version satisfies an \gls{mr} and the updated version violates it, the updated version demonstrably fails to preserve the expected behavioral relationship, thereby constituting a degradation of that property.
This does not imply that the original output is correct in an absolute sense, but rather that the updated version violates a necessary safety property that the original version satisfied.
Conversely, a transition from violation to satisfaction indicates improvement in that property, while other undetected faults may persist.
Importantly, although \gls{mr} satisfaction does not guarantee the absence of all faults, an \gls{mr} violation does confirm the presence of a safety-relevant problem, namely that the system fails to exhibit the expected behavioral relationship specified by the \gls{mr}.

Based on this interpretation, differences in \gls{mr} violations signal potential degradations warranting further investigation, or improvements that can provide confidence in the update.
The test cases and interpretable rules generated by \emph{\gls{ours}} provide practitioners with concrete, actionable evidence of how the updated system differs from its predecessor and under which conditions such divergences occur, supporting informed assessment.
Whether such differences warrant blocking a release ultimately requires human judgment, as it depends on domain-specific factors such as the severity of the observed failures and the practitioners' risk tolerance.
Accordingly, the outputs of \emph{\gls{ours}} are designed to serve as complementary evidence that practitioners can incorporate into their broader decision-making process, alongside other quality assurance activities.
}
\addedtextsec{
\subsubsection{Guidance on selecting MRs}

The selection of \glspl{mr} can influence the performance of \gls{ours}, both through the number of \glspl{mr} included in a search process and through the types of \glspl{mr} selected. Regarding the number of \glspl{mr}, using multiple \glspl{mr} in a single search process can improve search effectiveness when the \glspl{mr} share the same output relation. In \gls{ours}, a search process is guided by a fitness function that quantifies the difference in the extent of \gls{mr} violation between two system versions. Therefore, grouping several \glspl{mr} is meaningful only when their violations can be measured using the same output relation. In such cases, the search can explore a richer set of input perturbations while still relying on a coherent fitness signal. In our case study, the \glspl{mr} in \(GP1\) all expect a speed reduction, while the \glspl{mr} in \(GP2\) all expect invariance of the steering angle under their corresponding perturbations. Grouping such \glspl{mr} allows the perturbation population to include multiple types of environmental changes while evaluating them against the same expected steering behavior. This can help the search generate potentially more complex follow-up scenarios, thereby increasing the chance of identifying changes that induce behavioral differences between the two system versions. In contrast, grouping \glspl{mr} with incompatible output relations in the same search process can make the fitness signal less meaningful, since different relations may require different violation functions, thresholds, or output variables.

The type of \gls{mr} is also important. \gls{ours} is most suitable for \glspl{mr} that satisfy three practical requirements. First, the \gls{mr} should define a relation between two test inputs, namely a source scenario and a follow-up scenario. Some \glspl{mr} may involve three or more related inputs. Such \glspl{mr} cannot be directly represented as a pair consisting of one input perturbation and one output relation, and therefore they are not currently supported by our framework without modification. Second, the input perturbation associated with the \gls{mr} should be well-defined and feasible within the selected simulator or testing environment. Although \gls{ours} is not tied to a specific simulator, our case study uses Carla; therefore, the selected \glspl{mr} had to be implementable within Carla's capabilities. For example, an \gls{mr} requiring the replacement of buildings with trees would be difficult to apply in our setting if the simulator does not support such map-level modifications. More generally, whether Carla or another simulator is used, the \gls{mr}'s input perturbation must be executable through the available scenario configuration and simulation APIs.

Third, the output relation should be sufficiently well-defined to support a numerical measure of the extent of violation. This requirement is central to \gls{ours} because the search process is guided by a fitness value. If the expected output relation is ambiguous or purely qualitative, defining a numerical violation signal may become arbitrary. For example, an \gls{mr} stating that the \gls{ads} should ``drive more cautiously'' after a perturbation is difficult to use directly unless cautiousness is operationalized through measurable outputs such as speed reduction, acceleration bounds, braking intensity, minimum distance to obstacles, or lane deviation. In our framework, we consider three common types of output relations: decreasing, increasing, and invariance relations. A decreasing relation expects the value of an output variable in the follow-up scenario to be lower than its value in the source scenario by a specified threshold. An increasing relation expects the opposite, namely that the follow-up output should be higher than the source output by a specified threshold. An invariance relation requires that the output difference between the source and follow-up scenarios remain within an acceptable tolerance. For each relation type, the extent of violation can be computed by measuring how much the observed outputs deviate from the expected relation. This numerical formulation enables \gls{ours} to guide the search toward test cases that reveal stronger violation differences.

Based on these considerations, we recommend selecting \glspl{mr} that are pairwise, executable, and measurable. When multiple \glspl{mr} share the same output relation, grouping them can be beneficial because it expands the space of meaningful input perturbations while preserving a coherent fitness definition. When \glspl{mr} differ substantially in their output relations, they should preferably be evaluated in separate search processes or require a carefully designed multi-objective formulation.
}

\addedtextsec{
\subsection{Impact of Re-evaluation on False Alarms}
\label{sec:reevaluation-false-alarms}

To assess how many reported divergences may be artifacts of simulation instability, we analyzed the re-evaluation histories recorded for the evaluated solutions. We define a \emph{false positive} as a candidate divergence whose initial execution exceeds the re-evaluation threshold of 0.2, but whose value after repeated execution and median aggregation falls below that threshold. This operational definition captures false alarms caused by nondeterminism during simulation.

For each evaluated solution, we reconstructed the initial \gls{dt} divergence as the absolute difference between the initial \gls{mr} violation values observed for the reference and test versions. We then compared this value with the final divergence after re-executing threshold-exceeding cases three times
and aggregating them using the median. Table~\ref{tab:reevaluation-fp} reports the results for the primary \gls{ours} configuration.

\begin{table}[t]
\centering
\caption{Impact of re-evaluation on false positives for the
primary \gls{ours} configuration. \(Initial\) is the number of candidate alarms initially above the re-evaluation threshold, \(Retained\) is the number retained after
median aggregation, \(Dropped\) is the number dropped below the threshold or discarded as invalid, and \(FPR\) is the number of dropped candidates divided by the initial candidates.}
\label{tab:reevaluation-fp}
\begin{tabular}{lrrrrr}
\toprule
MR set & Evaluated & $Initial$ & $Retained$ & $Dropped$ & $FPR$ \\
\midrule
\(GP1\) & 1,068 & 337 & 296 & 41 & 12.2\% \\
\(GP2\) & 1,555 & 75 & 63 & 12 & 16.0\% \\
\midrule
Total & 2,623 & 412 & 359 & 53 & 12.9\% \\
\bottomrule
\end{tabular}
\end{table}

Overall, 359 of the 412 initially threshold-exceeding \gls{ours} alarms remained above the threshold after re-evaluation, while 53 were removed. Thus, the estimated false-positive rate is 12.9\% for the primary \gls{ours} configuration. Across all experimental runs, including baselines and ablations, 1,581 candidate alarms initially exceeded the threshold, 1,362 remained above it after re-evaluation, and 219 were removed,
yielding a similar rate of 13.9\%. At the version-specific \gls{mr} level, the corresponding rates were 11.0\% for \gls{ours} and 10.5\% across all configurations and baselines.
These results indicate that re-evaluation removes a significant number of unstable alarms while preserving the large majority of reproducible behavioral divergences. Of course, such percentages are expected to vary across simulators, but our approach to eliminating false positives is always applicable.

The dropped cases were mainly attributable to nondeterministic simulator outcomes, a situation we expect to be common. We therefore keep re-evaluation as part of the reporting pipeline to reduce runtime-induced false alarms before engineers inspect the generated test cases.
}

\subsection{Threats to Validity}

This section discusses potential threats to the validity of our results, following the classification into four categories: Internal, External, Construct, and Conclusion validity~\cite{wohlin2012experimentation}.

\subsubsection{Internal Validity}

Internal validity reflects how reliably a study establishes a cause-and-effect relationship between the experiment and its outcomes.
One potential threat is the sensitivity of population-based methods (i.e., \emph{\gls{ours}} and \emph{\gls{sga}}) to hyperparameter settings.
The chosen configurations may not be optimal, potentially affecting performance.
To address this, we adopted widely recommended hyperparameter values from the literature~\cite{mirjalili2018genetic}.
For parameters lacking established defaults, we selected values based on a pilot experiment.

Another internal validity concern arises from slight mismatches between the actual number of simulations executed and the allocated simulation budgets for the population-based methods.
We mitigated this issue by applying linear interpolation to align simulation budgets across different methods.
However, interpolation assumes a linear, continuous relationship between the metrics and the search budget.
This assumption may not hold if metrics exhibit abrupt or non-linear changes across budgets, which could lead to inaccuracies in the aligned results.
To reduce this risk, we validated the interpolation by analyzing metric trends and confirming that the linear assumption was reasonable within the tested budget range.

A further threat involves the correctness and reliability of the \glspl{mr} used to detect behavioral differences.
Since \glspl{mr} serve as the basis for identifying violations, any inaccuracies, such as overly strict conditions, could result in false positives or negatives.
To address this, we used well-established \glspl{mr} from prior studies and conducted a pilot experiment to validate them.
During this phase, we analyzed unexpected results and adjusted parameters to ensure the \glspl{mr} were calibrated correctly.
This refinement process helped ensure the reliability of the \glspl{mr} in the main experiments.

% A final internal validity threat arises from the accuracy of the surrogate models used in the feature contribution analysis.
% These models are intended to approximate the behavior of the \glspl{as} under test, and discrepancies between surrogate predictions and actual system behavior could bias the interpretation of feature contributions, potentially affecting the conclusions drawn in the root cause analysis.
% While directly using the actual system would avoid this issue, it is computationally expensive for large-scale studies.
% To mitigate this risk, we trained surrogate models with approximately optimal hyperparameters selected through randomized search~\cite{bergstra2012random} with cross-validation and assessed their performance using metrics such as \(R^2\) to ensure fidelity to the actual system.
% Although alternatives to \gls{gbdt}-based surrogates, such as deep neural networks, could also be employed, the specific choice of model is not the focus of this work.
% Importantly, the feature contribution analysis is model-agnostic, meaning that the overall approach remains valid as long as the surrogate provides a sufficiently accurate approximation of system behavior.

\subsubsection{External Validity}

External validity concerns the extent to which results can be generalized to other contexts, particularly their applicability beyond the specific simulation environment used in this study.
Due to the enormous effort and computational time required to prepare and run our experiments, a primary threat to external validity arises from our reliance on a single \gls{ads} (\textsc{InterFuser}) and a specific simulator (\textsc{Carla}) for the case study. For example, the substantial engineering effort required to enable seamless scenario execution using \textsc{Carla} and \textsc{InterFuser} constrained our ability to include additional platforms.
While this choice may limit generalizability, both technologies are widely recognized within the autonomous driving research community.
\textsc{Carla} is an open-source, high-fidelity simulator extensively used in academia and industry, and \textsc{InterFuser} ranked among the top-performing systems on the \textsc{Carla} \emph{Leaderboard} during the evaluation period.

% To broaden the scope of our findings, we plan to extend our approach in future work to other \glspl{as}, including mobile robotic platforms and unmanned aerial vehicles.
% Such extensions will help evaluate the robustness and limitations of our methodology across diverse real-world domains.

Another potential threat lies in the representativeness of the execution scenarios used.
If the scenarios do not sufficiently reflect the variability of real-world driving conditions, the findings may lack general applicability.
To mitigate this concern, we leveraged a variety of maps contributed by the \textsc{Carla} community, encompassing diverse urban layouts, road geometries, and traffic conditions.
We executed the \gls{ads} in these environments and systematically collected execution scenarios at five-second intervals to ensure coverage of different driving situations, including variations in traffic density, weather, and road types.
From the collected execution scenarios, we randomly sampled 800 scenarios, distributed equally across all maps, for use in our experiments.
This sampling strategy aimed to ensure broad scenario diversity and better reflect the range of conditions encountered in real-world autonomous driving, thereby improving the generalizability of our results.

\addedtextsec{
Another threat to external validity concerns the limited set of \gls{mr} groups considered in our empirical evaluation. In this study, we evaluated \gls{ours} using two \gls{mr} groups, \(GP1\) and \(GP2\), that capture common, practically relevant output-relation patterns in \gls{ads} testing. However, these groups do not cover all possible \glspl{mr} that may be proposed in the future. Therefore, our empirical results should not be interpreted as evidence that \gls{ours} will perform equally well for every possible future \gls{mr} category.
Nevertheless, \gls{ours} is not inherently restricted to \(GP1\) and \(GP2\). The method only requires that an \gls{mr} be expressible through two components: (i) an input perturbation, which defines how a source scenario is transformed into a follow-up scenario, and (ii) an output relation, which defines how the outputs of the two system versions should be compared and how the extent of violation should be measured. Once these components are specified, \gls{ours} can use the corresponding perturbation operators to generate follow-up scenarios and compute the differential fitness between system versions. Thus, extending \gls{ours} to new \glspl{mr} mainly requires implementing the associated scenario transformation and defining a measurable violation function for the expected output relation, while the core co-evolutionary search, differential fitness computation, constraint mechanism, and initialization strategy remain unchanged.
}

\subsubsection{Construct Validity}

Construct validity concerns the extent to which the measurements used in a study accurately reflect the theoretical constructs they are intended to represent.
In our work, a central concern is whether the \emph{\gls{ds}} metric appropriately captures each method's effectiveness in identifying differences in \gls{mr} violations and in exploring diverse regions of the search space.
Although this metric is designed to reflect these aspects, it may not fully encompass all dimensions of solution quality or diversity.

\subsubsection{Conclusion Validity}

Conclusion validity refers to the extent to which a research conclusion can be trusted.
A notable limitation in our experiments is the relatively small number of repetitions, i.e., 10 runs per method, due to the substantial computational costs involved.
To mitigate the risk of statistical error due to this small sample size, we report descriptive statistics along with their corresponding confidence intervals.
This approach helps convey both the central trends and the uncertainty in the results, providing a more reliable basis for interpreting our findings.

\section{Conclusion}\label{sec:conclusion}

In this paper, we presented \emph{\gls{ours}}, a novel automated testing method that integrates \acrfull{dt}, \acrfull{mt}, and advanced search-based testing to assess the overall impact of updates on \gls{as} behavior.
By leveraging \gls{mt}, \emph{\gls{ours}} effectively identifies test cases that expose behavioral discrepancies between system versions, which may lead to safety violations.
Its cooperative co-evolutionary approach efficiently explores the high-dimensional search space, directing the search toward test cases most likely to reveal significant behavioral differences.
The use of constraints and population-generation strategies favors test cases that reflect realistic operating conditions and remain applicable to practical scenarios.
The interpretability approach we propose complements this process by providing clear insights into the underlying causes of identified discrepancies, enabling developers to better understand and mitigate potential safety risks.

Our evaluation on the \textsc{Carla} simulator with the \textsc{Interfuser} \gls{ads} shows that \emph{\gls{ours}} consistently outperforms baseline methods across diverse budget and threshold settings, efficiently generating severe test cases that uncover important behavioral differences between system versions.
These findings indicate that \emph{\gls{ours}} offers an effective and efficient solution for \gls{as} differential testing in open contexts by decomposing the high-dimensional search space through cooperative co-evolution.
The realism of the generated scenarios, combined with the interpretability approach's explanatory power, makes \emph{\gls{ours}} a practical and actionable tool for developers aiming to improve the safety and reliability of evolving \glspl{as}.

\section*{Data Availability}
The replication package for our experiments, which includes the implementation of our approach, baseline methods, configuration files, questionnaire, and experimental results,
will be made available on Figshare upon acceptance of our paper.

\begin{acks}
  This work was supported by a research grant from Huawei Technologies Canada Co., Ltd,
  as well as the Canada Research Chair and Discovery Grant programs of the Natural Sciences and Engineering Research Council of Canada (NSERC).
  Lionel C. Briand's contribution was partially funded by the Research Ireland grant 13/RC/209.
\end{acks}

% \section*{Abbreviations} \small{ \printglossary[type=\acronymtype, nonumberlist] }

\bibliographystyle{ACM-Reference-Format}
\bibliography{main}

\clearpage
\appendix

\section{Experimental Results on \texorpdfstring{\(GP_2\)}{GP2}}\label{sec:gp2}

\begin{figure}[htbp]
    \centering
    \includegraphics[width=\linewidth]{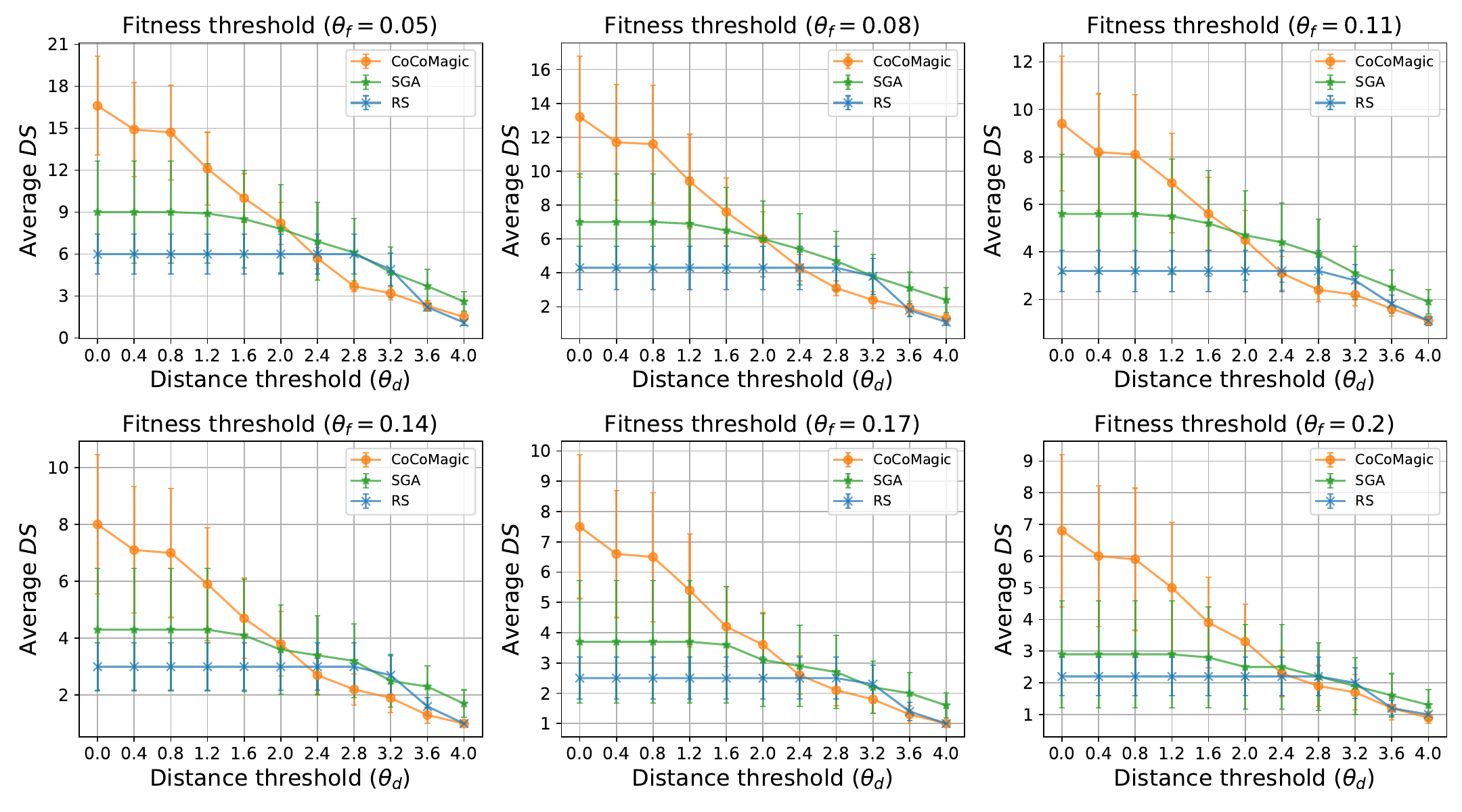}
    \caption{
        \emph{\Acrfull{ds}} vs.\ distance threshold (\(\theta_d\)) across different fitness thresholds (\(\theta_f\)).
        A higher \emph{\gls{ds}} value indicates that more distinct test cases have been discovered by the method.
        Each curve plots the average \emph{\gls{ds}} value at different \(\theta_d\) settings, under a specific fitness threshold \(\theta_f\).
        This figure reveals how each method balances the quantity and diversity of test cases as \(\theta_f\) varies.
    }\label{fig:ds_mr2}
    \Description{Distinct solutions vs.\ distance threshold across different fitness thresholds.}
\end{figure}

Based on \(GP_2\), \cref{fig:ds_mr2} compares the number of distinct solutions (\emph{\gls{ds}}) obtained by \emph{\gls{ours}}, \emph{\gls{sga}}, and \emph{\gls{rs}} over 10 runs. The x-axis shows the distance threshold \(\theta_d\), while the y-axis reports the average \emph{\gls{ds}}, with 95\% confidence intervals as error bars. Results are shown across multiple fitness thresholds \(\theta_f\).

\begin{figure}[htbp]
    \centering
    \includegraphics[width=.7\linewidth]{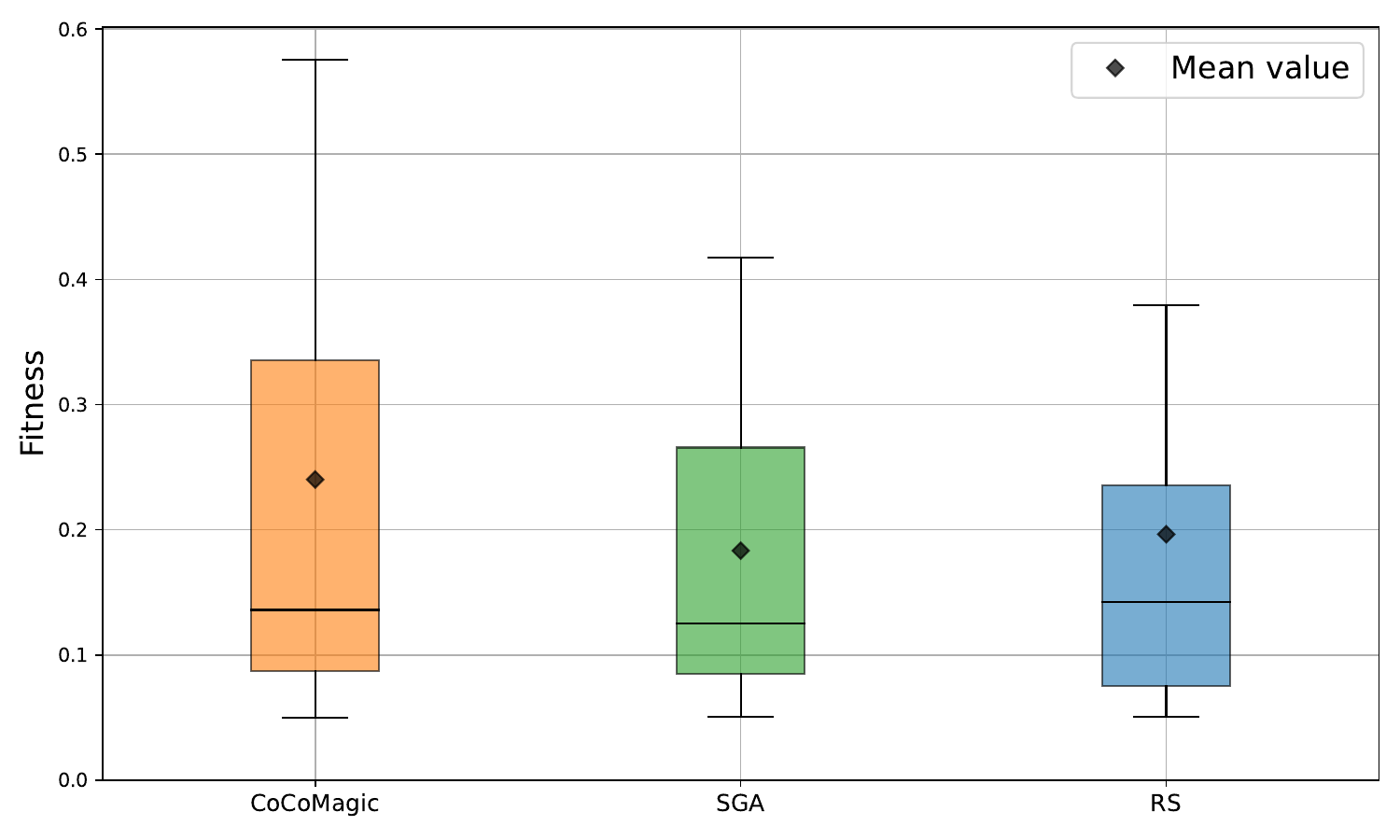}
    \caption{
        Fitness distribution of test cases generated by \emph{\gls{ours}}, \emph{\gls{sga}}, and \emph{\gls{rs}}.
        Higher fitness indicates that the discovered test cases induce greater differences in the violations exhibited by the two \glspl{ads}.
        Each box shows the fitness values for the test cases across 10 executions, providing insight into each method's peak effectiveness.
        This figure highlights the relative strength of each method in uncovering critical test cases with severe behavioral differences.
    }\label{fig:avg_fitness_mr2}
    \Description{Fitness distribution of test cases generated by CoCoMagic, SGA, and RS.}
\end{figure}

\Cref{fig:avg_fitness_mr2} presents the distribution of fitness values obtained by \emph{\gls{ours}}, \emph{\gls{sga}}, and \emph{\gls{rs}} across 10 runs. Higher median values and upper ranges indicate a stronger ability to generate test cases that induce more severe behavioral differences between the two \gls{ads} versions.

\begin{figure}[htbp]
    \centering
    \includegraphics[width=.7\linewidth]{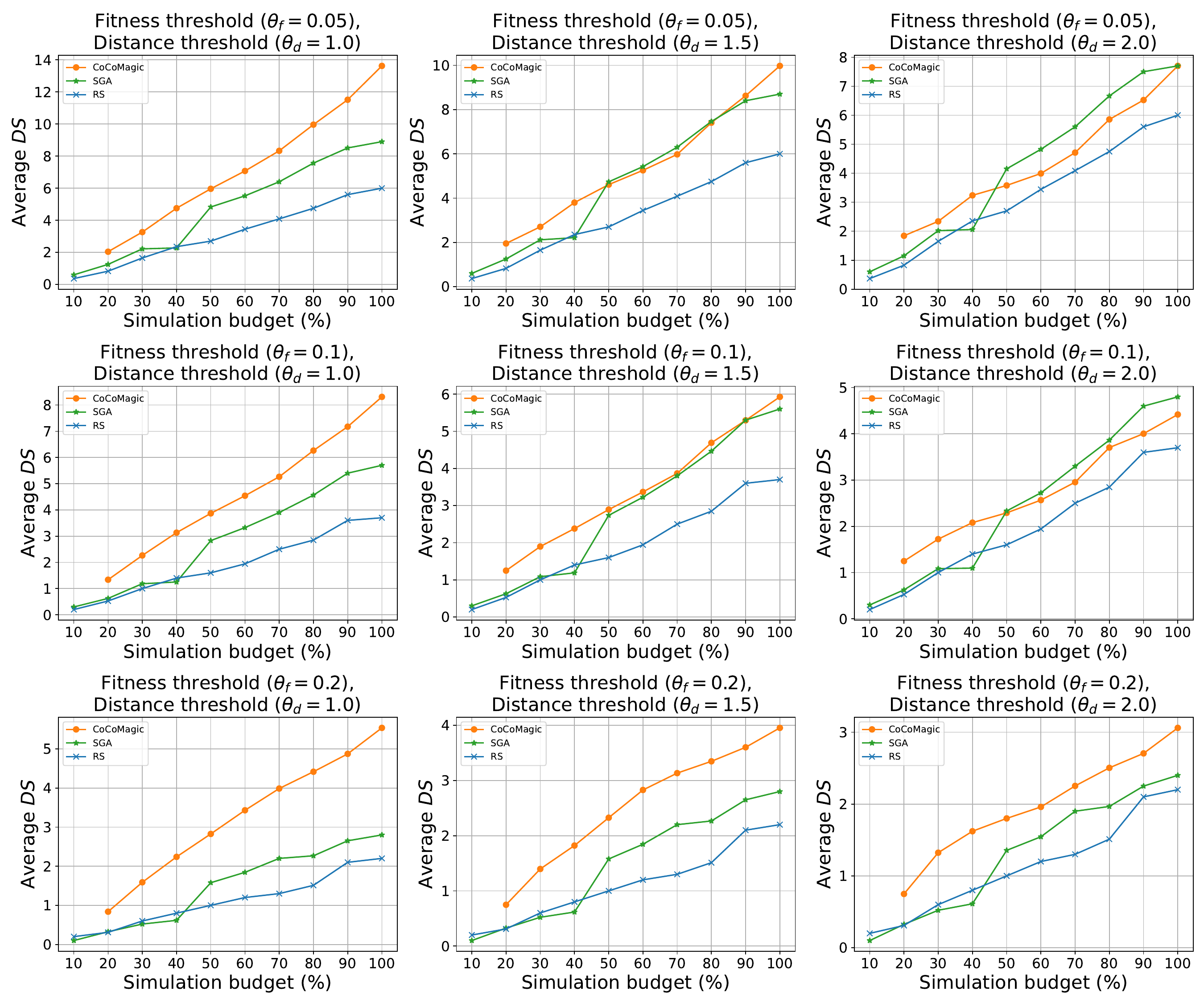}
    \caption{
        \emph{\Acrfull{ds}} progression across simulation budget levels. Each subplot illustrates how the \emph{\gls{ds}} value grows as more simulation resources are consumed, under specific combinations of fitness thresholds (\(\theta_f\)) and distance thresholds (\(\theta_d\)). The curves represent different methods, allowing a direct comparison of how quickly and effectively each approach uncovers severe and diverse behavioral discrepancies between \gls{ads} versions. Sharper increases and higher endpoints indicate greater efficiency in identifying test cases early in the search process.
    }\label{fig:ds_over_sim_mr2}
    \Description{Distinct solutions progression across simulation budget levels.}
\end{figure}

\Cref{fig:ds_over_sim_mr2} illustrates the progression of \emph{\gls{ds}} as the simulation budget increases for \(GP_2\). Each subplot corresponds to a specific combination of \(\theta_f\) and \(\theta_d\), with distinct curves for \emph{\gls{ours}}, \emph{\gls{sga}}, and \emph{\gls{rs}}. Similar to GP1, the plots suggest differences in efficiency as methods uncover severe and diverse behavioral discrepancies over time.

\begin{figure}[htbp]
    \centering
    \includegraphics[width=.41\linewidth]{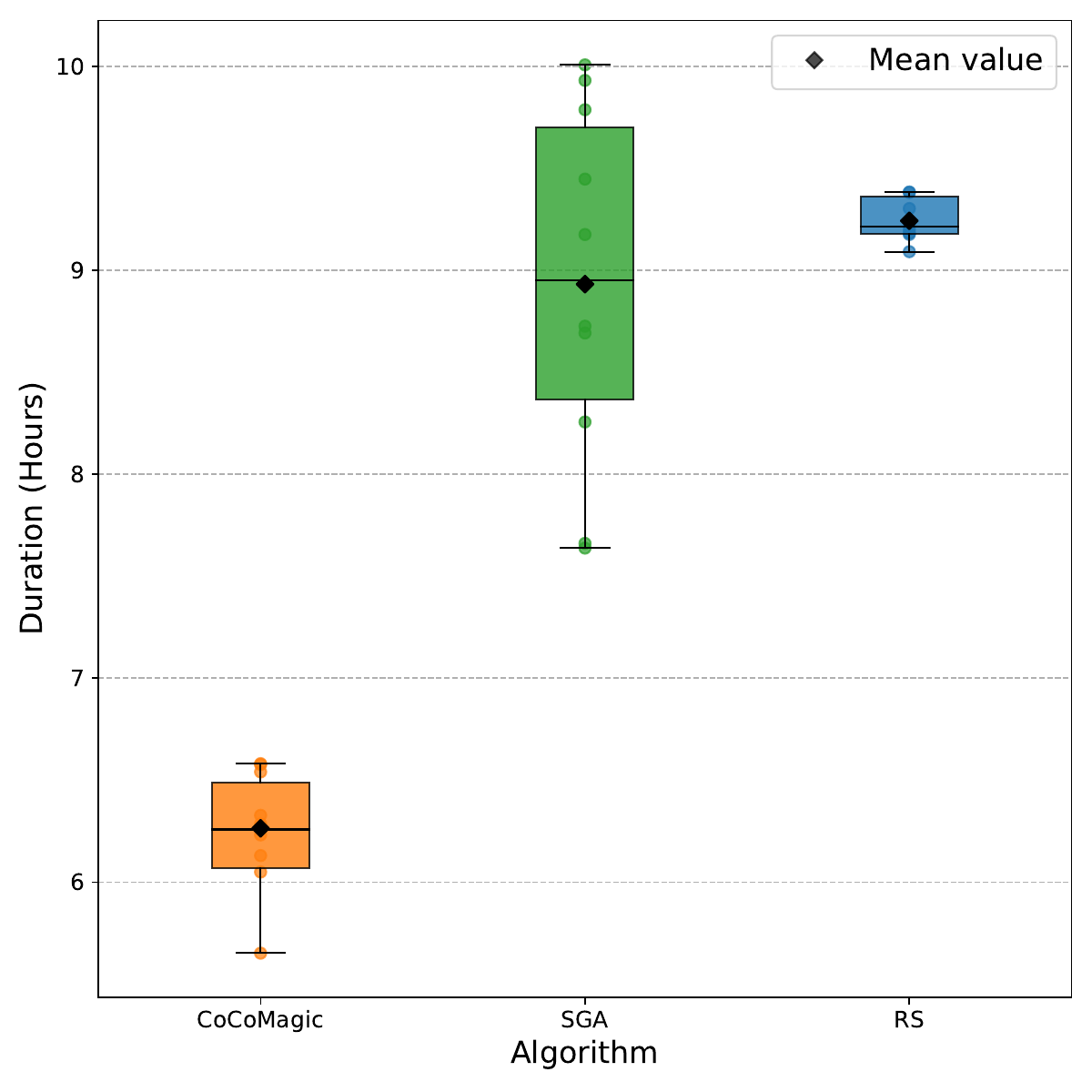}
    \caption{
        Distribution of execution times (in hours) for \emph{\gls{ours}}, \emph{\gls{sga}}, and \emph{\gls{rs}} under a fixed search budget. Lower execution time indicates greater computational efficiency in completing the search. Each box summarizes the distribution of execution times across 10 runs.
    }\label{fig:exec_time_mr2}
    \Description{Distribution of execution times (in hours) for CoCoMagic, SGA, and RS under a fixed search budget.}
\end{figure}

Finally, \Cref{fig:exec_time_mr2} presents the distribution of execution times (in hours) for \emph{\gls{ours}}, \emph{\gls{sga}}, and \emph{\gls{rs}} under a fixed search budget in \(GP_2\). Similar to \(GP_1\), each boxplot summarizes results across 10 runs, enabling a comparison of computational efficiency, where shorter times indicate faster completion.

Overall, the results for \(GP_2\) exhibit similar trends to those observed for \(GP_1\), reinforcing the consistency of our findings across different \gls{mr} groups.

\end{document}